\documentclass[12pt,a4paper,fleqn,oneside]{book}
\usepackage{graphicx}
\DeclareGraphicsExtensions{.jpg}
\usepackage[tbtags]{amsmath}
\usepackage{amssymb}
\usepackage{mathrsfs}
\usepackage{appendix}
\usepackage[pdftex,hyperindex,bookmarks=true,colorlinks=true,citecolor=red,linkcolor=blue]{hyperref}

\usepackage{fancyhdr}
\makeatletter
\def\cleardoublepage{\clearpage\if@twoside \ifodd\c@page\else%
    \hbox{}%
    \thispagestyle{empty}
    \newpage%
    \if@twocolumn\hbox{}\newpage\fi\fi\fi}
\makeatother

\newcommand{\fillblank}{\textsf}

\def\l{\lambda}

\def\s{\sigma} 

\def\th{\theta}
\def\ph{\phi} 
\def\m{\mu}

\def\r{\ref}
\def\p{\partial}
\def\no{\nonumber}
\def\f{\frac}
\def\S{\Sigma}
\def\D{\Delta}
\def\Om{\Omega}
\def\c{\cite}
\begin{document}


\setlength{\parskip}{1.0ex plus 0.5ex minus 0.5ex}
\frontmatter 

\begin{titlepage}
\begin{center}

 {\bf {\huge Thermodynamics of Black Holes: Semi-Classical Approaches \\ \vspace{0.5 cm} and Beyond}}
\vfill

\normalsize
{\Large Thesis submitted for the degree of}\\[2.2ex]
\textbf{\Large Doctor of Philosophy (Sc.)}\\[2ex]
{\Large in}\\[2ex]
\textbf{\Large Physics (Theory)}\\
{\Large by}\\[2ex]
\textbf{\Large Sujoy Kumar Modak}

\vfill

\vfill

\textbf{{\Large Department of Theoretical Sciences}}\\[2ex]
\textbf{\large{University of Calcutta}}\\
{\Large 2012}

\end{center}
\end{titlepage}


\begin{figure}[t]
  \begin{center}
    \includegraphics[width=\textwidth]{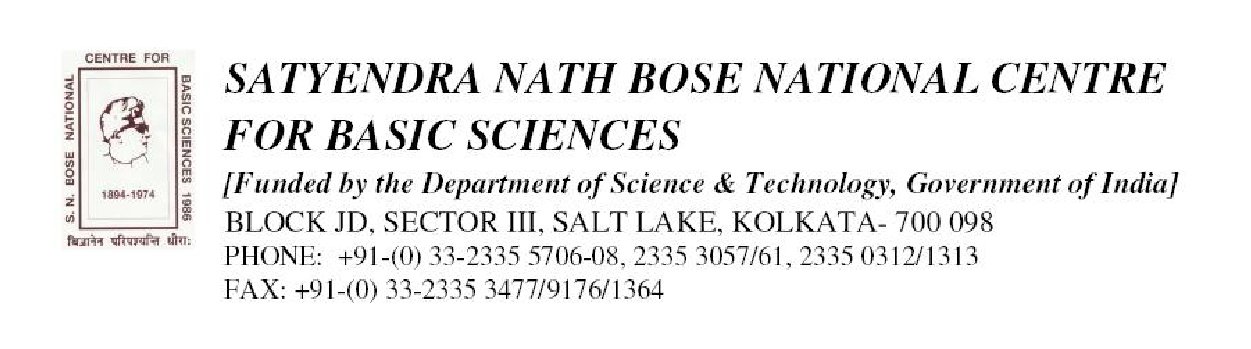}
  \end{center}

\end{figure}
\section*{\begin{center}Certificate from the supervisor\end{center}}

\thispagestyle{empty}


This is to certify that the thesis entitled \fillblank{``Thermodynamics of Black Holes: Semi-Classical Approaches and Beyond''} submitted by Mr. \fillblank{Sujoy Kumar Modak}, who got his name registered on \fillblank{March 18, 2010} for the award of \fillblank{Ph.D.~(Science)} degree in \fillblank{Physics (Theory)} of \fillblank{Calcutta University}, is absolutely based upon his own work under the supervision of \fillblank{Rabin Banerjee} at \fillblank{S. N.~Bose National Centre for Basic Sciences, Kolkata, India}, and that neither this thesis nor any part of it has been submitted for any degree/diploma or any other academic award anywhere before.

\vspace{2.0cm}

\hfill \begin{tabular}{@{}l@{}}
\fillblank{Rabin Banerjee}\\
Senior Professor\\
S. N.~Bose National Centre for Basic Sciences\\
JD Block, Sector 3, Salt Lake\\
Kolkata 700098, India\\\\
\end{tabular}


\thispagestyle{empty}
\vspace*{19.5 cm}




\hspace*{8.6 cm} $\mathcal{TO}$\\ \hspace*{10 cm}$\mathcal{MY \ PARENTS}$

\chapter*{Acknowledgements}
\thispagestyle{empty}

This thesis is an outcome of my last four and half years work at S. N. Bose National Centre, Kolkata. This journey was full of excitement and hard work. There are many individuals who have tremendous contributions in helping me to reach this milestone. I take this opportunity for thanking all of them.  

I start by expressing my sincere gratitude to Prof. Rabin Banerjee, my supervisor, for his extraordinary guidance throughout my work. He has always been open for scientific conversations. I thank him for our hours long discussions in countless occasions. His amazing enthusiasm and deep involvement among students certainly makes him very special.

I express my thanks and respect to Dr. Narendranath Mukherjee (Natu Babu), Emeritus Associate Professor, Dinhata College, for motivating me  to pursue research in physics during my undergraduate studies. Thanks to Dr. Madhusudan Ghosh and Dr. Chhanda Basu Choudhury for their academic help in my college days. I am indebted to my teachers Dr. Amitabha Mukhopadhyay, Prof. Dhruba Dasgupta, Prof. P. K. Mandal and Prof. Nikhilesh Kar of North Bengal University for their care and support. I am thankful to Prof. Subir Ghosh, ISI, Kolkata for fruitful discussions during our work. It was a nice experience to work with him.  

Dr. Amitabha Lahiri, Dr. Biswajit Chakraborty have always been supportive during my stay at S. N Bose Centre. I thank them for being very kind to me. I also thank Dr. Partha Guha, Dr. Samir Kumar Paul (theory division) for their comments during our academic interactions. It is my pleasure to appreciate fruitful academic interactions with Prof. Soumitra Sengupta, IACS.

I am thankful to Prof. Douglas Singleton, California State University, Fresno, for participating and helping me in writing a successful grant proposal to the American Physical Society. I am more than grateful to him for being an outstanding host during my research visit to USA. I also thank Prof. L. SriramKumar, Dr. S. Shankaranarayanan and Dr. Justin David for hosting me during my academic visits to HRI, IISER-TVM and IISc respectively.  

So far, the journey of my life is made cheerful by the presence of important friends. My childhood friends Indranil, Rajat, Rajdeep, Debasattam, Amit and Prasenjit have always been special to me and they have remarkable contribution in reaching me here. Thanks friends. Then during college I have had company of some bright minds, Mihir, Sandip (Chhote) and Anirban. I thank them for spending those three spectacular years together, surely best in my life. I thank my friends Biplab (Billu) and Nilanjan for always been with me during university days. 

My stay at this institute have been made comfortable and exciting by some of my very best friends. My batchmates Kapil, Prashant, Hirak, Soumojit, Ambika were always there for me. I can't forget Kapil's {\it singing} and {\it jokes}, Prashant's {\it `Gandhigiri'} and {\it `nehi'}, Hirak's {\it mouthorgan}, Soumojit's {\it `party time'} and Ambika's {\it meditation}. I thank all the members of SNB cricket and football teams for their love and support. Special thanks to Rudranil, Indrakshi, Swastika, Raka-di, Soma, Rajiv, Abhinav, Sumit for always being supportive. Mr. Sudip Garain and Amartya deserves lots of credit for arranging most of the collective student activities here. I would like to express my love and respect to all of my group mates. Seniors like Shailesh and Bibhas were always open to me for discussing every aspect of physics. Debraj has been always supportive and helpful. I can't find words for thanking him. My senior cum friend Saurav Samanta is undoubtedly the most intellectual person I have known. I enjoyed a lot while working with him. Dibakar, Biswajit and Arindam have added much energy to our group. I thank all of them.

I owe my deepest gratitude to each and every member of my family. My parents, my elder brother and sister, my brother and sister in laws, all have shown their love and support to me always. My niece Guddi and nephew Adrito have added so much happiness to my life. I thank my father and mother in laws for their care and understanding. Finally, I thank my wife Piyali and she knows why.


\chapter*{List of publications}
\thispagestyle{empty}


\begin{enumerate} \raggedright
\item \textbf{Noncommutative Schwarzschild Black Hole and Area Law.}\\
    Rabin Banerjee, Bibhas Ranjan Majhi and Sujoy Kumar Modak\\
    {\em Class.Quant.Grav.} {\bf 26}, 085010 (2009), e-Print: arXiv:0802.2176 [hep-th].

\item \textbf{Corrected Entropy of BTZ Black Hole in Tunneling Approach}\\
    Sujoy Kumar Modak\\
    {\em Phys.Lett.} {\bf B 671}, 167 (2009), e-Print: arXiv:0807.0959 [hep-th].
     
\item \textbf{Exact Differential and Corrected Area Law for Stationary Black Holes in Tunneling Method}\\
    Rabin Banerjee and Sujoy Kumar Modak \\
    {\em Jour. of High Energy Phys.} {\bf 0905}, 063 (2009), e-Print: arXiv:0903.3321 [hep-th].
    
\item \textbf{Quantum Tunneling, Blackbody Spectrum and Non-Logarithmic Entropy Correction for Lovelock Black Holes}\\
  Rabin Banerjee and Sujoy Kumar Modak \\
  {\em Jour. of High Energy Phys.} {\bf 0911}, 073 (2009), e-Print: arXiv:0908.2346 [hep-th].

\item \textbf{Voros Product, Noncommutative Schwarzschild Black Hole and Corrected Area Law}\\
  Rabin Banerjee, Sunandan Gangopadhyay and Sujoy Kumar Modak \\
  {\it Phys. Lett.} {\bf B 686}, 181 (2010), e-Print: arXiv:0911.2123 [hep-th].

\item \textbf{Killing Symmetries and Smarr Formula for Black Holes in Arbitrary Dimensions}\\
  Rabin Banerjee, Bibhas Ranjan Majhi, Sujoy Kumar Modak and Saurav Samanta\\
  {\em Phys. Rev.} {\bf D 82}, 124002 (2010), e-Print: arXiv:1007.5204 [gr-qc].

\newpage

\item \textbf{Glassy Phase Transition and Stability in Black Holes}\\
  Rabin Banerjee, Sujoy Kumar Modak and Saurav Samanta \\
  {\em Eur. Phys. Jour.} {\bf C}, 70, 317 (2010) e-Print: arXiv:1002.0466 [hep-th].

\item \textbf{Second Order Phase Transition and Thermodynamic Geometry of Kerr-AdS Black Hole} \\
  Rabin Banerjee, Sujoy Kumar Modak and Saurav Samanta \\
  {\em Phys. Rev.} {\bf D 84}, 064024 (2011), e-Print:arXiv:1005.4832 [hep-th]. 
  
\item \textbf{Effective Values of Komar Conserved Quantities and Their Applications}\\
  Sujoy Kumar Modak and Saurav Samanta \\
  {\em Int. J. Theo. Phys.} {\bf 51}, 1416 (2012) , e-Print: arXiv:1006.3445 [gr-qc].
  
\item \textbf{Classical Oscillator with Position-dependent Mass in a Complex Domain}\\
  Subir Ghosh and Sujoy Kumar Modak \\
 {\em Phys. Lett.} {\bf A 373}, 1212 (2009) [arXiv:0803.2531 [math-ph]].\\
 
\item \textbf{Thermodynamics of Phase Transitions in AdS Black Holes}\\
  Rabin Banerjee, Sujoy Kumar Modak and Dibakar Roychowdhuri\\
  {\it Communicated}, e-Print:arXiv:1106.3877 [gr-qc].   
   
\end{enumerate}
This thesis is based on the papers [1]-[9] whose reprints are attached at the end of the thesis.


\chapter*{}
\pagenumbering{roman}
\thispagestyle{empty}

\begin{center}
\uppercase{Thermodynamics of black holes: semi-classical approaches and beyond}
\end{center}



\tableofcontents



\mainmatter
\chapter{\label{chap:introduction} Introduction}
  
\section{Overview}
Gravitation is believed to be one of the four fundamental forces in Nature. And so far it is also one of the interesting but difficult subjects to understand. While the Standard Model is able to describe other three forces, namely, electromagnetic, strong and weak forces, it does not include gravitation. As a consequence, not surprisingly, lot of efforts have been devoted to understand gravitation both within and beyond the general theory of relativity. 

In general theory of relativity (GTR) of gravitation, a measure of the effect of gravity in a space-time is given by `curvature'. If there is matter, it makes the space-time `curved' and thus induces gravity. GTR is a classical theory of gravity and it is very successful to understand the large scale structure of space-time. However there remain two intriguing issues in GTR. One is the appearance of curvature singularity and the other is the presence of space-time horizons. The concept of space-time curvature is quantified by using the Riemann curvature tensor ($R^{\alpha\beta\mu\nu}$). In a region of space-time where gravity is negligible the norm of this tensor vanishes and space-time is considered to be Minkwoskian. Near a strong gravitating object (like massive star or black hole) its norm is non-zero and it gives us an estimate how strong is that gravity. However if at any point this norm diverges to infinity the geometry does not remain smooth and that point is considered to be a space-time singularity. In fact at the centre of a black hole one finds this singularity. As a consequence such a space-time is geodesically incomplete. Classical GTR cannot provide a satisfactory explanation of such a behaviour. 

Perhaps these are the reasons why physicists now look beyond the classical GTR to solve the above mentioned problems. There are three approaches which are widely studied. These are based on semi-classical, loop quantum and string theory methods. In this thesis I shall discuss semi-classical approaches in which gravity is treated inherently classical but the fields moving in the background are considered to be quantum in nature. This provides some very important new insights about spacetime horizons like black hole (event) and cosmic horizons. For example black holes are identified as perfect thermodynamical systems and they are associated with an entropy and a physical temperature. Black holes in Einstein gravity have an entropy equal to one fourth of its horizon area ($A_H$) and its temperature is proportional to the surface gravity ($\kappa$) at the event horizon. These are respectively given by, {\footnote{here $c$ is the speed of light in vacuum, $k_{B}$ is the Boltzman constant, $G$ is the gravitational constant and $\hbar$ is the reduced Planck constant.}}
\begin{eqnarray}
S=\frac{k_{B}c^3A_{H}}{4G\hbar}=\frac{4\pi G k_{B} M^2}{\hbar c}
\label{aby4}
\end{eqnarray}
and 
\begin{eqnarray}
T=\frac{\hbar c \kappa}{2\pi k_{B}}=\frac{\hbar c^3}{8\pi G Mk_{B}}.
\label{tkappa}
\end{eqnarray}
where in the second equality we have written the entropy and temperature for the Schwarzschild black hole which is the static, spherically symmetric solution of Einstein equation. The reason behind writing them explicitly is the following. Note that, in the above expressions for entropy and temperature (on the r.h.s of second equality), remarkably, three fundamental constants $G,c,\hbar$ are appearing simultaneously. These constants respectively represent gravity, relativity and quantum mechanics which are also believed to be the ingredients of quantum gravity. Thus, interestingly all these theories meet at a single platform of black hole thermodynamics and by studying this subject it is expected to gain further insights on quantum theory for gravitation. In some sense black holes might play an important role for this which is somewhat similar to what atoms did before the advent of quantum mechanics.  Moreover, one of the strong aspects of this approach is that irrespective of the microscopic or quantum theory of gravity the results found here would still be valid at a relatively low energy scale (than the Planck scale) where quantum gravity is very weak.

The thermodynamical aspect of black holes was first noticed when Bekenstein argued in favour of black hole entropy based on simple aspects of standard thermodynamics \cite{Beken1}-\cite{Beken3}. His point was that the entropy of the universe cannot decrease due to the capture of any object by black holes as that would violate the laws of usual thermodynamics. For making the total entropy of the universe at least unchanged, a black hole should gain same amount of entropy which is lost from the rest of the universe. This idea was supported by the work of Bardeen, Carter, Hawking \cite{Bardeen} which revealed that the ``laws of black hole mechanics'' were closely similar to the ``laws of thermodynamics'' provided black holes have some temperature. Later, Hawking showed that black holes radiate with a temperature (\ref{tkappa}) \cite{hk1}-\cite{hk3}, usually known as the Hawking temperature and thus the above ideas were given a solid mathematical ground. Finally, the analogy between the ``first law of black hole mechanics'' with the first law of thermodynamics gave a specific formula for the black hole entropy as one fourth of its horizon area, known as the famous Bekenstein-Hawking area law (\ref{aby4}). This can also be derived by using the Wald entropy formula \cite{viyer}, \cite{wald} involving the diffeomorphism-invariant Lagrangian constructed through a combination of curvature invariants. 

After Hawking's original derivation, several approaches were advanced to deduce the semi-classical Hawking temperature. Amongst them two most intuitive approaches were the tunneling \cite{Paddy, ng1} and anomaly \cite{chris, robin, bkk, bkk1} mechanisms. Temperature can be obtained in two different ways through the tunneling method. These are null geodesic method \cite{ng1} and Hamilton-Jacobi method \cite{Paddy}. Furthermore, the tunneling mechanism was used to go beyond the semi-classical approximation to find the corrections to the semi-classical Hawking temperature \cite{Majhibeyond}. Also it was powerful enough to reproduce the blackbody spectrum of radiation \cite{Majhiflux}.  

Although till now there is no microscopic description of black hole entropy, several approaches have shown that the semi-classical Bekenstein-Hawking entropy undergoes corrections. These approaches are mainly based on field theory \cite{Fursaev}, quantum geometry \cite{Partha}, statistical mechanics \cite{Das}, Cardy formula \cite{Carlip}, brick wall method \cite{Hooft} and tunneling method \cite{Majhitrace}. These corrections are very important since they play the dominant role in Planck scale where the effects of quantum gravity cannot be ignored. Therefore these corrections are an indirect way to understand the inherent features of the fundamental theory of gravity.

One of the important thermodynamic properties that black holes exhibit is phase transition. Phase transitions in black holes was first observed by Hawking and Page \cite{hp1} in Schwarzschild Anti-de Sitter space-time. It was shown that above a certain (minimum) temperature the thermal radiation in AdS space can collapse to form a black hole. If the mass of the black hole is low it is unstable and it absorbs radiation from thermal AdS space and increases its own mass. When the mass value reaches a critical point a phase transition takes place which makes the black hole thermodynamically stable. Thereafter several works \cite{hp2}-\cite{hp14} highlight phase transition in other black holes through various approaches.

\section{Outline of the thesis}
This thesis is based on my works \cite{mod1}-\cite{mod9} which are focussed to study various aspects of black hole physics. This includes the study of black holes from various viewpoints. Let us now mention some notable facets of these studies. In two of my papers \cite{mod6}, \cite{mod9} we look into the issue of generalised Smarr formula for arbitrary dimensional black holes in Einstein-Maxwell gravity. We not only derive this formula for these black holes, but also demonstrate that such a formula can be expressed in the form of a dimension independent identity $K_{\chi^{\mu}}=2ST$ (where the l.h.s is the Komar conserved charge corresponding to the null Killing vector $\chi^{\mu}$ and in the r.h.s $S,~T$ are the semi-classical entropy and temperature of a black hole) defined at the black hole event horizon. We also highlight the role of exact differentials in computations involving black hole thermodynamics \cite{mod2}, \cite{mod3}. In fact results like the first law of black hole thermodynamics and semi-classical entropy are obtained without using the laws of black hole mechanics, as is usually done. The blackbody radiation spectrum for higher dimensional black holes is also computed by using a density matrix technique of tunneling mechanism by considering both event and cosmological horizons \cite{mod4}. We also provide the modifications to the semi-classical Hawking temperature and Bekenstein-Hawking entropy due to various effects \cite{mod1}, \cite{mod2}, \cite{mod3}, \cite{mod5}. These modifications are mainly found due to higher order (in $\hbar$) effects to the WKB ansatz used for the quantum tunneling formalism \cite{mod2}, \cite{mod3}, \cite{mod5} and non-commutative effects \cite{mod1},\cite{mod5}.  Finally, in \cite{mod7}, \cite{mod8} we discuss phase transition phenomena in black holes. We formulate a new scheme based on Clausius-Clapeyron and Ehrenfest's equations to exhibit and classify phase transitions in black holes in analogy to what is done in standard thermodynamics.
 
The summary of each chapter of this thesis is given below.

In chapter \ref{chap:thermodynamic signature},  we calculate the effective Komar conserved quantities for the Kerr-Newman and arbitrary dimensional charged
Myers-Perry spacetime. The results thus obtained give an effective value of the mass and angular momentum of these charged and rotating black holes which are distinct from their respective asymptotic expressions. Using these results, at the event horizon, we derive a new identity $K_{\chi^{\mu}} = 2ST$ where the left hand side is the Komar conserved quantity corresponding to the null Killing vector $\chi^{\mu}$ while in the right hand side $S,~T$ are the black hole entropy and Hawking temperature. From this identity we derive the generalized Smarr formula connecting the macroscopic parameters $(M, J, Q)$ of the black hole with its surface gravity and horizon area. The consistency of this new formula is established by an independent algebraic approach. Finally, we provide the first law of black hole mechanics for these spacetimes. 

In chapter \ref{chap:hawking-effect}, we adopt the tunneling method for discussing Hawking effect. Mainly we follow two variants, one is based on the principle of detailed balance and the other uses a density matrix type analysis. While the first method identifies the Hawking temperature, it does not say anything about the radiation spectrum. The second method directly provides the radiation spectrum with the known temperature.

We consider the tunneling of both scalar particles and fermions to compute Hawking temperature. The idea is to solve the Klein-Gordon or Dirac equation in the curved spacetime background by using a WKB type approach. The solutions thus found correspond to ingoing and outgoing modes. These modes are used to calculate the respective absorption/transmission probabilities. Finally, by using the principle of detailed balance, the Hawking temperature is identified. Going beyond the semi-classical limit we also consider higher order terms (in $\hbar$) in WKB ansatz. It generates some higher order corrections to semi-classical temperature with some unknown coefficients. We discuss more about these coefficients in the next chapter (\ref{chap-entropy}).

Then using a density matrix method, we show that black holes emit scalar particles and fermions with a perfect blackbody spectrum. The temperature is given by the semi-classical Hawking temperature. This result is valid for both black hole (event) horizon and cosmological horizon of arbitrary dimensional static black holes. In the presence of higher order corrections to the WKB ansatz the modified radiation spectrum retains its blackbody nature. However,  the temperature receives some higher order in $\hbar$ corrections. This  corrected temperature corresponding to the modified spectrum  yields the semi-classical Hawking temperature at the lowest order (in $\hbar$).
   
Chapter \ref{chap-entropy} gives a new and conceptually simple approach to obtain the "first law of black hole thermodynamics". It is based on a basic thermodynamical property that entropy ($S$) for any stationary black hole is a state function implying that $dS$ must be an exact differential. Using this property we obtain some conditions which are analogous to Maxwell's relations in ordinary thermodynamics. From these conditions we explicitly calculate the semiclassical Bekenstein-Hawking entropy, considering the most general metric represented by the Kerr-Newman spacetime in $3+1$ dimensions and BTZ spacetime in 2+1 dimensions. We then extend our method to find the corrected entropy of stationary black holes. For that we use the expressions of the corrected Hawking temperature found in chapter \ref{chap:hawking-effect} using tunneling method beyond the semi-classical approximation. Using this corrected Hawking temperature we compute the corrected entropy, based on properties of exact differentials. By using an infinitesimal scale transformation to the metric the connection of the coefficient of the leading (logarithmic) correction with the trace anomaly of the stress tensor is established . We explicitly calculate this coefficient for stationary black holes with various metrics, emphasising the role of Komar integrals discussed in chapter \ref{chap:thermodynamic signature}.   

In the next chapter (\ref{chap-leading-correction}) we study thermodynamics of noncommutativity-inspired Schwarzschild black hole. Unlike the standard noncommutative (NC) geometry where a NC effect is introduced through the coordinate sector, here, it is introduced through the matter sector. This approach was first worked out in \cite{nico5,Smail}. The black hole solution is found by solving the Einstein equation, in which the left hand side (coordinate sector) is kept unchanged, but on the right hand side the energy momentum tensor represents a NC fluid with inhomogeneous pressure. The static, spherically symmetric black hole solution is referred as NC Schwarzschild black hole. One important property of this black hole solution is that it is free from singularity, {\it i.e.}, the scalar curvature does not diverge at $r=0$, instead the black hole has a de-Sitter core for small $r$. We further discuss these issues in chapter \ref{chap-leading-correction}. Then we study thermodynamic properties, like temperature, entropy/area law of this black hole. These expressions are found to be modified from the usual Schwarzschild spacetime due to the presence of NC parameter ($\theta$). Interestingly, Hawking temperature does not diverge as mass tends to zero, rather, it vanishes for a minimum remnant mass. The semi-classical area law is computed by using a graphical analysis for all orders in $\theta$. In the lowest order of $\theta$ we have the well known area law, but, from the next to leading order correction the functional form of this law is violated. There exist some extra terms like the exponential and error functions in addition of the $\frac{A}{4}$ term. Finally, we also find a general expression of entropy where both noncommutative and quantum corrections are present up to the leading order. Using the method used in chapter \ref{chap-entropy} we also calculate the coefficient of the leading (log) correction to the entropy which is found to be modified from the standard Schwarzschild example. All the results, in an appropriate limit, agree with the commutative counterpart. 
   
In chapter \ref{chap-phase-transition} we study phase transitions in black holes. We exploit fundamental concepts of thermodynamics to identify and classify black hole phase transitions. We use the scheme based on Clausius-Clapeyron and Ehrenfest relations which are satisfied for first and second order phase transitions respectively. We first derive these relations for black holes. Then we systematically study rotating Kerr black holes defined in anti-deSitter (AdS) spacetime. 

For the Kerr-AdS black holes we identify a phase transition from the smaller to higher mass branches. We do not find any discontinuity in the first order derivatives of Gibbs energy ($G$) (entropy or angular momentum) and thus the chance of a first order phase transition is ruled out. However, all second order derivatives of $G$ (specific heat, volume expansion coefficient and compressibility) are found to be discontinuous and infinitely diverging at the phase transition point. This opens up the possibility of a second order phase transition. To confirm this, we check the validity of Ehrenfest's relations. It is found that despite the infinite divergences in individual quantities both Ehrenfest relations hold and consequently this transition is indeed second order. Then we use thermodynamic state space geometry approach and study phase transitions in Kerr-AdS black holes. This particular approach is pioneered by the works of Wienhold \cite{Weinhold} and Ruppeiner \cite{Ruppeiner:1983zz, Ruppeiner:1995zz}. We also build a close connection between the results found in this approach with those found by using the Ehrenfest's scheme.

Finally, chapter \ref{chap:conclusions} contains conclusions and future outlook.

\chapter{\label{chap:thermodynamic signature}Killing symmetries, conserved charges and generalized Smarr formula}

A spacetime encoded with symmetries is known to have conserved physical entities. These symmetries are characterised by Killing vectors ($\xi_{\mu}$) which satisfy the Killing equation $\nabla_{\mu}\xi_{\nu}+\nabla_{\nu}\xi_{\mu}=0$. For a given spacetime metric one needs to solve this equation to obtain all independent vector fields which obey this equation. Then the integral curves on these vector fields define symmetry directions of the spacetime. Also depending upon the number of these vectors one can associate an equal number of conserved quantities. These are known as Komar conserved quantities \cite{komar,komar2}. In this chapter we shall discuss the importance of such conserved charges and deduce the mass formula of arbitrary dimensional charged, rotating black holes. We shall also highlight a dimension independent interpretation of the mass formula for such spacetimes. However before addressing these issues we provide a brief introduction to Komar conserved charges from the existing literature. This would also serve as a motivation for the subsequent analysis. 

For an axially symmetric stationary spacetime one has multiple Killing vectors. Any $N+1$ dimensional spacetime with one timelike and one spacelike Killing vector, given by $\xi^{\mu}_{(t)}=(1,0,0,0\cdot\cdot\cdot)$ and $\xi^{\mu}_{(\phi)}=(0,0,0,1,\cdot\cdot\cdot)$, has the following two conserved quantities \cite{komar,komar2}
\begin{eqnarray}
K_{\xi^{\mu}_{(t)}}=-\frac{1}{8\pi}\int_{\partial\Sigma} \xi^{\mu;\nu}_{(t)}d^{N-1}\Sigma_{\mu\nu}
\label{meff}
\end{eqnarray}
and
\begin{eqnarray}
K_{\xi^{\mu}_{(\phi)}}=-\frac{1}{8\pi}\int_{\partial\Sigma} \xi^{\mu;\nu}_{(\phi)}d^{N-1}\Sigma_{\mu\nu}.
\label{jeff}
\end{eqnarray}
respectively. Here $\Sigma$ is a $N$ dimensional spatial volume element and its boundary $\partial\Sigma$, where above expressions are defined, is $N-1$ dimensional spatial hypersurface. These expressions, when evaluated at asymptotic infinity, give distinct black hole parameters (mass $M$ or angular momentum $J$) upto some normalisation constant. 

For example in the case of (3+1) dimensional Kerr-Newman black hole where the results at asymptotic infinity are $\displaystyle\lim_{r\rightarrow\infty}K_{\xi^{\mu}_{(t)}}=M$ (the normalisation $(8\pi)^{-1}$ is chosen so that the result matches with the Newtonian mass in the weak field approximation) and $\displaystyle\lim_{r\rightarrow\infty}K_{\xi^{\mu}_{(\phi)}}=-2J$. Based on these results one can now define a Komar mass ($M$) and Komar angular momentum ($J$) for (3+1) dimensional Kerr-Newman black hole in the following way
\begin{eqnarray}
M_{(3+1)}=-\frac{1}{8\pi }\displaystyle\int_{\partial\Sigma(r\rightarrow\infty)} \xi^{\mu;\nu}_{(t)}d^{2}\Sigma_{\mu\nu}
\label{meffkn}
\end{eqnarray}
and
\begin{eqnarray}
J_{(3+1)}=\frac{1}{16\pi }\displaystyle\int_{\partial\Sigma(r\rightarrow\infty)} \xi^{\mu;\nu}_{(\phi)}d^{2}\Sigma_{\mu\nu}.
\label{jeffkn}
\end{eqnarray}
These definitions are often found in text books \cite{carrol,wald}. However, note that while the normalisations between (\ref{meff}) and (\ref{meffkn}) do not differ they are not same for (\ref{jeff}) and (\ref{jeffkn}). This difference is required for the correct identification of the black hole angular momentum ($J$). 

Incidentally, when we are dealing with higher dimensions the normalisation for the black hole mass which should match with the Newtonian mass in the weak gravity limit also needs to be changed from its usual value as appears in (\ref{meff}). These identifications finally define the Komar mass and angular momentum for the $N+1$ dimensional charged, rotating Myers-Perry black hole \cite{myers},
\begin{eqnarray}
M_{(N+1)}=-\frac{(N-1)}{16\pi (N-2)}\int_{\partial\Sigma(r\rightarrow\infty)} \xi^{\mu;\nu}_{(t)}d^{N-1}\Sigma_{\mu\nu}
\label{meffknh}
\end{eqnarray}
which reduces to (\ref{meffkn}) for $N=3$ and,
\begin{eqnarray}
J_{(N+1)}=\frac{1}{16\pi }\int_{\partial\Sigma(r\rightarrow\infty)} \xi^{\mu;\nu}_{(\phi)}d^{N-1}\Sigma_{\mu\nu}.
\label{jeffknh}
\end{eqnarray}  

Thus we see that the Komar integrals describe physical entities at asymptotic infinity. But to evaluate these integrals considering this limit is not necessary. In fact one can calculate the respective values of these integrals without using such approximation, i.e., by staying at any arbitrary distance outside the black hole event horizon. Such a calculation does not suffer  any conceptual problem. As it would be discussed in the latter parts of this chapter, the effective values of Komar integrals are indeed very useful to derive some key results of black hole physics with some new insights.

\section{\label{rnhd}Algebraic approach to Smarr formula}
 In 1973, using an algebraic approach, Smarr \cite{smarr} showed that the mass of Kerr-Newman black hole can be written as a sum of three terms; surface energy, rotational energy and electromagnetic energy. In this section we shall generalise that approach to derive a mass formula for $N+1$ dimensional Reissner-Nordstrom (R-N) black holes. We also point out the problem of generalising this approach for higher dimensional rotating spacetimes.

The metric for the $N+1$ dimensional R-N black hole is easily found by setting $a=0$ in (\ref{metric}). Similarly all other physical entities like the horizon radius, horizon area, Hawking temperature and black hole entropy for this spacetime are obtained from the respective expressions of Appendix \ref{appendix2A1}. 

The horizon condition, $\Delta|_{r=r_+}=0$ (for $a=0$), follows from (\ref{delta}),
\begin{eqnarray}
r_+^{N-2}=\frac{m}{2}+\sqrt{\frac{m^2}{4}-q^2}
\label{rnhor}
\end{eqnarray}
Substituting this result in the expression of horizon area (\ref{hora1}) (with $a=0$), we find 
\begin{eqnarray}
A_{H}=\frac{4\pi \bar{A}_{N-3}}{N-2}\left(\frac{m}{2}+\sqrt{\frac{m^2}{4}-q^2}\right)^{\f{N-1}{N-2}}.
\label{rnarea}
\end{eqnarray}
Now inverting this relation we obtain $m$ in terms of $A_{H}$ and $q$, given by
\begin{eqnarray}
m=\left(\frac{(N-2)A_{H}}{4\pi \bar{A}_{N-3}}\right)^{\frac{N-2}{N-1}}+q^2\left(\frac{4\pi \bar{A}_{N-3}}{(N-2)A_{H}}\right)^{\frac{N-2}{N-1}}.
\label{rnmass}
\end{eqnarray} 
The differential form of this relation yields
\begin{eqnarray}
dm &=&t~dA_{H}+\tilde\ph ~dq\label{rnmadss}
\end{eqnarray}
where,
\begin{eqnarray}
t &=& \f{\p m}{\p A_H} = \f{(N-2)}{(N-1)\bar{A}_{N-3}}T\label{t}\\
\tilde\ph &=& \f{\p m}{\p q} = \f{2q}{(N-2)r_+^{N-2}}\label{p1}
\end{eqnarray}
and the Hawking temperature $T$ is defined in (\r{temp}) with $a=0$. From (\r{rnmass}) we find that $m$ is a homogeneous function of degree $\frac{N-2}{N-1}$ in ($A_{H}$, $q^\f{N-1}{N-2}$) respectively. Therefore making use of Euler's theorem on homogeneous functions it follows that
\begin{eqnarray}
m=\frac{(N-1)}{(N-2)}~tA_{H}+\tilde\ph~ q.
\label{rnsmarr1}
\end{eqnarray}
Now let us express this relation in terms of physical mass ($M$), charge ($Q$), temperature ($T$) and electric potential ($\Phi$) by using (\r{m}), (\r{q}), (\r{temp}) and (\r{phi}) respectively. Here $\Phi$ is the timelike part of the gauge potential $A_{\mu}=-\f{Q}{(N-2)r^{N-2}}(1,0,0,0,\cdot\cdot\cdot)$ (also follows from (\r{phi}) with $a=0$) and this can be calculated as follows. The $N+1$ dimensional R-N black hole being spherically symmetric, the global timelike Killing vector is $\xi^{\mu}=(1,0,0,0,\cdot\cdot\cdot)$. Therefore the scalar potential is found to be
\begin{eqnarray}
\Phi=\xi^{\mu}A_{\mu}\big|_{r=\infty}-\xi^{\mu}A_{\mu}\big|_{r=r_+}=0-\left(-\f{Q}{(N-2)r_+^{N-2}}\right)=\f{Q}{(N-2)r_+^{N-2}}.
\label{rnpot}
\end{eqnarray}
Now exploiting (\r{m}), (\r{q}), (\r{temp})and (\r{rnpot}) and using (\r{a}), (\r{an-3}) one gets the desired result for the Smarr formula from (\ref{rnsmarr1}), given by
\begin{eqnarray}
M-Q\Phi=\f{N-1}{N-2}\f{\kappa A_{H}}{8\pi}.
\label{rnsmarr}
\end{eqnarray}
This is a new result. Note that for $Q=0$ it yields the correct result for the $N+1$ dimensional Schwarzschild black hole \cite{myers} and for $N=3$ it reduces to the well-known result for 3+1 dimensional R-N black hole \c{poisson}. 

Although the above algebraic approach successfully yields Smarr formula for the R-N black holes, unfortunately this cannot be generalized to the rotating case. The reason for this is that we cannot write the horizon radius as a function of mass and angular momentum, mimicking (\r{rnhor}). A similar problem arises for the charged, rotating case (\r{metric}). In the subsequent part of this chapter we develop a new technique, using the concept of Killing symmetries and effective Komar conserved quantities, which solves this problem.

\section{\label{kintkn}Effective values of Komar conserved charges for the Kerr-Newman black hole}
Before addressing the issue of higher dimensions in this section we evaluate the Komar integrals for the Kerr-Newman black holes at a finite radial distance outside the event horizon. This will enable us to pursue an analogous treatment for the charged Myers-Perry spacetime in arbitrary dimensions.



Kerr-Newman spacetime being axially symmetric it has two Killing vectors and corresponding conserved charges are given by (\ref{meff}) and (\ref{jeff}). In a coordinate free notation these are given by \cite{cohen2,wald,dadhich1,dadhich2}
\begin{eqnarray}
K_{\xi^{\mu}_{(t)}}=-\frac{1}{8\pi}\int_{\partial \Sigma}{}^*d\sigma,
\label{eeff}
\end{eqnarray}
and
\begin{eqnarray}
K_{\xi^{\mu}_{(\phi)}}=-\frac{1}{8\pi}\int_{\partial \Sigma}{}^*d\eta
\label{peff}
\end{eqnarray}
respectively, where the timelike and spacelike Killing one forms are defined as
\begin{eqnarray}
\sigma =\xi_{(t)\mu}dx^{\mu}=g_{0\mu}dx^{\mu}=g_{00}(r,\theta)dt+g_{03}(r,\theta)d\ph \label{tonf}\\
\eta =\xi_{(\phi)\mu}dx^{\mu}= g_{\mu 3}dx^{\mu} = g_{03}(r,\theta)dt+g_{33}(r,\theta)d\phi.
\label{111}
\end{eqnarray}

For the integral involving Komar mass (\ref{eeff}) we provide the final result, as calculated in \cite{cohen2}, in the following form
\begin{eqnarray}
K_{\xi^{\mu}_{(t)}}= M-\frac{Q^2}{2r}-\frac{Q^2(r^2+a^2)}{2ar^2}\tan^{-1}{\frac{a}{r}}.
\label{ms}
\end{eqnarray}
Let us now proceed with the evaluation of (\ref{peff}) without using any asymptotic approximation. Differentiating (\ref{111}) we find the following two form
\begin{eqnarray}
d\eta=\frac{\partial g_{03}}{\partial r}dr\wedge dt+\frac{\partial g_{03}}{\partial\theta}d\theta\wedge dt+\frac{\partial g_{33}}{\partial r}dr\wedge d\ph+\frac{\partial g_{33}}{\partial\theta}d\theta\wedge d\phi.
\label{deta}
\end{eqnarray}
Instead of working with $dt,dr,d\theta,d\phi$ let us introduce the following orthonormal one forms \cite{cohen2}
\begin{eqnarray}
\hat{\varkappa}_0=-\left(\frac{\Delta\Sigma}{A}\right)^{\frac{1}{2}}dt;~\hat{\varkappa}_1=\left(\frac{\Sigma}{\Delta}\right)^{\frac{1}{2}}dr;~\hat{\varkappa}_2=\Sigma^{\frac{1}{2}}d\theta ;~\hat{\varkappa}_3=\left(\frac{A{\textrm{sin}}^2\theta}{\Sigma}\right)^{\frac{1}{2}}\left(d\ph-\Omega dt\right)\nonumber
\label{ford}
\end{eqnarray}
such that the metric (\ref{2.1}) takes the Minkwoskian form 
\begin{eqnarray}
ds^2=-\hat{\varkappa}_0^2+\hat{\varkappa}_1^2+\hat{\varkappa}_2^3+\hat{\varkappa}_3^2.
\end{eqnarray}
 Using the inverse relations 
\begin{eqnarray}
dt&=&-\left(\frac{A}{\Delta\Sigma}\right)^{\frac{1}{2}}\hat{\varkappa}_0 ;~dr=\left(\frac{\Delta}{\Sigma}\right)^{\frac{1}{2}}\hat{\varkappa}_1;~d\theta=\left(\frac{1}{\Sigma}\right)^{\frac{1}{2}}\hat{\varkappa}_2\nonumber\\
d\phi&=&\left(\frac{\Sigma}{A{\textrm{sin}}^2\theta}\right)^{\frac{1}{2}}\hat{\varkappa}_3-\Omega\left(\frac{A}{\Delta\Sigma}\right)^{\frac{1}{2}}\hat{\varkappa}_0\nonumber
\end{eqnarray}
we write (\ref{deta}) as,
 \begin{eqnarray}
d\eta=\lambda_{10}\hat{\varkappa}_1\wedge\hat{\varkappa}_0+\lambda_{20}\hat{\varkappa}_2\wedge\hat{\varkappa}_0+\lambda_{13}\hat{\varkappa}_1\wedge\hat{\varkappa}_3+\lambda_{23}\hat{\varkappa}_2\wedge\hat{\varkappa}_3
\label{eta11}
\end{eqnarray}
where
\begin{eqnarray}
\lambda_{10}&=&-\frac{\partial g_{03}}{\partial r}\frac{A^{\frac{1}{2}}}{\Sigma}-\frac{\partial g_{33}}{\partial r}\Omega\frac{A^{\frac{1}{2}}}{\Sigma};~\lambda_{20}=\frac{\partial g_{03}}{\partial \theta}\frac{1}{\Sigma}\left(\frac{A}{\Delta}\right)^{\frac{1}{2}}-\frac{\partial g_{33}}{\partial \theta}\frac{\Omega}{\Sigma}\left(\frac{A}{\Delta}\right)^{\frac{1}{2}}\nonumber\\
\lambda_{13}&=&\frac{\partial g_{33}}{\partial r}\frac{1}{{\textrm{sin}}\theta}\left(\frac{\Delta}{A}\right)^{\frac{1}{2}};~\lambda_{23}=\frac{\partial g_{33}}{\partial \theta}\frac{1}{{\textrm{sin}}\theta}\left(\frac{1}{A}\right)^{\frac{1}{2}}
\label{lam}
\end{eqnarray}
 The dual of (\ref{eta11}) is 
\begin{eqnarray}
^* d\eta=-\lambda_{10}\hat{\varkappa}_2\wedge\hat{\varkappa}_3+\lambda_{20}\hat{\varkappa}_1\wedge\hat{\varkappa}_3+\lambda_{13}\hat{\varkappa}_2\wedge\hat{\varkappa}_0-\lambda_{23}\hat{\varkappa}_1\wedge\hat{\varkappa}_0
\label{zzz}
\end{eqnarray}
Using (\ref{ford}), above equation is written in usual coordinate two form as,
\begin{eqnarray}
^* d\eta =\delta_{rt}dr\wedge dt+\delta_{\theta t}d\theta\wedge dt+\delta_{r\phi}dr\wedge d\phi+\delta_{\theta\phi}d\theta\wedge d\phi
\label{eta1}
\end{eqnarray}
where
\begin{eqnarray}
\delta_{rt}&=&-\lambda_{20}\Omega\left(\frac{A{\textrm{sin}}^2\theta}{\Delta}\right)^{\frac{1}{2}}+\lambda_{23}\Sigma\left(\frac{1}{A}\right)^{\frac{1}{2}} \nonumber\\
\delta_{\theta t}&=&-[\lambda_{13}\Sigma\left(\frac{\Delta}{A}\right)^{\frac{1}{2}}- \lambda_{10}\Omega\sqrt{A}{\textrm{sin}}\theta]\nonumber\\
\delta_{r\phi}&=&\lambda_{20}\left(\frac{A{\textrm{sin}}^2\theta}{\Delta}\right)^{\frac{1}{2}}\nonumber\\
\delta_{\theta\phi}&=&-\lambda_{10}\sqrt{A}{\textrm{sin}}\theta
\label{delta}
\end{eqnarray}

To calculate the effective Komar angular momentum (\ref{peff}) one needs to choose an appropriate boundary surface ($\partial\Sigma$). It is the boundary of a spatial three volume ($\Sigma$) characterised by a constant $r$ and $dt=-\frac{g_{03}}{g_{00}}d\phi$. Under this condition (\ref{eta1}) is simplified as
\begin{eqnarray}
^*d\eta =-\frac{g_{03}}{g_{00}}\delta_{\theta t}d\theta\wedge d\phi+\delta_{\theta\phi}d\theta\wedge d\phi
\label{aaaa}
\end{eqnarray}
Now substituting (\ref{aaaa}) in (\ref{peff}), we find the expression of Komar angular momentum as,
\begin{eqnarray}
K_{\xi^{\mu}_{(\phi)}}=\frac{1}{8\pi}\int{\frac{g_{03}}{g_{00}}\delta_{\theta t}d\theta d\phi}-\frac{1}{8\pi}\int{\delta_{\theta\phi}d\theta d\phi}
\label{komas1}
\end{eqnarray}
Moving along a closed contour, the first term of the right hand side gives the shift of time between the initial and the final events. Since we are performing an integration over simultaneous events this term must be subtracted from (\ref{komas1}) \cite{cohen1,cohen2}.  So we write the above equation as,
\begin{eqnarray}
K_{\xi^{\mu}_{(\phi)}}=\frac{1}{8\pi}\int \lambda_{10}\sqrt{A}{\textrm{sin}}\theta d\theta d\phi
\label{komas}
\end{eqnarray}
where we have used the relation (\ref{delta}). Substituting $\lambda_{10}$ from (\ref{lam}), we write the above expression as
\begin{eqnarray}
K_{\xi^{\mu}_{(\phi)}}=\frac{1}{8\pi}\int\left(-\Omega\frac{\partial g_{33}}{\partial r}-\frac{\partial g_{03}}{\partial r}\right)\frac{A{\textrm{sin}}\theta}{\Sigma}d\theta d\phi
\label{}
\end{eqnarray}
Using the metric coefficients
\begin{eqnarray}
g_{03}&=&\frac{a(Q^2-2Mr){\textrm{sin}}^2\theta}{r^2+a^2{\textrm{cos}}^2\theta}\\
g_{33}&=&\frac{{\textrm{sin}}^2\theta\left((a^2+r^2)(r^2+a^2{\textrm{cos}}^2\theta)-a^2(Q^2-2Mr){\textrm{sin}}^2\theta\right)}{r^2+a^2{\textrm{cos}}^2\theta}
\label{gs}
\end{eqnarray}
we find,
\begin{eqnarray}
K_{\xi^{\mu}_{(\phi)}}&=&\frac{1}{4}\int\frac{2a}{(r^2+a^2{\textrm{cos}}^2\theta)^2}[r^3(2Q^2-3Mr)+ra^2(Q^2-Mr)+\nonumber\\&&a^2(a^2M+r(Q^2-Mr)){\textrm{cos}}^2\theta]{\textrm{sin}}^3\theta d\theta
\end{eqnarray} 
After performing the integration we find the Komar conserved quantity corresponding to the spacelike one form $\eta$, given by, 
\begin{eqnarray}
K_{\xi^{\mu}_{(\phi)}}=-\left(2aM+\frac{rQ^2}{2a}-\frac{aQ^2}{2r}-\frac{Q^2(a^2+r^2)^2}{2a^2r^2}{\textrm{tan}}^{-1}(\frac{a}{r})\right)
\label{jeff1}
\end{eqnarray}
Taking the asymptotic limit one finds $\displaystyle\lim_{r\rightarrow\infty}~K_{\eta}=-2Ma$. This is different from the angular momentum of Kerr-Newman black hole $J=Ma$. The anomalous factor $-2$ appears here due to the use of the Boyer-Lindquist coordinate in defining the Kerr-Newman spacetime \cite{katz}. Therfore in order to derive the correct expression of angular momentum the normalisation of (\ref{peff}) should be changed accordingly. Thus our result (\ref{jeff1}) is divided by $-2$ to get the correct result of effective angular moentum, which yields,
\begin{eqnarray}
J_{\textrm{eff}}=Ma+\frac{rQ^2}{4a}-\frac{aQ^2}{4r}-\frac{Q^2(a^2+r^2)^2}{4a^2r^2}{\textrm{tan}}^{-1}(\frac{a}{r})
\label{jeffc}
\end{eqnarray}

Thus we find that at finite radial distance mass (\ref{ms}) and angular momentum (\ref{jeffc}) of the Kerr-Newman black hole get modified. All extra contributions appearing here are due to the presence of electric charge outside the event horizon of a black hole. Other than the usual Coulomb's field there is also a coupling between the charge and angular momentum of the black hole. At infinity and/or charge-less limit they reproduce well known results.

\section{\label{kintmp}Effective values of Komar conserved charges for charged Myers-Perry black holes} 

In this section we extend the previous analysis to derive effective values of (\ref{eeff}) and (\ref{peff}) by considering the arbitrary dimensional charged, rotating Myers-Perry spacetime (\ref{metric}). The timelike and the spacelike one forms for the spacetime metric (\r{metric}) are respectively given by
 $\s =\xi_{(t)\mu}dx^{\mu}=g_{0\m}dx^{\m}=g_{00}~dt+g_{03}~d\ph$,      $~~\eta=\xi_{(\ph)\mu}dx^{\mu}=g_{3\m}dx^{\m}=g_{30}~dt+g_{33}~d\ph$.
This yields the following two forms
\begin{eqnarray}
d\s &=&\p_rg_{00}~dr\wedge dt+\p_{\theta}g_{00}~d\th \wedge dt+\p_rg_{03}~dr\wedge d\ph+\p_{\th}g_{03}~d\th\wedge d\ph
\label{r1}\\
d\eta &=&\p_rg_{03}~dr\wedge dt+\p_{\theta}g_{03}~d\th \wedge dt+\p_rg_{33}~dr\wedge d\ph+\p_{\th}g_{33}~d\th\wedge d\ph
\label{tform}
\end{eqnarray}

The orthonormal basis for the spacetime metric (\r{metric}) is given by
\begin{eqnarray}
\varkappa_0&=&-\sqrt{\left(\f{g_{03}^2}{g_{33}}-g_{00}\right)}~dt~;~\varkappa_1=\sqrt{g_{11}}dr\nonumber\\
\varkappa_2&=&\sqrt{g_{22}}d\th ~;~\varkappa_3=\sqrt{g_{33}}(d\ph+\f{g_{03}}{g_{33}}dt)\label{vark}\\
\varkappa_4&=&r\cos\th d\chi_1 ~;\varkappa_5=r\cos\th\sin\chi_1 d\chi_2\nonumber\\
\varkappa_i&=&r\cos\th\sin\chi_1\cdot\cdot\cdot\sin\chi_{i-4}d\chi_{i-3} \  \ ({\textrm{where}} \ 6\le i\le N+1)\nonumber
\end{eqnarray}
in which the metric (\r{metric}) takes the Minkwoskian form
\begin{eqnarray}
ds^2=-\varkappa^2_0+\varkappa_1^2+\varkappa_2^2+\cdot\cdot\cdot+\varkappa_N^2
\label{metricvar}
\end{eqnarray}
Using the inverse relations,
\begin{eqnarray}
dt&=&-\f{1}{\sqrt{\f{g_{03}^2}{g_{33}}-g_{00}}}\varkappa_0 ~;~dr=\f{1}{\sqrt{g_{11}}}\varkappa_1\nonumber\\
d\th &=&\f{1}{\sqrt{g_{22}}}\varkappa_2 ~;~d\ph=\f{1}{\sqrt{g_{33}}}\left(\varkappa_3+\f{g_{03}}{\sqrt{g_{00}g_{33}+g_{03}^2}}\varkappa_0\right)\\
d\chi_1&=&(r\cos\th)^{-1} \varkappa_4\nonumber\\
d\chi_2&=&(r\cos\th\sin\chi_1)^{-1}\varkappa_5\nonumber\\
d\chi_{i-3}&=&(r\cos\th\sin\chi_1\cdot\cdot\cdot\sin\chi_{i-4})^{-1}\varkappa_i \  \ ({\textrm{where}} \ 6\le i\le N+1)\nonumber
\label{vark1}
\end{eqnarray}
(\r{r1}) is written as
\begin{eqnarray}
d\s=\Lambda_{10}\varkappa_1\wedge\varkappa_0+\Lambda_{20}\varkappa_2\wedge\varkappa_0
+\Lambda_{13}\varkappa_1\wedge\varkappa_3+\Lambda_{23}\varkappa_2\wedge\varkappa_3
\label{aaa}
\end{eqnarray}
where
\begin{eqnarray}
\Lambda_{10}&=&-\f{1}{\sqrt{g_{03}^2-g_{00}g_{33}}}\left(\sqrt{\f{g_{33}}{g_{11}}}\p_rg_{00}+\f{g_{03}}{\sqrt{g_{11}g_{33}}}\p_rg_{03}\right) \label{lam10} \\
\Lambda_{20}&=& -\f{1}{\sqrt{g_{03}^2-g_{00}g_{33}}}\left(\sqrt{\f{g_{33}}{g_{22}}}\p_{\th} g_{00}+\f{g_{03}}{\sqrt{g_{22}g_{33}}}\p_{\th}g_{03}\right) \\
\Lambda_{13}&=&\f{1}{\sqrt{g_{11}g_{33}}}\p_rg_{03}  \\
\Lambda_{23}&=& \f{1}{\sqrt{g_{22}g_{33}}}\p_{\th}g_{03} \\
\end{eqnarray}
The Hodge dual of (\r{aaa}) is
\begin{eqnarray}
{}^*d\s&=&\left(-\Lambda_{10}\varkappa_2\wedge\varkappa_3+\Lambda_{20}\varkappa_1\wedge\varkappa_3
+\Lambda_{13}\varkappa_2\wedge\varkappa_0-\Lambda_{23}\varkappa_1\wedge\varkappa_0\right)
\wedge\varkappa_4\wedge\varkappa_5\wedge\cdot\cdot\cdot\nonumber
\end{eqnarray}
Using (\r{vark}), the above expression is written in the usual coordinates as,
\begin{eqnarray}
{}^*d\s&=&(-\Lambda_{10}\sqrt{g_{22}g_{33}}d\th\wedge d\ph+[-\Lambda_{10}g_{03}\sqrt{\f{g_{22}}{g_{33}}}-\Lambda_{13}\sqrt{\f{g_{22}}{g_{33}}}(g_{03}^2-g_{00}g_{33})^{\f{1}{2}}]d\th\wedge dt\nonumber\\
&&+\Lambda_{20}\sqrt{g_{11}g_{33}}dr\wedge d\ph+[\Lambda_{20}g_{03}\sqrt{\f{g_{11}}{g_{33}}}+\Lambda_{23}\sqrt{\f{g_{11}}{g_{33}}}(g_{03}^2-g_{00}g_{33})^{\f{1}{2}}]dr\wedge dt
)\wedge\nonumber\\&&(r\cos\th d\chi_1)\wedge(r\cos\th\sin\chi_1 d\chi_2)\wedge\cdot\cdot\cdot
\end{eqnarray}
At this point we need to choose an appropriate boundary ($\partial\Sigma$) characterised by a constant $r$ and $dt=-\frac{g_{03}}{g_{00}}d\phi$. This choice of surface reduces (\ref{meff}) to
\begin{eqnarray}
K_{\xi^{\mu}_{(t)}}&=&\f{1}{8\pi G}\int \Lambda_{10}\sqrt{g_{22}g_{33}}(r\cos\th)^{N-3}
d\th d\ph d\chi_1\sin\chi_1 d\chi_2\cdot\cdot\cdot+\nonumber\\
&&\f{1}{8\pi G}\int\frac{g_{03}}{g_{00}}\left[\Lambda_{10}g_{03}\sqrt{\f{g_{22}}{g_{33}}}+\Lambda_{13}\sqrt{\f{g_{22}}{g_{33}}}(g_{03}^2-g_{00}g_{33})^{\f{1}{2}}\right]\nonumber\\&&(r\cos\th)^{N-3}d\th d\ph d\chi_1\sin\chi_1d\chi_2\cdot\cdot\cdot\nonumber
\end{eqnarray}
The second term on the right hand side measures the time shifting, when one moves along a closed contour. Since the calculation is perfomed over simultaneous events we subtract this term\cite{cohen1,cohen2} from the above integral to obtain,
\begin{eqnarray}
K_{\xi^{\mu}_{(t)}}&=&\f{1}{8\pi G}\int \Lambda_{10}\sqrt{g_{22}g_{33}}(r\cos\th)^{N-3}d\th d\ph d\chi_1\sin\chi_1d\chi_2\cdot\cdot\cdot
\end{eqnarray}
Making use of (\r{lam10}) we write the above equation as, 
\begin{eqnarray}
K_{\xi^{\mu}_{(t)}} &=&-\f{1}{8\pi G}\int r^{N-3}\cos^{N-3}\th\Lambda(r,\th)d\th d\ph d\chi_1\cdot\cdot\cdot\sin\chi_{N-4}d\chi_{N-3}\label{meff2}
\end{eqnarray}
where $\Lambda$ is defined as
\begin{eqnarray}
\Lambda(r,\th) &=&\sqrt{g_{22}g_{33}}\Lambda_{10}=\left(g_{33}\p_r g_{00}-g_{03}\p_r g_{03}\right)\sqrt{\f{g_{22}}{g_{33}(-g_{00}g_{33}+g_{03}^2)}},
\label{Lr}
\end{eqnarray}
Similar consideration for the other Killing vector will give the relation 
\begin{eqnarray}
K_{\xi^{\mu}_{(\ph)}} &=& -\f{1}{8\pi G}\int r^{N-3}\cos^{N-3}\th\Psi(r,\th)d\th d\ph d\chi_1\cdot\cdot\cdot\sin\chi_{N-4}d\chi_{N-3}\label{jeff3}
\end{eqnarray}
where $\Psi$ is given by
\begin{eqnarray}
\Psi(r,\th) &=&-\left(g_{33}\p_r g_{03}-g_{03}\p_r g_{33}\right)\sqrt{\f{g_{22}}{g_{11}(g_{03}^2-g_{00}g_{33})}}.
\label{psi}
\end{eqnarray}
To find the effective values of $K_{\xi^{\mu}_{(t)}}$ and $K_{\xi^{\mu}_{(\ph)}}$ the above integrations (\ref{meff2}, \r{jeff3}) need to be performed over all the angular variables ($\th,~\phi,~\chi_1,\cdot\cdot\chi_{N-3}$). 

Let us first separate different parts of the integral (\r{meff2}) and (\r{jeff3}) in the following manner 
\begin{eqnarray}
K_{\xi^{\mu}_{(t)}}=\frac{1}{8\pi G}\int_{\ph =0}^{2\pi}d\ph ~\int_{\th =0}^{\pi}{r^{(N-3)}\cos^{N-3}\th}~\Lambda(r,\th)d\th\nonumber\\
\int_{\chi_1,\chi_2,\cdot\cdot\cdot,\chi_{N-3} =0}^{\pi}~d\chi_1 \sin\chi_1\cdot\cdot\cdot\sin{\chi_{N-4}}~d\chi_{N-3},
\label{meff3}\\
K_{\xi^{\mu}_{(\ph)}}=\frac{1}{8\pi G}\int_{\ph =0}^{2\pi}d\ph ~\int_{\th =0}^{\pi}{r^{(N-3)}\cos^{N-3}\th}~\Psi(r,\th)d\th\nonumber\\
\int_{\chi_1,\chi_2,\cdot\cdot\cdot,\chi_{N-3}=0}^{\pi}~d\chi_1 \sin\chi_1\cdot\cdot\cdot\sin{\chi_{N-4}}~d\chi_{N-3}.
\end{eqnarray}
Performing the integrations over the azimuthal angle ($\ph$) and angular variables ($\chi_1,\chi_2,\cdot\cdot\cdot\chi_{N-3}$), we find
\begin{eqnarray}
K_{\xi^{\mu}_{(t)}}=\frac{\bar{A}_{N-3}}{4G}~\int_{\th =0}^{\pi}{r^{(N-3)}\cos^{N-3}\th}~\Lambda(r,\th)d\th
\label{meff4}\\
K_{\xi^{\mu}_{(\ph)}}=\frac{\bar{A}_{N-3}}{4G}~\int_{\th =0}^{\pi}{r^{(N-3)}\cos^{N-3}\th}~\Psi(r,\th)d\th.
\end{eqnarray}
where $\bar{A}_{N-3}$ has been defined in (\r{an-3}). Now an explicit integration over the polar angle ($\th$) gives the desired results for the effective values of different Komar conserved quantities for arbitrary dimensional charged Myers-Perry black holes. These are given by 
\begin{eqnarray}
K_{\xi^{\mu}_{(t)}}=\frac{\bar{A}_{N-3}}{G}\left(\frac{m}{2}-\frac{q^2}{r^{N-2}}-\frac{q^2a^2(a^2+r^2)}{Nr^{N+2}}~2F_{1}(2,\frac{N}{2};\frac{N}{2}+1;-\frac{a^2}{r^2})\right),
\label{meff5}
\end{eqnarray}
and
\begin{eqnarray}
K_{\xi^{\mu}_{(\ph)}} &=& -\frac{a\bar{A}_{N-3}}{4Gr^{N+4}}(K_1-K_2)\label{jeff5}\\
{\textrm{where,}}\nonumber\\
K_{1} &=& \frac{4mNr^{N+4}+2q^2r^2[a^4(N-2)^2+a^2(N-4)(N-2)r^2-4(N-1)r^4]}{N(N-2)}\\
K_{2} &=& \frac{2a^2(N-2)q^2(a^2+r^2)^2~ 2F_{1}(1,\frac{N+2}{2};\frac{N+4}{2};-\frac{a^2}{r^2})}{N+2}.
\end{eqnarray}
In the above expressions $2F_{1}$ is a hypergeometric function,
\begin{eqnarray}
2F_1(b,c;d;-\f{a^2}{r^2})=1+\f{b.c}{1!d}(-\f{a^2}{r^2})+\f{b(b+1).c(c+1)}{2!d(d+1)}(-\f{a^2}{r^2})^2+\cdot\cdot\cdot .
\end{eqnarray}

In the limiting case $N=3$ (\ref{meff5}) and (\ref{jeff5}) successfully reproduce the results of Kerr-Newman spacetime obtained in (\ref{ms}) and (\ref{jeff1}). Also note that for a finite $r$, the effective values of $K_{\xi^{\mu}_{(t)}}$ and $K_{\xi^{\mu}_{(\ph)}}$ differ from their asymptotic values only due to presence of electric charge ($q$) in black holes. The extra contributions come in two different ways, one is proportional to the electric charge and the other is a coupling between the charge and the reduced angular momentum parameter ($a$). The second one can be termed as a gravito-electric effect. In the asymptotic limit all contributions due to the electric charge drop out and we find,  
\begin{eqnarray}
\lim_{r\rightarrow\infty}~K_{\xi^{\mu}_{(t)}} &=&\frac{m\bar{A}_{N-3}}{2G}\label{asykt}\\ \lim_{r\rightarrow\infty}~K_{\xi^{\mu}_{(\ph)}} &=&-\frac{ma\bar{A}_{N-3}}{(N-2)G}.
\label{asymkp}
\end{eqnarray}
Comparing these relations with (\r{m}), (\r{j}) and using the relations (\r{a}), (\r{an-3}) one finds,
\begin{eqnarray}
\lim_{r\rightarrow\infty}~K_{\xi^{\mu}_{(t)}} &=& \frac{2(N-2)}{(N-1)}M\label{asykt1}\\ \lim_{r\rightarrow\infty}~K_{\xi^{\mu}_{(\ph)}} &=& -2J\label{asmkp1}.
\label{compr}
\end{eqnarray}
From these two equations it is now clear that only in (3+1) dimensions the asymptotic value of the Komar conserved quantity, corresponding to the Killing vector $\xi^{\mu}_{(t)}=(1,0,0,0)$, gives the correct value of the black hole mass ($M$). For any other higher dimension the value of $K_{\xi^{\mu}_{(t)}}$ differs from the black hole mass by a  dimension dependent numerical factor. The other conserved quantity $K_{\xi^{\mu}_{(\ph)}}$ differs by $-2$ factor from the angular momentum ($J$) for all spacetime dimensions greater than or equal to four.

In the subsequent part of this chapter we mainly focus on Myers-Perry spacetime and make use of (\ref{meff5}) and (\ref{jeff5}) to study some important properties of black hole event horizons. Since the results of Kerr-Newman case can be obtained just be considering $N=3$ limit we do not consider this case separately.

\section{\label{ident}The identity $K_{\chi^{\mu}}=2ST$ and Smarr mass formula}


In this section we shall make use of the effective expressions of Komar conserved charges and study some important properties of black hole event horizons. The motivation of doing this is the fact that event horizons are only special cases of Killing horizons. To see this recall that for a vector $\chi^{\mu}$ to be Killing at the event horizon one must have $(\nabla_{\mu}\chi_{\nu}+\nabla_{\mu}\chi_{\nu})_{r_+}=0$. Also, if we call the event horizon to be a Killing horizon this implies $\chi^{\mu}\chi_{\mu}|_{r=r_+}=0$, i.e. it should be null at the event horizon. For the spacetime metric (\r{metric}) although both $\xi^{\mu}_{(t)}$ and $\xi^{\mu}_{(\phi)}$ satisfy the first condition they violate the second one. Only a specific linear combination of these vectors, given by, $\chi^\mu=\xi^{\mu}_{(t)}+\Omega_{H}\xi^{\mu}_{(\ph)}$ (where $\Omega_H$ is the angular velocity at the event horizon) satisfies both conditions. It can be verified that the use of $\chi^{\mu}$ leads to the correct result for the surface gravity $\kappa=\sqrt{-\f{1}{2}(\nabla^{\mu}\chi^{\nu})(\nabla_{\mu}\chi_{\nu})}\big|_{r=r_+}=\f{F'{(r_+)}}{2}$ (\r{surface}) which is related to the Hawking temperature (\ref{temp}) through the relation $T=\f{\kappa}{2\pi}$. Thus the black hole event horizon is a Killing horizon of the Killing vector $\chi^{\mu}$. 

Because of such a Killing vector one can again associate a Komar conserved quantity at the event horizon, which can be finally brought to the following form   
\begin{eqnarray}
K_{\chi^{\mu}}=K_{\xi^{\mu}_{(t)}}+\Omega_{H}~K_{\xi^{\mu}_{(\ph)}}
\label{kev}
\end{eqnarray}
Now it is easy to calculate this quantity by using the horizon condition $\Delta|_{r_+}=0$ and by using equations (\r{ang}), (\r{meff5}) and (\r{jeff5}). This result is found to be
\begin{eqnarray}
K_{\chi^{\mu}}=2 \left(\f{\pi r_+^{N-3}(r_+^2+a^2)\bar{A}_{N-3}}{N-2}\right)\left(\f{(N-4)(r_+^2+a^2)+2r_+^2-(N-2)q^2r_+^{2(3-N)}}{4\pi r_+(r_+^2+a^2)}\right)
\label{lhs}
\end{eqnarray}
where we have used the following identities involving the hypergeometric functions, 
\begin{eqnarray}
&&\f{Na^2}{(N+2)r^2}2F_{1}(1,\f{N+2}{2};\f{N+4}{2};-\f{a^2}{r^2})+2F_{1}(1,\f{N}{2};\f{N+2}{2};-\f{a^2}{r^2})=1
\label{iden1}\\
&&2F_{1}(2,\f{N}{2};\f{N+2}{2};-\f{a^2}{r^2})=\f{N}{2(1+\f{a^2}{r^2})}+(1-\f{N}{2})2F_{1}(1,\f{N}{2};\f{N+2}{2};-\f{a^2}{r^2})
\label{iden2}
\end{eqnarray}
It is very interesting to note that, using (\r{entr}) and (\r{temp}), equation (\r{lhs}) can be written as 
\begin{eqnarray}
K_{\chi^{\mu}}=2ST,
\label{2st}
\end{eqnarray}
i.e. the Komar conserved quantity at the event horizon corresponding to the null Killing vector is twice the product of Hawking temperature and black hole entropy. A remarkable feature of the above relation is that it is independent of the number of dimensions $N$. In the following we discuss the significance of this identity in two different contexts.

First, it is possible to make a correspondence of (\r{2st}) with a relation,
\begin{eqnarray}
E=2ST,
\label{e=2st}
\end{eqnarray}
discussed in the literature \c{paddy1}-\c{majhi}. Here $E$ is the conserved Noether charge corresponding to a diffeomorphic transformation of the spacetime metric which may not be a black hole metric. It can be of static Rindler type spacetime having a timelike Killing horizon. However we see that an analogous relation (\ref{2st}) is also found for a black hole spacetime which is nonstatic as well. A comparison between (\ref{e=2st}) and (\ref{2st}) together with (\ref{kev}) show that it is possible to identify $E$ as a particular combination of properly normalised Komar conserved quantities.

Secondly, the significance of (\r{2st}) in arbitrary dimension can also be understood by deriving the mass formula for black holes. For that first note that, the right  hand side of (\r{2st}) can be written in terms of surface gravity ($\kappa$) and horizon area ($A_H$) of the black hole as $2(\f{A_H}{4})(\f{\kappa}{2\pi})$. Now we want to rewrite the left hand side of (\r{2st}) as a combination of three seperate terms, one invoving $m$, one involving $ma$ and the other involving $q^2$. To do that we use (\r{kev}), (\r{meff5}), (\r{jeff5}) together with the identities (\r{iden1}), (\r{iden2}) and the expression of $\Omega_H$ (\r{ang}). This gives
\begin{eqnarray}
\f{\bar{A}_{N-3}}{G}\left(\f{m}{2}-\f{ma\Omega_H}{N-2}-\f{q^2r_+^{2-N}[a^2(N-3)+r_+^2(N-2)]}{(N-2)(a^2+r_+^2)}\right)=2\f{\kappa A_H}{8\pi}
\label{}
\end{eqnarray}
Making use of the relations (\r{m}), (\r{j}) and (\r{q}) we write the above relation as,
\begin{eqnarray}
\f{2M(N-2)}{(N-1)}-2\Omega_H J-\f{2Q^2r_+^{4-N}}{(N-1)(a^2+r_+^2)}-\f{2Q^2r_+^{2-N}a^2(N-3)}{(N-1)(N-2)(a^2+r_+^2)}=2\f{\kappa A_H}{8\pi}
\label{smarr}
\end{eqnarray}

The scalar potential $\Phi$ at the event horizon is given by the timelike component of (\r{phi}). This is found by contracting $A_{\mu}$ with the Killing field, $\chi^{\mu}=\xi^{\mu}_{(t)}+\Omega_H\xi^{\mu}_{(\ph)}=(1,0,0,\Omega_H,0,\cdot\cdot\cdot)$ corresponding to the metric (\r{metric}) and is given by
\begin{eqnarray}
\Phi=\chi^{\mu}A_{\mu}\big|_{r=\infty}-\chi^{\mu}A_{\mu}\big|_{r=r_+}=\f{Q}{(N-2)r_+^{N-4}(r_+^2+a^2)}
\label{ph}
\end{eqnarray}
Using the above relation, (\r{smarr}) is written as,
\begin{eqnarray}
\f{M(N-2)}{(N-1)}-\Omega_H J-Q\Phi\left(\f{(N-2)}{N-1}+\f{a^2(N-3)}{r_+^2(N-1)}\right)=\f{\kappa A_H}{8\pi}
\label{sm1}
\end{eqnarray}
Structurally this is very similar to the Smarr formula. However there is only one unfamiliar term (involving $a^2$) arising at the left hand side of (\r{sm1}). Let us now recall that the spacetime (\ref{metric}) is a solution of the Einstein-Maxwell equation only for slowly rotating case (upto linear order in $a$) \cite{aliev}. Therefore we can drop this unfamiliar term and rewrite (\r{sm1}) in the following way, 
\begin{eqnarray}
M(N-2)=(N-1)\Omega_H J+(N-2)Q\Phi +(N-1)\f{\kappa A_H}{8\pi}.
\label{sm2}
\end{eqnarray}
This is the cherished Smarr mass formula for the $N+1$ dimensional charged Myers-Perry black hole.

One can also show that (\r{sm2}) is indeed compatible with the first law of black hole thermodynamics in the following way. Looking at the metric (\r{metric}) we can express the following three basic variables in terms of their dimensionless {\it primed} counterparts, given by
\begin{eqnarray}
m=\l^{N-2} m',~a=\l a',~q=\l^{N-2} q'~,
\label{sm3}
\end{eqnarray}
where $\l$ is a dimensionful quantity having the dimension of length. Now using (\r{sm3}) it is easy to check that the horizon radius can be expressed as $r_+=\l r_+'$. Exploiting these four basic transformations in (\r{m}), (\r{j}), (\r{q}) and (\r{hora1}) we obtain the following relations
\begin{eqnarray}
M=\l^{N-2} M',~J=\l^{N-1}J',~Q=\l^{N-2} Q',~A_{H}=\l^{N-1}A_{H}'~ .
\label{sm4}
\end{eqnarray}
Now if we treat $M$, $J$, $A_{H}$ and $Q$ as independent variables then (\r{sm2}) is a homogeneous equation of degree $(N-2)$ in $M$, $(N-1)$ in $J$, $(N-2)$ in $Q$ and $(N-1)$ in $A_H$. Therefore one can now exploit Euler's theorem on homogeneous functions to extract the differential form of Smarr formula,
\begin{eqnarray}
dM=\f{\kappa}{8\pi}~dA_{H}+\Omega_{H} ~dJ+\Phi ~dQ.
\end{eqnarray}   
This is the desired result of the first law of black hole thermodynamics for rotating, charged black holes in arbitrary dimensions, compatible with (\r{sm2}).

In 3+1 dimensions 
the resulting relation is the Smarr formula for the Kerr-Newman black hole. This was originally given by Smarr \c{smarr} and also by Bardeen, Carter and Hawking \c{Bardeen}. 
Moreover (\r{sm2}) is consistent with the result for the R-N black hole as derived in (\r{rnsmarr}). Finally, this result also gives the correct Smarr formula for the arbitrary dimensional (charge-less) Myers-Perry black hole\c{myers}.

\section{\label{discts}Discussions} 
In this chapter we performed a detailed study to explore various important features of charged, rotating black holes in arbitrary dimensions. We discussed the thermodynamical properties of Kerr-Newman and charged Myers-Perry black holes. Using an algebraic approach, motivated by the work of Smarr, we found a mass formula for $N+1$ dimensional Reissner-Nordstrom black hole. However this method could not be generalized to the rotating case for higher dimensions. We therefore developed an alternative scheme which involved the knowledge of conserved quantities of the respective spacetime. These conserved quantities were related to various Killing symmetries and were found by evaluating the Komar integrals. The Komar integrals were evaluated at the boundary of a spatial hypersurface in a spacetime. The exact nature of this surface was $r=$ constant with time synchronised events. This gave a freedom to calculate these conserved quantities explicitly on any such surface, which need not be the asymptotic surface or event horizon of a black hole. These quantities were then used to find the Komar conserved quantity ($K_{\chi^{\mu}}$) at the event horizon corresponding to the null Killing vector $\chi^{\mu}$. This lead to the remarkable dimension independent identity, $K_{\chi^{\mu}}=2ST$. Here ($S$) and $T$ were the black hole entropy and Hawking temperature respectively. The profound nature of this identity was that although each individual term was dimension dependent they together satisfy an identity which is completely dimension independent. As an application of this identity we derived the generalised Smarr formula for the Kerr-Newman and charged Myers-Perry black holes. This formula was also shown to be compatible with the first law of black hole thermodynamics. A connection of this identity for black hole event horizon was discussed with a similar relation $E=2ST$ ($E$ is the Noether charge corresponding to diffeomorphic transformation of the metric coefficient) valid for any timelike Killing horizon which may or may not be associated with a black hole.

\begin{subappendices}
\chapter*{Appendices}
\section{\label{appendix2A1}Glossary of formuale for charged Myers-Perry black hole}
\renewcommand{\theequation}{2A.\arabic{equation}}
\setcounter{equation}{0}
The spacetime metric for the $N+1$ dimensional charged Myers-Perry black hole in Boyer-Lindquist type coordinates, with one spin parameter ($a$), is given by \c{aliev,dianyan}
\begin{eqnarray}
ds^2&=&-\left(1-\f{m}{r^{N-4}\Sigma}+\frac{q^2}{r^{2(N-3)}\Sigma}\right)dt^2+\f{r^{N-2}\Sigma}{\D}dr^2+\S d\th^2-\f{2a(mr^{N-2}-q^2)\sin^2\th}{r^{2(N-3)}\S}dtd\ph\no\\
&&+\left(r^2+a^2+\f{a^2(mr^{N-2}-q^2)\sin^2\th}{r^{2(N-3)}\S}\right)\sin^2\th d\ph^2+r^2\cos^2\th d\Om^2_{N-3},
\label{metric}
\end{eqnarray}
with the following identifications,
\begin{eqnarray}
\D&=&r^{N-2}(r^2+a^2)-mr^2+q^2r^{4-N},
\label{delta}\\
\S&=&r^2+a^2\cos^2\th,
\label{sigma}
\end{eqnarray}
\begin{eqnarray}
d\Om^2_{N-3}=d\chi_1^2+\sin^2\chi_1[d\chi^2_2+\sin^2\chi_2(\cdot\cdot\cdot d\chi^2_{N-3})].
\label{omega}
\end{eqnarray}
The electromagnetic potential one form for the spacetime (\r{metric}) is 
\begin{eqnarray}
A=A_{\mu}dx^{\mu}=-\f{Q}{(N-2)r^{N-4}\Sigma}(dt-a\sin^2\theta d\phi).
\label{phi}
\end{eqnarray}
In appropriate limits this metric reproduces the $N+1$ dimensional spherically symmetric, static Schwarzscild, Reissner-Nordstrom \c{tangher} and axially symmetric, rotational Myers-Perry spcetime \c{myers}.

It must be stated that the metric (\ref{metric}) is a solution of the Einstein-Maxwell equation only for linear order in $a$ \c{aliev}. However   in our subsequent analysis we keep all terms involving $a$. Finally it will be shown that the arbitrary dimensional result for the Smarr formula, compatible with black hole thermodynamics, can be recovered only if terms linear in $a$ are retained.

The determinant ($g$) of the metric (\r{metric}) gives
\begin{eqnarray}
\sqrt{-g}=\sqrt{\gamma}\S r^{N-3}\sin\th\cos^{N-3}\th,
\label{deter}
\end{eqnarray}
where $\gamma$ is the determinant of the metric (\r{omega}). The parameters $m,~a,~q$ are related with the physical mass ($M$), angular momentum ($J$) and charge ($Q$) through the relations given by
\begin{eqnarray}
M &=&\f{A_{N-1}(N-1)}{16\pi G}m
\label{m}\\
J &=&\f{A_{N-1}}{8\pi G}ma
\label{j}\\
Q &=&\pm\sqrt{\f{(N-2)(N-1)A_{N-1}}{8\pi G}}q
\label{q}
\end{eqnarray}
Here $A_{N-1}$ is the area of the unit sphere in $N-1$ dimensions 
\begin{eqnarray}
A_{N-1}=\int_{0}^{2\pi}d\ph\int_0^\pi \sin\th\cos^{N-3}\th d\th\huge\prod_{i=1}^{N-3}\int_{0}^{\pi}\sin^{(N-3)-i}\chi_{i}d\chi_{i}=\f{2\pi^{N/2}}{\Gamma(N/2)}.
\label{a}
\end{eqnarray}
The position of the event horizon is represented by the largest root of the polynomial $\Delta_{r=r_+}=0$. The angular velocity at the event horizon is given by
\begin{eqnarray}
\Omega_{H}=\f{a}{r_+^2+a^2}.
\label{ang}
\end{eqnarray}

Let us now calculate the area of the event horizon for the spacetime metric (\r{metric}). This is given by the standard definition of horizon area by integrating the positive square-root of the induced metric ($\eta$) on the $N-1$ dimensional spatial angular hypersurface ($\Sigma$) for fixed $r=r_+$  
\begin{eqnarray}
A_{H}=\int_{\Sigma(r=r_+)}\sqrt{|\eta|}~d^{N-1}\Sigma .
\label{hora}
\end{eqnarray}
The result of the integration yields,
\begin{eqnarray}
A_{H}=\f{4\pi r_+^{N-3}(r_+^2+a^2)\bar{A}_{N-3}}{N-2}.
\label{hora1}
\end{eqnarray}
Here $\bar{A}_{N-3}$ is the area of $N-3$ dimensional angular hypersurface represented by the angular variables $\chi_j$-s
\begin{eqnarray}
\bar{A}_{N-3}=\huge\prod_{i=1}^{N-3}\int_{0}^{\pi}\sin^{(N-3)-i}\chi_{i}d\chi_{i}=\frac{\pi^{\frac{N}{2}-1}}{\Gamma(\frac{N}{2}-1)}.
\label{an-3}
\end{eqnarray}
Therefore the semiclassical black hole entropy, as given by the Bekenstein-Hawking formula, yields 
\begin{eqnarray}
S=\f{A_H}{4}=\f{\pi r_+^{N-3}(r_+^2+a^2)\bar{A}_{N-3}}{N-2}.
\label{entr}
\end{eqnarray}

To find the Hawking temperature we shall take help of Hawking's periodicity argument \c{Hawking3}. Such analysis can be simplified by considering the near horizon effective metric of (\ref{metric}). Near the event horizon the effective theory is driven by the two dimensional ($t-r$) metric \cite{carlip}-\cite{kumet}. In our case the two dimensional metric is given by (see appendix A) (\r{effective1}). Now the metric (\ref{effective1}) has singularity at $r=r_+$. To remove this singularity consider the following transformation
\begin{eqnarray}
x = \frac{1}{\kappa} F^{1/2}(r)
\label{trans}
\end{eqnarray}
where $\kappa$, the surface gravity of the black hole (\ref{metric}), is given by
\begin{eqnarray}
\kappa = \frac{F'(r_+)}{2}=\f{(N-4)(r_+^2+a^2)+2r_+^2-(N-2)q^2r_+^{2(3-N)}}{2 r_+(r_+^2+a^2)}.
\label{surface}
\end{eqnarray}
Substituting (\ref{trans}) in (\ref{effective1}) and then euclideanizing (i.e. $t=i\tau$)  we obtain,
\begin{eqnarray}
ds^2_{eff} = (\kappa x)^2 d\tau^2 + \Big(\frac{2\kappa}{F'(r)}\Big)^2dx^2.
\label{effective3}
\end{eqnarray}
This metric is regular at $x = 0$ and $\tau$ is regarded as an angular variable with period $2\pi/\kappa$, which is regarded as the inverse of Hawking temperature (in units of $\hbar = 1$). Hence the explicit expression for Hawking temperature is given by,
\begin{eqnarray}
T= \frac{\kappa}{2\pi} = \f{(N-4)(r_+^2+a^2)+2r_+^2-(N-2)q^2r_+^{2(3-N)}}{4\pi r_+(r_+^2+a^2)}.
\label{temp}
\end{eqnarray}
In appropriate limits, (\r{entr}) and (\r{temp}) reproduce the entropy and Hawking temperature for all other black holes in $N+1$ dimensions. 



\section{\label{appendix2B}Glossary of formulae for Kerr-Newman black hole}
\renewcommand{\theequation}{2B.\arabic{equation}}
\setcounter{equation}{0}
The spacetime metric of the Kerr-Newman black hole in Boyer-Linquist coordinates ($t,~r,~\theta,~\phi$) is given by,
\begin{equation}
ds^{2} =g_{00}dt^{2}+g_{11}dr^{2}+ 2g_{03}dtd\phi +g_{33}d\phi ^{2}+g_{22}d\theta ^{2}
\label{2.1}  
\end{equation}
with the electromagnetic vector potential, 
$$A_{a} =-\frac{Qr}{\Sigma (r,\theta)}[(dt)_{a}-a\sin ^{2}\theta (d\phi )_{a}]$$ and, 
\begin{align}
g_{00}& =\frac{-\Delta (r)+a^{2}\sin ^{2}\theta }{\Sigma (r,\theta )} \\
g_{11}& =\frac{\Sigma (r,\theta )}{\Delta (r)},  \notag \\
g_{03}& =\frac{-a\sin ^{2}\theta (r^{2}+a^{2}-\Delta (r))}{\Sigma
(r,\theta )}  \notag \\
g_{33}& =\frac{(r^{2}+a^{2})^{2}-\Delta (r)a^{2}\sin ^{2}\theta }{%
\Sigma (r,\theta )}\sin ^{2}(\theta )  \notag \\
g_{22}&=\Sigma (r,\theta ) =r^{2}+a^{2}\cos ^{2}\theta  \notag \\
\Delta (r)& =r^{2}+a^{2}+Q^{2}-2Mr  \notag\\
a=\frac{J}{M} \notag
\end{align}
The Kerr-Newman metric represents the most general class of stationary black hole solution of Einstein-Maxwell equations having all three parameters Mass $(M)$, Angular momentum $(J)$ and Charge $(Q)$. All other known stationary black hole solutions are encompassed by this three parameter solution.\\
(i)For $ Q=0$ it gives the rotating Kerr solution, (ii) $ J= 0$ leads to the Reissner-Nordstrom black hole, and (iii) for both  $ Q=0$ and $J=0$ the standard Schwarzschild solution is recovered.\\
For the non-extremal Kerr-Newman black hole the location of outer ($r_+$, event) and inner ($r_-$) horizons are given by setting $g^{rr}=0=g_{tt}$ or equivalently $\Delta =0$, which gives   
\begin{equation}
r_{\pm}=M\pm \sqrt{M^{2}-a^{2}-Q^{2}}.
\label{2.2}
\end{equation}
The angular velocity of the event horizon, which follows from the general expression of angular velocity for any rotating black hole, is given by
\begin{eqnarray}
\Omega_{\textrm H}= \Big[-\frac{g_{\phi t}}{g_{\phi \phi}}-\sqrt{{(\frac{g_{t\phi}}{g_{\phi\phi}})^2}-\frac{g_{tt}}{g_{\phi \phi}}}\Big]_{r=r_+}= \frac{a}{r^2_+ + a^2}.
\label{angv}
\end{eqnarray}
The electric potential at the event horizon is given by,
\begin{eqnarray}
\Phi_{\textrm H}= \frac{r_+ Q}{r^2_+ + a^2}.
\label{epot}
\end{eqnarray}
The area of the event horizon is given by,
\begin{eqnarray}
A= \int_{r_+} {{\sqrt{g_{\theta\theta}g_{\phi\phi}}}d\theta d\phi}= 4\pi (r^2_+ + a^2)
\label{area}
\end{eqnarray}
and semi-classical entropy is given by one fourth of its horizon area
\begin{equation}
S=\frac{A}{4\hbar}=\frac{\pi(r_+^2+a^2)}{\hbar}.
\label{equation 8}
\end{equation}
The semiclassical Hawking temperature in terms of surface gravity ($\kappa$) of the Kerr-Newman black hole is given by
\begin{eqnarray}
T_{\textrm H}=\frac{\hbar\kappa}{2\pi}=\frac{\hbar}{2\pi}\frac{(r_+ -M)}{(r^2_+ +a^2)}.
\label{hawktemp}
\end{eqnarray}
Using (\ref{2.2}), (\ref{angv}), (\ref{epot}) and (\ref{hawktemp}) one can find the following quantities,
\begin{eqnarray}\frac{1}{T_{\textrm H}}=\frac{2\pi}{\hbar}\left(\frac{2M{[M+(M^2-\frac{J^2}{M^2}-Q^2)^{1/2}}]-Q^2}{{(M^2-\frac{J^2}{M^2}-Q^2)^{1/2}}}\right),
\label{4.5}\end{eqnarray}
\begin{eqnarray}
-\frac{\Omega_{\textrm H}}{T_{\textrm H}}=-\frac{2\pi J}{\hbar M}\left(\frac{1}{(M^2-\frac{J^2}{M^2}-Q^2)^{1/2}}\right),
\label{4.6}
\end{eqnarray}
\begin{eqnarray}
-\frac{\Phi_{\textrm H}}{T_{\textrm H}}=-\frac{2\pi Q[M+(M^2-\frac{J^2}{M^2}-Q^2)^{1/2}]}{\hbar(M^2-\frac{J^2}{M^2}-Q^2)^{1/2}}.
\label{4.7}
\end{eqnarray}

\section{\label{appendix2C}Dimensional Reduction for Charged Myers--Perry Black Hole}
\renewcommand{\theequation}{2C.\arabic{equation}}
\setcounter{equation}{0}
 We consider the complex scalar action under the background metric (\ref{metric}):
\begin{eqnarray}
{\cal{A}} &=& -\int d^{N+1}X \varphi^* (\nabla_\mu + iA_\mu)(\nabla^\mu-iA^\mu)\varphi
\label{new1}
\end{eqnarray}
where $d^{N+1}X = \sqrt{-g} dt dr d\theta d\phi d\chi_1 ...... d\chi_{N-3}$ with $\sqrt{-g}$ given by (\ref{deter}). Now substituting the expansion for $\varphi$ in terms of spherical harmonics
\begin{eqnarray}
\varphi = \sum_{m_1,m_2,...,m_{N-1}}\varphi_{m_1,m_2,...,m_{N-1}}(t,r)Y_{m_1,m_2,...,m_{N-1}}(\theta,\phi,\chi_1,....,\chi_{N-3})
\label{new2}
\end{eqnarray}
and then using 
\begin{eqnarray}
&&dr^* = \frac{(r^2+a^2)r^{N-2}}{\Delta}dr
\label{dim12}
\nonumber
\\
&&\hat{L}_\phi Y_{m_1,m_2,...,m_{N-1}}(\theta,\phi,\chi_1,....,\chi_{N-3}) = m_2 Y_{m_1,m_2,...,m_{N-1}}(\theta,\phi,\chi_1,....,\chi_{N-3})
\label{dim2}
\end{eqnarray}
we obtain,
\begin{eqnarray}
{\cal{A}}&=& \sum_{m'_1,m'_2,...,m'_{N-1}}\varphi^*_{m'_1,m'_2,...,m'_{N-1}}(t,r)Y^*_{m'_1,m'_2,...,m'_{N-1}}(\theta,\phi,\chi_1,....,\chi_{N-3})\nonumber\\&&\int dt~dr^* d\Omega_{N-1}\sqrt{\gamma}r^{N-3}~\sin\theta~\cos^{N-3}\theta 
\Big[(r^2+a^2) (\partial_t+iA_t)^2
- \partial_{r^*}(r^2+a^2)\partial_{r^*} 
\nonumber
\\
&+& \frac{\Delta}{(r^2+a^2)r^{N-2}}({\hat G} + {\hat L}^2_{(\theta,\chi_1,....,\chi_{N-3})})\Big]\nonumber\\&&\sum_{m_1,m_2,...,m_{N-1}}\varphi_{m_1,m_2,...,m_{N-1}}(t,r)Y_{m_1,m_2,...,m_{N-1}}(\theta,\phi,\chi_1,....,\chi_{N-3})
\label{dim1.01}
\end{eqnarray}
where,
\begin{eqnarray}
A_t = -eV(r) - m_2\Omega(r); V(r) = \frac{Q}{(N-2)r^{N-4}(r_+^2+a^2)}; \Omega = \frac{a}{r^2+a^2}
\label{dim3}
\end{eqnarray}
and ${\hat G}$,  ${\hat L}^2_{(\theta,\chi_1,....,\chi_{N-3})}$ are operators whose explicit forms are not necessary and $e$ is the charge of the complex scalar field. Since near the horizon $\Delta\rightarrow 0$, the scalar action reduces to
\begin{eqnarray}
{\cal{A}} &\simeq& \sum_{m'_1,m'_2,...,m'_{N-1}}\varphi^*_{m'_1,m'_2,...,m'_{N-1}}(t,r)Y^*_{m'_1,m'_2,...,m'_{N-1}}(\theta,\phi,\chi_1,....,\chi_{N-3})\nonumber\\&&\int dt~dr^*d\Omega_{N-1} \sqrt{\gamma}r^{N-3}~\sin\theta~\cos^{N-3}\theta 
\Big[(r^2+a^2) (\partial_t+iA_t)^2
- \partial_{r^*}(r^2+a^2)\partial_{r^*} 
\Big]\nonumber\\&&\sum_{m_1,m_2,...,m_{N-1}}\varphi_{m_1,m_2,...,m_{N-1}}(t,r)Y_{m_1,m_2,...,m_{N-1}}(\theta,\phi,\chi_1,....,\chi_{N-3})
\label{dim1.02}
\end{eqnarray}
Now using the orthonormal condition 
\begin{eqnarray}
&&\int d\theta d\phi d\chi_1 .... d\chi_{N-3} \sin\theta \cos^{N-3}\theta \sin\chi_1 .... \sin\chi_{N-4}
\nonumber
\\
&&Y^*_{m_1,m_2,...,m_{N-1}}(\theta,\phi,\chi_1,....,\chi_{N-3}) Y_{m'_1,m'_2,...,m'_{N-1}}(\theta,\phi,\chi_1,....,\chi_{N-3})
\nonumber
\\
&=&\delta_{m'_1,m_1}.......\delta_{m'_{N-1},m_{N-1}}
\label{new3}
\end{eqnarray}
we obtain,
\begin{eqnarray}
{\cal{A}} &\simeq & \sum_{m_1,m_2,...,m_{N-1}}\varphi^*_{m_1,m_2,...,m_{N-1}}(t,r)\int dt~dr^* r^{N-3}
\nonumber
\\
&&\Big[(r^2+a^2) (\partial_t+iA_t)^2
- \partial_{r^*}(r^2+a^2)\partial_{r^*} 
\Big]\varphi_{m_1,m_2,...,m_{N-1}}(t,r)
\label{dim1.03}
\end{eqnarray}
Reverting back to the original $r$ coordinate,
\begin{eqnarray}
{\cal{A}} &\simeq & 
-\sum_{m_1,m_2,....,m_{N-1}} \int dt~dr H(r_+) \varphi^*_{m_1,m_2,....,m_{N-1}}(t,r)\Big[-\frac{(r^2+a^2)r^{N-2}}{\Delta} (\partial_t+iA_t)^2 
\nonumber
\\
&+& \partial_r\frac{\Delta}{(r^2+a^2)r^{N-2}}\partial_r \Big]\varphi_{m_1,m_2,....,m_{N-1}}(t,r)
\label{dim4}
\end{eqnarray}
Here $H$ is a function of $r_+$ whose explicit expression is not required for our analysis.
This shows that each partial wave mode of the fields can be described near the horizon as a (1 + 1) dimensional complex scalar field with two U(1) gauge potentials $V (r)$,$\Omega(r)$ and the dilaton field $\psi = H(r_+)$. It should be noted that the above action for each $m_1,m_2,....,m_{N-1}$ can also be obtained from a complex scalar field action in the background of metric 
\begin{eqnarray}
ds^2_{eff} = -F(r)dt^2 + \frac{dr^2}{F(r)},
\label{effective1}
\end{eqnarray}
where
\begin{eqnarray}
F(r) = 1 - \frac{m}{r^{N-4}(r^2+a^2)} + \frac{q^2}{r^{2(N-3)}(r^2+a^2)}
\label{effective2}
\end{eqnarray}
with the dilaton field $\psi$. A detailed study on dimensional reduction can be found in \cite{robin,kumet,iso1}. Subsequently we use this effective metric to derive the Hawking temperature (\ref{temp}).
\end{subappendices}

\chapter{\label{chap:hawking-effect} Quantum tunneling and the Hawking effect}


In pathbreaking works Hawking \cite{hk1}-\cite{hk3}, considering quantum fields moving in a curved spacetime, first showed that black holes behave exactly like a blackbody. They emit radiation with a pure thermal spectrum, provided the backreaction effect on the metric is neglected. This is known as Hawking effect. After this derivation some alternative suggestions have also been put forward. There are two popular approaches. One is based on the gravitational anomaly- either the trace anomaly \cite{chris} or the divergence anomaly  \cite{robin}, \cite{bkk}. The other involves the quantum tunneling mechanism \cite{Paddy}, \cite{ng1}. An attempt has also been made made to connect these two approaches \cite{subir}-\cite{Majhiconnect}. In this chapter we adopt the quantum tunneling method for discussing Hawking effect. In this picture one can derive both Hawking temperature and blackbody spectrum. There are two variants in which temperature can be derived. The radiation spectrum is found by following a density matrix method first introduced in \cite{Majhiflux}. These ideas have been profusely researched in the last one decade \cite{gran2}-\cite{bhatt}.

In the tunneling picture a pair creation takes place inside the black hole event horizon. One mode goes towards the centre of the black hole and is completely lost. There is another mode which come across the event horizon from inside to outside. However this path is classically forbidden and can only be defined in terms of complex coordinates.   Using this  method of complex path (also known as Hamilton-Jacobi (WKB) type approach) we calculate the semi-classical Hawking temperature. This method, however, is unable to tell us anything about the radiation spectrum. Therefore we also consider a density matrix method and derive the blackbody radiation spectrum not only for black holes but also for cosmological horizons. In addition to that we discuss a general formalism where one can go beyond the semi-classical framework. Incidentally in this case we generate some higher order corrections (in $\hbar$) to the semi-classical Hawking temperature. A similar conclusion is drawn by finding the modified blackbody spectrum by going beyond the semi-classical approximation.   

This chapter is organised in the following manner. In section-\ref{qthj} we discuss tunneling of scalar particles and fermions and derive semi-classical Hawking temperature considering the Kerr-Newman spacetime. The next section (\ref{qtbeyht}) is used for computation of temperature of a black hole beyond the semi-classical limit. Using a density matrix technique we calculate the blackbody spectrum of Hawking radiation in section-\ref{qtbs}. Finally in the last section (\ref{qtbeybs}) modified blackbody spectrum is derived where we show that the thermal nature does not get changed, rather, temperature gets some higher order corrections. There are three appendices presented at the very end of this chapter where we provide maximal extension of spacetimes through the black hole (event) and cosmic horizons (appendices \ref{appendix3A} and  \ref{Appendix3B})) and also describe a method of identifying various modes of particle creation  (appendix \ref{Appendix3C}).  

\section{\label{qthj}Derivation of semi-classical Hawking temperature}
In the derivation of Hawking temperature through tunneling mechanism we shall consider two cases of scalar and fermion particles separately. Both these approaches yield an identical result for Hawking temperature which is reassuring. In our analysis we shall focus on Kerr-Newman spacetime but this approach is general enough to include all such space-time metrics where $r-t$ sector can be decoupled from the angular parts in the near horizon limit. 

For the case of Kerr-Newman (KN) spacetime, in order to isolate the $r-t$ sector, we first rewrite the original metric (\ref{2.1}) in the following form,
\begin{align}
ds^{2}  &  =(g_{00}-\frac{g_{03}^2}{g_{33}})dt^{2}+g_{33}(d\phi+\frac{g_{03}}{g_{33}}dt)^2 + g_{11}dr^2+g_{22}d\theta^2
\label{2.3nh}
\end{align}
As tunneling mechanism is a local phenomena and defined infinitesimally close to the event horizon we take the near horizon form of this metric and redefine the angular part in such a way that the $r-t$ sector becomes isolated. This redefinition only changes the notion of the total energy for the tunneling particle (as we shall see) and does not affect the thermodynamical entities of the black hole. In the case of (\ref{2.3nh}) this issue is very subtle since the metric coefficients not only depend on $r$ but also on $\theta$. However, because of the presence of an ergosphere, $-g_{00}(r,\theta)$ in (\ref{2.1}) is positive at the horizon for two specific values of $\theta$, say, $\theta_0=0$ or $\pi$. For these two values of $\theta$ the ergosphere and the event horizon coincide. Physically tunneling of any particle through the horizon of the Kerr-Newman black hole is allowed only for these two specific values of $\theta$. 

When we take the near horizon limit of the metric (\ref{2.3nh}) the value of $\theta$ is first fixed to $\theta_0$. Then the metric near the horizon for fixed $\theta=\theta_{0}$\ is given by,       
\begin{equation}
ds^{2}=-F(r_{+},\theta_{0})(r-r_{+})dt^{2}+\frac{dr^{2}}{\tilde g(r_{+},\theta_{0})(r-r_{+})}+g_{33}(r_{+},\theta_{0})(d\phi-\Omega_{\textrm{H}}dt)^{2} 
\label{2.4}
\end{equation}
where,
\begin{eqnarray}
F &=& -\left[\partial_{r}\left(-g_{00}+\frac{g_{03}^2}{g_{33}}\right)\right]_{r_+}\nonumber\\
\tilde g &=& \left[\partial_{r}(g_{11}^{-1})\right]_{r_+}\nonumber\\
\Omega_{\textrm H}&=& -\frac{g_{03}}{g_{33}}\big|_{r_+}=\frac{a}{r_{+}^{2}+a^{2}}.
\label{2.5}
\end{eqnarray}
Here $\Omega_{\textrm{H}}$ is the angular velocity at event horizon. Finally, a coordinate transformation
\begin{eqnarray} 
d\chi=d\phi-\Omega_{\textrm H}dt\implies\chi=\phi-\Omega_{\textrm H} t
\label{2.51}
\end{eqnarray}
 will take the metric (\ref{2.4}) into an effective 2 dimensional $r-t$ sector, given by %
\begin{equation}
ds^{2}=-F(r_{+},\theta_{0})(r-r_{+})dt^{2}+\frac{dr^{2}}{\tilde g(r_{+},\theta_{0})(r-r_{+})}+g_{33}(r_{+},\theta_{0})d\chi^{2}, 
\label{2.6}
\end{equation}
 where the `$r-t$' sector is isolated from the angular part $d\chi^2$ and has the following form,
\begin{eqnarray}
ds^2=-f(r)dt^2+{\frac{1}{g(r)}}{dr^2}, 
\label{2.7ht}
\end{eqnarray}
where
\begin{eqnarray}
f(r)=F(r_{+},\theta_{0})(r-r_{+})
\label{2.71}\\
g(r)=\tilde g(r_{+},\theta_{0})(r-r_{+}). 
\nonumber
\end{eqnarray}
Also note that, as inside the ergosphere no observer can be static, the redefinition of angular coordinate in (\ref{2.51}) is necessary for taking the near horizon limit. Physically it ensures  that the observer is in a nonstatic frame of reference. Now we shall consider the new reduced metric and study scalar and fermion tunneling separately.

\subsection{{Scalar Particle tunneling}}
The massless scalar particle in spacetime (\ref{2.6}) is governed by the Klein-Gordon equation
\begin{equation}
-\frac{1}{\sqrt{-g}}{\partial_\mu[g^{\mu\nu}\sqrt{-g}\partial_{\nu}]\Phi}=0.
\label{2.8sp}
\end{equation}  
 In the tunneling approach we are concerned about the radial trajectory, so that only the $r-t$ sector (\ref{2.7ht}) of the metric (\ref{2.6}) is relevant.

To solve (\ref{2.8sp}) under the effective background metric (\ref{2.6}) we therefore start with the following standard WKB ansatz for $\Phi$ as
\begin{eqnarray}
\Phi(r,t)=\exp[-\frac{i}{\hbar}{{\cal S}(r.t)}], 
\label{2.9ssp}
\end{eqnarray}
where,
\begin{eqnarray}
{\cal S}(r,t)={\cal S}{_0}(r,t)+  \sum_i{\hbar^i {\cal S}_i(r,t)}.
\label{2.10ssa}
\end{eqnarray}
Substituting $\Phi$ in (\ref{2.8sp}) yields,
\begin{eqnarray}
&&\frac{i}{\sqrt{f(r)g(r)}}\Big(\frac{\partial S}{\partial t}\Big)^2 - i\sqrt{f(r)g(r)}\Big(\frac{\partial S}{\partial r}\Big)^2 - \frac{\hbar}{\sqrt{f(r)g(r)}}\frac{\partial^2 S}{\partial t^2} + \hbar \sqrt{f(r)g(r)}\frac{\partial^2 S}{\partial r^2}
\nonumber
\\
&&+ \frac{\hbar}{2}\Big(\frac{\partial f(r)}{\partial r}\sqrt{\frac{g(r)}{f(r)}}+\frac{\partial g(r)}{\partial r}\sqrt{\frac{f(r)}{g(r)}}\Big)\frac{\partial S}{\partial r}=0.
\label{2.9a}
\end{eqnarray}
Considering the semi-classical limit, i.e. only considering the terms of zeroth order in  $\hbar$ in both sides of the above equation, we get
\begin{eqnarray}
\frac{\partial S_0}{\partial t}=\pm \sqrt{f(r)g(r)}\frac{\partial S_0}{\partial r}.
\label{2.11pd}
\end{eqnarray}
This first order partial differential has a structure similar to the Hamilton-Jacobi (HJ) equation. Realising this one can now consider a general form of the semi-classical action for the original Kerr-Newman spacetime (\ref{2.1}), given by
\begin{eqnarray}
{\cal S}_0(r,t,\theta,\phi)=K_{\xi^{\mu}_{(t)}}~t+ K_{\xi^{\mu}_{(\phi)}}~\phi+ \tilde {\cal S}_0(r,\theta).
\label{2.12}
\end{eqnarray}
Here $K_{\xi^{\mu}_{(t)}}$ and $K_{\xi^{\mu}_{(\phi)}}$ are the Komar conserved quantities \cite{komar,komar2} corresponding to the two Killing vectors $\xi^{\mu}_{(t)}$ (timelike) and $\xi^{\mu}_{(\phi)}$ (spacelike) and represented by the equations (\ref{eeff}) and (\ref{peff}).
 
In the near horizon approximation for fixed $\theta=\theta_0$ and using (\ref{2.51}) one can isolate the semi-classical action (\ref{2.12}) for the `$r-t$' sector as,
\begin{eqnarray}
{\cal S}_0(r,t)=\omega t+ \tilde {\cal S}_0(r),
\label{2.13}
\end{eqnarray}
 where
\begin{eqnarray}
\omega=(K_{\xi^{\mu}_{(t)}}+\Omega_{H}K_{\xi^{\mu}_{(\phi)}})\big|_{r_+}
\label{2.13a}
\end{eqnarray}
is identified as the total energy of the tunneling particle at the event horizon.

To find the solution for $S_0(r,t)$, let us put (\ref{2.13}) in the partial differential equation (\ref{2.11pd}) and integrate to obtain
\begin{eqnarray}
\tilde{{\cal S}_0}(r) =  \pm \omega\int_C\frac{dr}{\sqrt{f(r)g(r)}}
\label{2.17}
\end{eqnarray}
The + (-) sign indicates that the particle is ingoing (outgoing) \footnote {See appendix \ref{Appendix3C} for further details.}. Using the above expression one can write (\ref{2.13}) as
\begin{eqnarray}
{\cal S}_0(r,t)=(\omega t \pm \omega\int_C\frac{dr}{\sqrt{f(r)g(r)}}) .
\label{2.18}
\end{eqnarray}
The solution for the ingoing and outgoing particle of the Klein-Gordon equation under the background metric (\ref{2.7ht}) follows from (\ref{2.9ssp}) and (\ref{2.10ssa}),
\begin{eqnarray}
\Phi_{{\textrm {in}}}= {\textrm{exp}}\Big[-\frac{i}{\hbar}\Big(\omega t +\omega\int_C\frac{dr}{\sqrt{f(r)g(r)}}\Big)\Big]
\label{2.19}
\end{eqnarray} 
and
\begin{eqnarray}
\Phi_{{\textrm {out}}}= {\textrm{exp}}\Big[-\frac{i}{\hbar}\Big(\omega t  -\omega\int_C\frac{dr}{\sqrt{f(r)g(r)}}\Big)\Big].
\label{2.20}
\end{eqnarray} 
The paths for the ingoing and outgoing particle crossing the event horizon are not same. The ingoing particle can cross the event horizon classically, whereas, the outgoing particle trajectory is classically forbidden. The metric coefficients for `$r-t$' sector alter sign at the two sides of the event horizon. Therefore, the path in which tunneling takes place has an imaginary time coordinate (${\textrm {Im}}~t$). The ingoing and outgoing probabilities are now given by,
\begin{eqnarray}
P_{{\textrm{in}}}=|\Phi_{{\textrm {in}}}|^2= {\textrm{exp}}\Big[\frac{2}{\hbar}\Big(\omega{\textrm{Im}}~t +\omega{\textrm{Im}}\int_C\frac{dr}{\sqrt{f(r)g(r)}}\Big)\Big]
\label{2.21}
\end{eqnarray}
and
\begin{eqnarray}
P_{{\textrm{out}}}=|\Phi_{{\textrm {out}}}|^2= {\textrm{exp}}\Big[\frac{2}{\hbar}\Big(\omega{\textrm{Im}}~t -\omega{\textrm{Im}}\int_C\frac{dr}{\sqrt{f(r)g(r)}}\Big)\Big].
\label{2.22}
\end{eqnarray}
Since in the classical limit ($\hbar\rightarrow 0$) $P_{{\textrm {in}}}$ is unity, one has,
\begin{eqnarray}
{\textrm{Im}}~t = -{\textrm{Im}}\int_C\frac{dr}{\sqrt{f(r)g(r)}}.
\label{2.23}
\end{eqnarray}
The presence of this imaginary time component is due to the connection between the two patches (in Kruskal-Szekeres coordinates) exterior and interior of the event horizon. This connection is only possible by considering a contribution coming from the imaginary time coordinate. The value of this contribution is $2\pi i M$ \cite{oth3} which exactly coincides with (\ref{2.23}) evaluated for the Schwarzschild case with $f(r)= g(r)= (1- \frac{2M}{r})$.

As a result the outgoing probability for the tunneling particle becomes,
\begin{eqnarray}
P_{{\textrm{out}}}={\textrm{exp}}\Big[-\frac{4}{\hbar}\omega{\textrm{Im}}\int_C\frac{dr}{\sqrt{f(r)g(r)}}\Big].
\label{2.24}
\end{eqnarray}
The principle of ``detailed balance'' \cite{Paddy} for the ingoing and outgoing probabilities states that,
\begin{eqnarray}
P_{{\textrm{out}}}= {\textrm {exp}}\Big(-\frac{\omega}{T_{\textrm {bh}}}\Big)P_{\textrm{in}}={\textrm{exp}} \Big(-\frac{\omega}{T_{\textrm bh}}\Big)
\label{2.25}
\end{eqnarray}
where in the second equality we have considered $P_{\textrm{in}}=1$. Comparing (\ref{2.24}) and (\ref{2.25}) semiclassical Hawking temperature for the Kerr-Newman black hole is now given by
\begin{eqnarray}
T_{\textrm H} = \frac{\hbar}{4}\Big({\textrm{Im}}\int_C\frac{dr}{\sqrt{f(r)g(r)}}\Big)^{-1}
\label{2.27}
\end{eqnarray}
Using the expressions of $f(r)$ and $g(r)$ form (\ref{2.71}) it follows that,
\begin{eqnarray}
T_{\textrm H} = \frac{\hbar\sqrt{F(r_+,\theta_0)\tilde g(r_+,\theta_0)}}{4\pi}=\frac{\hbar}{2\pi}\frac{(r_+ -M)}{(r^2_+ +a^2)},
\label{2.28}
\end{eqnarray}
which is the familiar result of semi-classical Hawking temperature for KN spacetime. An identical result is also obtained by studying tunneling of fermions. We now discuss this issue.

\subsection{Fermion tunneling}


In this subsection we discuss Hawking effect through the tunneling of fermions. The Dirac equation for massless fermions is given by
\begin{eqnarray}
i\gamma ^{\mu }D_{\mu }\psi  =0,
\label{3.1}
\end{eqnarray} 
 where the covariant derivative is defined as,
\begin{eqnarray}
D_{\mu } =\partial _{\mu }+\frac{1}{2}i\Gamma _{\text{ \ }\mu }^{\alpha \text{ \ }%
\beta }\Sigma _{\alpha \beta }\nonumber\\
\Gamma^{\alpha~\beta}_{~\mu}= g^{\beta\nu}\Gamma^{\alpha}_{\mu\nu}
\label{3.2}
\end{eqnarray} 
and
\begin{eqnarray}
\Sigma _{\alpha \beta } =\frac{1}{4}i[\gamma _{\alpha },\gamma _{\beta }]
\label{3.3}
\end{eqnarray} 
The $\gamma^{\mu }$ matrices satisfy the anticommutation relation $\{\gamma ^{\mu },\gamma ^{\nu}\}=2g^{\mu \nu }\times {\bf 1}$.

We are concerned only with the radial trajectory and for this it is useful to work with the metric (\ref{2.7ht}). Using this one can write (\ref{3.1}) as
\begin{eqnarray}
i\gamma^{\mu}\partial_{\mu}\psi- \frac{1}{2}\left(g^{tt}\gamma^{\mu}\Gamma_{\mu t}^{r}-g^{rr}\gamma^{\mu}\Gamma_{\mu r}^{t}\right)\Sigma_{rt}\psi=0
\label{3.4}
\end{eqnarray}
The nonvanishing connections which contribute to the resulting equation are
\begin{eqnarray}
\Gamma^{r}_{tt}= \frac{f'g}{2}; \Gamma^{t}_{tr}= \frac{f'}{2f}.
\label{3.5}
\end{eqnarray}
Let us define the $\gamma$ matrices for the `$r-t$' sector as
\begin{eqnarray}
\gamma ^{t} &=&\frac{1}{\sqrt{f(r)}}\gamma ^{0},~~~~~\gamma^{r}=\sqrt{g(r)}\gamma ^{3},
\label{3.6}
\end{eqnarray}
where $\gamma^0$ and $\gamma^3$ are members of the standard Weyl or chiral representation of $\gamma$ matrices \cite{fermion,fermion2} in Minkwoski spacetime, expressed as
\begin{eqnarray}
\gamma ^{0} &=&\left( 
\begin{array}{cc}
0 & I \\ 
-I & 0%
\end{array}%
\right) \
 \text{ \ \ \ }\gamma ^{3}=\left( 
\begin{array}{cc}
0 & \sigma ^{3} \\ 
\sigma ^{3} & 0%
\end{array}%
\right).
\label{3.7}
\end{eqnarray}

Using (\ref{3.3}), (\ref{3.5}) and (\ref{3.6}) the equation of motion (\ref{3.4}) is simplified as,
\begin{eqnarray}
i\gamma^t\partial_t\psi+i\gamma^r\partial_r\psi+\frac{f'(r)g(r)}{2f(r)}\gamma^{t}\Sigma_{rt}\psi=0,
\label{3.8}
\end{eqnarray}
where 
\begin{eqnarray}
\Sigma_{rt}=\frac{i}{2}\sqrt\frac{f(r)}{g(r)}{\left(\begin{array}{c c c c}
1 & 0 & 0 & 0\\
0 & -1 & 0 & 0\\
0 & 0 & -1 & 0\\
0 & 0 & 0 & 1
\end{array}\right).}
\label{3.9}
\end{eqnarray}

The spin up (+ ve `$r$' direction) and spin down (- ve `$r$' direction) ansatz for the Dirac field have the following forms respectively,  
\begin{eqnarray}
\psi_\uparrow(t,r)= \left(\begin{array}{c}
A(t,r) \\
0 \\
B(t,r) \\
0
\end{array}\right){\textrm{exp}}\Big[-\frac{i}{\hbar}I_\uparrow (t,r)\Big]
\label{3.91}
\end{eqnarray}  
and
\begin{eqnarray}
\psi_\downarrow(t,r)= \left(\begin{array}{c}
0 \\
C(t,r) \\
0 \\
D(t,r) \\
\end{array}\right){\textrm{exp}}\Big[-\frac{i}{\hbar}I_\downarrow (t,r)\Big].
\label{3.10}
\end{eqnarray}
 Here $I_{\uparrow}(r,t)$ is the action for the spin up case and will be expanded in powers of $\hbar$. We shall perform our analysis within the semiclassical framework and only for the spin up case since the spin down case is fully analogous. On substitution of the ansatz (\ref{3.91}) in (\ref{3.8}) and simplifying, we get the following two nonzero equations,
\begin{eqnarray}
B(t,r)[\partial_t{I_{\uparrow}(r,t)}+\sqrt{fg}\partial_rI_{\uparrow}(r,t)]=0
\label{3.11}
\end{eqnarray}
and
\begin{eqnarray}
A(t,r)[\partial_t{I_{\uparrow}(r,t)}-\sqrt{fg}\partial_rI_{\uparrow}(r,t)]=0.
\label{3.12}
\end{eqnarray}
Now let us expand all the variables in the `$r-t$' sector in powers of $\hbar$, as
\begin{eqnarray}
I_{\uparrow}(r,t)=I(r,t)=I_0(r,t)+\displaystyle\sum_i \hbar^i I_i(r,t)\nonumber\\
A(r,t)=A_0(r,t)+\displaystyle\sum_i \hbar^i A_i(r,t)\label{3.13}\\
B(r,t)=B_0(r,t)+\displaystyle\sum_i \hbar^i B_i(r,t).\nonumber
\end{eqnarray}
In the semiclassical limit we perform calculations by considering the lowest order terms ($\hbar^0$) in the above expansion. Substituting these terms from (\ref{3.13}) into (\ref{3.11}) and (\ref{3.12}) and equating the coefficients of $\hbar^0$ in both sides we find 
\begin{eqnarray}
B_0(r,t)\left(\partial_t I_0(r,t)+\sqrt{fg}~\partial_rI_0(r,t)\right)=0\nonumber\\
A_0(r,t)\left(\partial_t I_0(r,t)-\sqrt{fg}~\partial_rI_0(r,t)\right)=0.
\label{3.14}
\end{eqnarray}
Thus we have the following sets of solutions, respectively, for $B_0\neq0$ and $A_0\neq0$,
\begin{eqnarray}
{\textrm{Set-I}:}~~~ \partial_t I_0(r,t)+\sqrt{fg}~\partial_rI_0(r,t)=0   
\label{3.15sc}
\end{eqnarray}
\begin{eqnarray}
{\textrm{Set-II}:}~~~ \partial_t I_0(r,t)-\sqrt{fg}~\partial_rI_0(r,t)=0.   
\label{3.16sc}
\end{eqnarray}
Similar to the scalar particle tunneling here also one can separate the semiclassical action for the `$r-t$' sector as
\begin{eqnarray}
I_0(r,t)=\omega t +W_0(r),
\label{3.17}
\end{eqnarray}
Substituting (\ref{3.17}) in (\ref{3.15sc}) and (\ref{3.16sc}) and integrating we get
\begin{eqnarray}
W_0^{\pm}(r)=\pm\omega\int_C\frac{dr}{\sqrt{f(r)g(r)}}
\label{3.18}
\end{eqnarray}
and subsequently
\begin{eqnarray}
I_0(r,t)=\left(\omega t \pm\omega\int_C\frac{dr}{\sqrt{f(r)g(r)}}\right),
\label{3.19}
\end{eqnarray}
where + (-) sign implies that the particle is ingoing (outgoing). This is an exact analogue of the scalar particle tunneling case (\ref{2.18}) and ultimately it leads to an identical expression of semi-classical Hawking temperature as given by (\ref{2.28}) by exactly mimicking the steps discussed in the case of scalar particle example. 

The agreement of the final result of temperature by using two different kinds of particles enhances the acceptability of the tunneling mechanism in discussing Hawking effect. In the calculations in last two subsections we have not considered the higher order terms in $\hbar$. These were confined within known semi-classical framework. However it is possible to include the higher order (in $\hbar$) terms in the computation by going beyond this semi-classical approximation. In the next section we shall address this issue and generate higher order corrections to Hawking temperature.

\section{\label{qtbeyht}Derivation of corrected Hawking temperature beyond semi-classical limit}
In this section we build on the ideas of tunneling mechanism discussed in last section and present a formulation to include higher order terms in WKB ansatz. Both scalar and fermion tunneling will be discussed which eventually generate higher order corrections to the semi-classical Hawking temperature. This approach was first introduced in \cite{Majhibeyond} for spherically symmetric, static black holes. Our motivation is to include the nonspherically symmetric, nonstatic black holes into the general formalism.

\subsection{Scalar particle tunneling}
As we have already mentioned that by going beyond the semi-classical framework we mean to include higher order in $\hbar$ terms coming with the WKB ansatz (\ref{2.9ssp}) and (\ref{2.10ssa}). The idea is to substitute (\ref{2.9ssp}) into the field equation (\ref{2.8sp}) considering the full expansion (\ref{2.10ssa}). Then one can isolate the coefficients of different powers of $\hbar$ from two sides which ultimately can be reduced to the following mathematical structures {\footnote{where to find the coefficient of $\hbar^1$ we have used the relation obtained for $\hbar^0$ case and so on.}} 
\begin{eqnarray}
\hbar^0~:~&&\frac{\partial S_0}{\partial t}=\pm \sqrt{f(r)g(r)}\frac{\partial S_0}{\partial r},
\nonumber
\\
\hbar^1~:~&&\frac{\partial S_1}{\partial t}=\pm \sqrt{f(r)g(r)}\frac{\partial S_1}{\partial r},
\nonumber
\\
\hbar^2~:~&&\frac{\partial S_2}{\partial t}=\pm \sqrt{f(r)g(r)}\frac{\partial S_2}{\partial r},
\nonumber
\\
.
\nonumber
\\
.
\nonumber
\\
.
\nonumber
\end{eqnarray} 
and so on. Note that the $n$-th order solution is expressed by,
\begin{eqnarray}
\frac{\partial S_n}{\partial t}=\pm \sqrt{f(r)g(r)}\frac{\partial S_n}{\partial r},
\label{2.11a}
\end{eqnarray}
where ($n=~0,~i;~i= 1,2,...) $.

 The solution for all ${\cal S}_i(r,t)$' s, is similar to (\ref{2.13}), modulo a proportionality factor, since they satisfy generically identical equations as (\ref{2.11a}). The most general form of action including the contribution from all orders of $\hbar$ is then given by
\begin{eqnarray}
{\cal S}(r,t)=(1+\sum\gamma_i \hbar^i) {\cal S}_0(r,t),
\label{2.14}
\end{eqnarray}
It is clear that the dimension of $\gamma_i$ is equal to the dimension of $\hbar^{-i}$. Let us now perform the following dimensional analysis to express these $\gamma_i$' s in terms of dimensionless constants. In (3+1) dimensions in the unit of $G= c= \kappa_B= \frac{1}{4\pi\epsilon_0}= 1$, $\sqrt\hbar$ is proportional to Planck length ($l_p$), Planck mass ($m_p$) and Planck charge ($q_p$) {\footnote {$l_p=\sqrt{{\frac{\hbar G}{c^3}}} , m_p=\sqrt{\frac{\hbar c}{G}} , q_p= \sqrt{c\hbar 4\pi\epsilon_0 }.$ }}. Therefore for the Kerr-Newman black hole, where $l_p,~m_p$ and $q_p$ are replaced by $r_+$, $M$ and $Q$, respectively, the most general term having dimension of $\hbar$ can be expressed in terms of these black hole parameters as 
\begin{equation}
H_{\textrm {KN}}(M,J,Q)=  {a_1r^2_{+}}+a_2 {Mr_+}+a_3{M^2}+a_4{r_+ Q}+a_5{MQ}+a_6{Q^2}. 
\label{2.15}
\end{equation} 
Using this the action in (\ref{2.14}) now takes the form
\begin{eqnarray}
{\cal S}(r,t)=(1+\sum\frac{\beta_i\hbar^i}{H_{\textrm {KN}}^{i}}) {\cal S}_0(r,t).
\label{2.16}
\end{eqnarray}
Upon substitution of this action (together with (\ref{2.18})) into (\ref{2.9ssp}) we can now find the ingoing and outgoing modes exactly similar to (\ref{2.19}) and (\ref{2.20}) with an extra factor $(1+\sum\frac{\beta_i\hbar^i}{H_{\textrm {KN}}^{i}})$ in the power of exponential function. Now it is trivial to folllow the rest of the analysis (as provided for the semiclassical case) to calculate the corrected Hawking temperature. This is found to be 
\begin{eqnarray}
T_{\textrm {bh}}=T_{\textrm H}\Big(1+\sum_i\beta_i\frac{\hbar^i}{H_{\textrm {KN}}^i}\Big)^{-1}.
\label{2.26}
\end{eqnarray}

Thus it is seen that the inclusion of higher order terms in $\hbar$ modifies the temperature of a black hole. Although this derivation is carried out for the KN spacetime it should be mentioned that all other black holes which are only special cases (Schwarzschild, Reissner-Nordstrom, Kerr) are already included in this general formalism. In each case we find that higher order corrections contain some unknown parameters which cannot be fixed within the tunneling formalism. However as we shall see, most of these unknown parameters can be fixed by considering the state function nature of black hole entropy and also by using the trace anomaly of the stress/energy-momentum tensor. These issues are discussed in chapters 4 and 5.

\subsection{Fermion tunneling}
One can also extend the previous analysis of calculating corrected Hawking temperature by considering fermions. To do that we consider the full expansion of the ansatz (\ref{3.91},\ref{3.13}) by using the set of equations (\ref{3.11},\ref{3.12}). Rearranging terms coming with different powers of $\hbar$ now we have the following sets of solutions, respectively, for $B_a{\textrm {'s}}\neq0$ and $A_a{\textrm {'s}}\neq0$,
\begin{eqnarray}
{\textrm{Set-I}:}~~~ \partial_t I_a(r,t)+\sqrt{fg}~\partial_rI_a(r,t)=0   
\label{3.15}
\end{eqnarray}
\begin{eqnarray}
{\textrm{Set-II}:}~~~ \partial_t I_a(r,t)-\sqrt{fg}~\partial_rI_a(r,t)=0.   
\label{3.16}
\end{eqnarray}
As a result, similar to the scalar particle tunneling, here also one can write the most general solution for $I(r,t)$, given by
\begin{eqnarray}
I(r,t)=(1+{\displaystyle\sum_i{\gamma_i\hbar^i}})\left(-\omega t \pm\omega\int_C\frac{dr}{\sqrt{f(r)g(r)}}\right).
\label{3.191}
\end{eqnarray}
This is an exact analogue of the scalar particle tunneling case (\ref{2.14}) and it is trivial to check this will lead to an identical expression of corrected Hawking temperature as given by (\ref{2.26}) and (\ref{2.28}) by exactly mimicking the steps discussed there.

\section{\label{qtbs}Derivation of blackbody spectrum}

In the last section we have calculated Hawking temperature by using the Hamilton-Jacobi type approach. In this section using a density matrix technique we calculate the radiation spectrum for black holes. This approach was first provided in \cite{Majhiflux} for the event horizons of spherically symmetric, static black holes in (3+1) dimensions. Later this method was used for other black holes \cite{kumet} as well as for Rindler type metrics \cite{debraj}. In our work we shall generalise this to arbitrary dimensions, for cosmological horizons and also go beyond the semi-classical framework to find modified blackbody spectrum.  

Let us first consider an arbitrary dimensional spherically symmetric, static spacetime represented by a generic metric
\begin{eqnarray}
ds^2=-g(r)dt^2+\frac{1}{g(r)}dr^2+r^2d\Omega_{D-2}^2
\label{metrichd}
\end{eqnarray}
The positions of the horizons can be found by solving $g(r=r_h)=0$. For the asymptotically flat and AdS black holes there is only one event horizon given by the largest root of this equation. For the dS case the largest root gives the position of cosmic horizon and the second largest root is the black hole event horizon.

We start our analysis by considering the massless scalar particle governed by the Klein-Gordon equation (\ref{2.8sp}) in the background of the spacetime metric (\ref{metrichd}).     
Since in our analysis we shall be dealing with the radial trajectory it is enough to consider the $r-t$ sector of the metric (\ref{metrichd}) to solve (\ref{2.8sp}). Choosing the standard (WKB) ansatz for $\Phi$ given by (\ref{2.9ssp},\ref{2.10ssa}) it follows from (\ref{2.8sp},\ref{metrichd}) 
\begin{eqnarray}
&&\frac{i}{{g(r)}}\Big(\frac{\partial {\cal S}}{\partial t}\Big)^2 - ig(r)\Big(\frac{\partial {\cal S}}{\partial r}\Big)^2 - \frac{\hbar}{g(r)}\frac{\partial^2 {\cal S}}{\partial t^2} + \hbar g(r)\frac{\partial^2 {\cal S}}{\partial r^2}
\nonumber
\\
&&+ \hbar{\frac{\partial g(r)}{\partial r}}\frac{\partial {\cal S}}{\partial r}=0.
\label{2.3}
\end{eqnarray}
Taking the semiclassical limit ($\hbar\rightarrow 0$), we obtain the first order partial differential equation,
\begin{eqnarray}
\frac{\partial {\cal S}_0}{\partial t}=\pm {g(r)}\frac{\partial {\cal S}_0}{\partial r}.
\label{semi}
\end{eqnarray}
This is nothing but the semi-classical Hamilton-Jacobi equation. We choose the semi-classical action for a scalar field moving under the background metric (\ref{metrichd}) in the same spirit as usually done in the semi-classical Hamilton-Jacobi theory. Looking at the time translational symmetry of the spacetime (\ref{metrichd}) we take the form of the semi-classical action as
\begin{eqnarray}
{\cal S}_{0}(r,t)=\Omega t+{\cal S}_{0}(r),
\label{actionform}
\end{eqnarray} 
where $\Omega$ is the conserved quantity corresponding to the time translational Killing vector field. It is identified as the effective energy experienced by the particle. Now substituting this in (\ref{semi}) one can easily find
\begin{eqnarray}
{\cal S}_{0}(r)=\pm\Omega\int\frac{dr}{g(r)},
\label{s0}
\end{eqnarray}
Using (\ref{s0}) in (\ref{actionform}) we finally find the semi-classical action as
\begin{eqnarray}
{\cal S}_{0}(r,t)=\Omega(t\pm\int\frac{dr}{g(r)}).
\label{action2}
\end{eqnarray} 
Now within the semi-classical limit we have the solution for the scalar field (\ref{2.9ssp}), 
\begin{eqnarray} 
\Phi(r,t) &=& e^{[-{\frac{i}{\hbar}}\Omega(t\pm\int\frac{dr}{g(r)})]}\nonumber\\
          &=& e^{-\frac{i}{\hbar}\Omega(t\pm r^*)} 
\label{soln}
\end{eqnarray}
expressed in terms of the tortoise coordinate $r^*=\int\frac{dr}{g(r)}$. We shall require this solution in our analysis to find the radiation spectrum for black holes.


\subsection{\label{bseh}Blackbody spectrum for the black hole (event) horizon}
The first step in the analysis is to find a coordinate system which is regular at the event horizon. We do not need to know the global behaviour of the spacetime. If we can find a coordinate system in which the metric (\ref{metrichd}) is defined both inside and outside the event horizon the purpose is solved. In such a coordinate system we can readily connect the right and left moving modes defined inside and outside the event horizon {\footnote{We use the subscript ``$in$'' for modes inside the event horizon. Since the observer stays outside the event horizon we use the subscript ``$obs$'' for modes outside the event horizon.}}. In appendix (\ref{appendix3A}) a Kruskal-like extension of (\ref{metrichd}) is done by concentrating on the behavior at the black hole event horizon only. In appendix (\ref{Appendix3B}) we show how to identify the left and right moving modes inside and outside the event horizon. 

Now we consider a situation where $n$ number of non-interacting virtual pairs are created inside the black hole horizon. The left and right moving modes inside the black hole horizon, found in (\ref{soln}), are then given by (\ref{conv3}) and (\ref{conv4}) respectively. Here the null coordinates are defined in (\ref{null2}) with $r^*$ given by (\ref{tortoise3}). Likewise the left and right moving modes outside the event horizon are given by (\ref{conv1}) and (\ref{conv2}) respectively. In order to connect the two patches of the coordinate system we first choose the following set from the transformations (\ref{connect2})
\begin{eqnarray}
t_{in} &=& t_{obs}-\frac{i\pi}{2\kappa}\nonumber\\
r^*_{in} &=&  r^*_{obs}+\frac{i\pi}{2\kappa},
\label{connecttt1}
\end{eqnarray}
so that
\begin{eqnarray}
u_{in} &=& u_{obs}-\frac{i\pi}{\kappa}\nonumber\\
v_{in} &=& v_{obs}.
\label{connecttt2}
\end{eqnarray}
We shall discuss about the other choice shortly in this section. With the above choice the inside and the outside modes are connected by,
\begin{eqnarray}
\Phi_{in}^{(R)} &=& {e^{-\frac{\pi\Omega}{\hbar\kappa}}}\Phi_{{obs}}^{{(R)}} \nonumber\\
\Phi_{in}^{(L)} &=& {\Phi_{{obs}}^{{(L)}}},
\label{connect3}
\end{eqnarray} 
where (\ref{conv1}), (\ref{conv2}), (\ref{conv3}), (\ref{conv4}) and (\ref{connecttt2}) have been used.

Any physical state corresponding to $n$ number of virtual pairs inside the black hole event horizon, when observed by an observer outside the horizon, is given by,
\begin{eqnarray}
|\Psi\rangle = N \displaystyle\sum_n |n^{(L)}_{\textrm{in}}\rangle\otimes|n^{(R)}_{\textrm{in}}\rangle
 = N \displaystyle\sum_n e^{-\frac{\pi n\Omega}{\hbar\kappa}}|n^{(L)}_{\textrm{obs}}\rangle\otimes|n^{(R)}_{\textrm{obs}}\rangle,
\label{state}
\end{eqnarray}
where $N$ is a normalisation constant. Here we have used the transformations (\ref{connect3}). Now using the normalization condition $\langle\Psi|\Psi\rangle=1$ and considering $n=0,1,2...$ for bosons or $n=0,1$ for fermions, one obtains
\begin{eqnarray}
N_{\textrm{(boson)}} &=& \Big(1-e^{-\frac{2\pi\Omega}{\hbar\kappa}}\Big)^{\frac{1}{2}}\label{norm1}\\
N_{\textrm{(fermion)}} &=& \Big(1+e^{-\frac{2\pi\Omega}{\hbar\kappa}}\Big)^{-\frac{1}{2}}
\label{norm2}
\end{eqnarray}
Therefore the physical states for them, viewed by an external observer, are given by
\begin{eqnarray}
|\Psi\rangle_{(\textrm{boson})}= \Big(1-e^{-\frac{2\pi\Omega}{\hbar\kappa}}\Big)^{\frac{1}{2}} \displaystyle\sum_n e^{-\frac{\pi n\Omega}{\hbar\kappa}}|n^{(L)}_{\textrm{obs}}\rangle\otimes|n^{(R)}_{\textrm{obs}}\rangle
\label{boson}
\\
|\Psi\rangle_{(\textrm{fermion})}= \Big(1+e^{-\frac{2\pi\Omega}{\hbar\kappa}}\Big)^{-\frac{1}{2}} \displaystyle\sum_n e^{-\frac{\pi n\Omega}{\hbar\kappa}}|n^{(L)}_{\textrm{obs}}\rangle\otimes|n^{(R)}_{\textrm{obs}}\rangle.
\label{fermion}
\end{eqnarray}
The density matrix operator for the bosons can be constructed as
\begin{eqnarray}
{\hat\rho}_{(\textrm{boson})}&=&|\Psi\rangle_{(\textrm{boson})}\langle\Psi|_{(\textrm{boson})}
\nonumber
\\
&=&\Big(1-e^{-\frac{2\pi\Omega}{\hbar\kappa}}\Big) \displaystyle\sum_{n,m} e^{-\frac{\pi (n+m)\Omega}{\hbar\kappa}} |n^{(L)}_{\textrm{obs}}\rangle\otimes|n^{(R)}_{\textrm{obs}}\rangle\langle m^{(R)}_{\textrm{obs}}|\otimes\langle m^{(L)}_{\textrm{obs}}|
\label{density}
\end{eqnarray}
Since the left going modes inside the horizon do not reach the outer observer we can take the trace over all such ingoing modes. This gives the reduced density operator for the right moving modes as
\begin{eqnarray}
{\hat{\rho}}^{(R)}_{(\textrm{boson})}= \Big(1-e^{-\frac{2\pi\Omega}{\hbar\kappa}}\Big) \displaystyle\sum_{n} e^{-\frac{2\pi n\Omega}{\hbar\kappa}}|n^{(R)}_{\textrm{obs}}\rangle \langle n^{(R)}_{\textrm{obs}}|
\label{densityright}
\end{eqnarray}
The average number of particles detected at asymptotic infinity, given by the expectation value of the number operator $\hat{n}$, is now given by
\begin{eqnarray}
\langle n\rangle_{(\textrm{boson})} &=& {\textrm{trace}}({\hat{n}} {\hat{\rho}}^{(R)}_{(\textrm{boson})})\nonumber\\
                       &=& \frac{1}{e^{\frac{2\pi\Omega}{\hbar\kappa}}-1},
\label{expectation}
\end{eqnarray}
which is nothing but the Bose-Einstein distribution of particles corresponding to the Hawking temperature 
\begin{eqnarray}
T_{\textrm H}=\frac{\hbar\kappa}{2\pi}=\frac{\hbar g'(r_h)}{4\pi}.
\label{semihawk}
\end{eqnarray}
The same methodology, when applied on fermions, gives the Fermi-Dirac distribution with the correct Hawking temperature.

Now it may be worthwhile to mention that one could have chosen the other set of transformations in (\ref{connect1}) with opposite relative sign between two terms at the right hand side, as chosen earlier (\ref{connecttt1}), so that, 
\begin{eqnarray}
t_{in} &=& t_{obs}+\frac{i\pi}{2\kappa}\nonumber\\
r^*_{in} &=&  r^*_{obs}-\frac{i\pi}{2\kappa}.
\label{alter}
\end{eqnarray}
For this choice also the inside and outside coordinates (\ref{transformation}), (\ref{transformation2}) are connected. However this is an unphysical solution. To see this note that, use of (\ref{alter}), gives 
\begin{eqnarray}
\Phi^L_{in} &=& \Phi^L_{obs}\nonumber\\
\Phi^R_{in} &=& e^\frac{\pi\Omega}{\hbar\kappa}\Phi^R_{obs}.
\label{alter1}
\end{eqnarray} 
Therefore the probabilities, that the ingoing (left-moving) modes can go inside the event horizon ($P^{R}$) and the outgoing (right-moving) modes can go outside the event horizon, as observed from outside,  are given by
\begin{eqnarray}
P^L=|\Phi^L_{in}|^2=|\Phi^L_{obs}|=1\nonumber\\
P^R=|\Phi^R_{in}|^2=e^\frac{2\pi\Omega}{\hbar\kappa}|\Phi^R_{obs}|^2=e^\frac{2\pi\Omega}{\hbar\kappa}.
\label{alter2}
\end{eqnarray} 
In the classical limit ($\hbar\rightarrow 0$), there is absolutely no chance that any mode can cross the event horizon from inside, therefore one must have $P^R=0$. But we can see from (\ref{alter2}) that it is diverging and therefore the choice (\ref{alter}) is unphysical. It may be worthwhile to mention that, interestingly, for the cosmological horizon a choice similar to (\ref{alter}) will be physical whereas the other choice similar to (\ref{connecttt1}) is unphysical. This issue will be discussed in the next section.

\subsection{Blackbody spectrum for the cosmological horizon}
To find the radiation spectrum for the cosmological horizon we first need to perform two tasks. One is the Kruskal-like extension of the space-time (\ref{metric}) just around the cosmological horizon which is done in appendix (\ref{Appendix3B}). The other requirement is to identify the left and right moving modes outside and inside the cosmological horizon which is discussed in appendix (\ref{Appendix3C}) {\footnote{We use the subscript ``$obs$'' for modes inside the cosmological horizon, such that $r_h<r<r_c$, since an observer can stay only in this region. The subscript ``$out$'' is used for modes outside the cosmological horizon.}}. 

We start by choosing the following set of relations from the transformation (\ref{connect1}) between the time and radial coordinate systems {\footnote{One can again choose the opposite relative signs between the quantities at the right hand side of (\ref{cosconnect1}). With this choice the probability for a right-moving mode to cross the cosmological horizon from inside is $P^R=1$. The probability for the left-moving mode to cross the cosmological horizon from the outside, observed from inside the horizon, is then given by $P^L=e^\frac{-2\pi\Omega}{\hbar\kappa}$. Since, for the cosmological horizon, $\Omega$ is the BBM energy which is negative \cite{BBM}, $P^L$ diverges in the classical limit ($\hbar\rightarrow 0$). Therefore (\ref{cosconnect1}) is the only physical choice for this case.}}
{\begin{eqnarray}
t_{out} &=& t_{obs}+\frac{i\pi}{2\kappa}\nonumber\\
r^*_{out} &=&  r^*_{obs}-\frac{i\pi}{2\kappa}.
\label{cosconnect1}
\end{eqnarray} 
The left and right moving modes inside the cosmological horizon (outside the event horizon) are given by (\ref{conv5}) and (\ref{conv6}) respectively, whereas, (\ref{conv7}) and (\ref{conv8}) give left and right moving modes outside the cosmological horizon respectively. Using (\ref{cosconnect1}) the modes inside and outside the cosmological horizon can be connected as
\begin{eqnarray}
\Phi_{out}^{(R)} &=& \Phi_{{obs}}^{{(R)}} \nonumber\\
\Phi_{out}^{(L)} &=& {e^{\frac{\pi\Omega}{\hbar\kappa}}}{\Phi_{{obs}}^{{(L)}}}
\label{cosconnect3}
\end{eqnarray}
The physical state representing $n$ number of non-interacting pairs, created outside the cosmological horizon, when viewed from inside the horizon, is given by
\begin{eqnarray}
|\Psi\rangle = N \displaystyle\sum_n |n^{(R)}_{\textrm{out}}\rangle\otimes|n^{(L)}_{\textrm{out}}\rangle
 = N \displaystyle\sum_n e^{\frac{\pi n\Omega}{\hbar\kappa}}|n^{(R)}_{\textrm{obs}}\rangle\otimes|n^{(L)}_{\textrm{obs}}\rangle.
\label{cosstate}
\end{eqnarray} 
Here the normalization constant $N$ can be found from $\langle\Psi|\Psi\rangle=1$ and the physical state for bosons and fermions, turns out to be,
\begin{eqnarray}
|\Psi\rangle_{(\textrm{boson})}= \Big(1-e^{\frac{2\pi\Omega}{\hbar\kappa}}\Big)^{\frac{1}{2}} \displaystyle\sum_n e^{\frac{\pi n\Omega}{\hbar\kappa}}|n^{(R)}_{\textrm{obs}}\rangle\otimes|n^{(L)}_{\textrm{obs}}\rangle,
\label{cosboson}
\\
|\Psi\rangle_{(\textrm{fermion})}= \Big(1+e^{\frac{2\pi\Omega}{\hbar\kappa}}\Big)^{-\frac{1}{2}} \displaystyle\sum_n e^{\frac{\pi n\Omega}{\hbar\kappa}}|n^{(R)}_{\textrm{obs}}\rangle\otimes|n^{(L)}_{\textrm{obs}}\rangle
\label{cosfermion}
\end{eqnarray}   
respectively. The density operator for the bosons is now constructed as 
\begin{eqnarray}
{\hat\rho}_{(\textrm{boson})}&=&|\Psi\rangle_{(\textrm{boson})}\langle\Psi|_{(\textrm{boson})}
\nonumber
\\
&=&\Big(1-e^{\frac{2\pi\Omega}{\hbar\kappa}}\Big) \displaystyle\sum_{n,m} e^{\frac{\pi (n+m)\Omega}{\hbar\kappa}} |n^{(R)}_{\textrm{obs}}\rangle\otimes|n^{(L)}_{\textrm{obs}}\rangle\langle m^{(R)}_{\textrm{obs}}|\otimes\langle m^{(L)}_{\textrm{obs}}|.
\label{cosdens}
\end{eqnarray}
Since in this case right-moving modes are going outside the cosmological horizon, these are completely lost. We take the the trace over all such right-moving modes to find the reduced density operator for the left-moving modes, given by 
\begin{eqnarray}
{\hat{\rho}}^{(L)}_{(\textrm{boson})}= \Big(1-e^{\frac{2\pi\Omega}{\hbar\kappa}}\Big) \displaystyle\sum_{n} e^{\frac{2\pi n\Omega}{\hbar\kappa}}|n^{(L)}_{\textrm{obs}}\rangle \langle n^{(L)}_{\textrm{obs}}|.
\label{cosdenright}
\end{eqnarray}
In the case of cosmological horizon the particles are not observed at asymptotic infinity, rather in a region in between the event and the cosmological horizon. The average number of particles which is detected by an observer in this region is now given by, 
\begin{eqnarray}
\langle n\rangle_{(\textrm{boson})} &=& {\textrm{trace}}({\hat{n}} {\hat{\rho}}^{(L)}_{(\textrm{boson})})\nonumber\\
                       &=& \frac{1}{e^{-\frac{2\pi\Omega}{\hbar\kappa}}-1}.
\label{cosexp}
\end{eqnarray}
This is again a Bose-Einstein distribution of particles corresponding to the new Hawking temperature 
\begin{eqnarray}
T_{\textrm c}=-\frac{\hbar\kappa}{2\pi}=-\frac{\hbar g'(r_c)}{4\pi}.
\label{semicos}
\end{eqnarray}
Note that the temperature has the same value (\ref{semihawk}) as found for the black hole (event) horizon but with a sign difference. This temperature together with the BBM energy \cite{BBM} of the dS spacetime make the first law of thermodynamics valid for the cosmological horizon.


\section{\label{qtbeybs}Derivation of modified blackbody spectrum beyond semi-classical limit}
We now develop a general framework to find the corrections to the semi-classical Hawking radiation for a general class of metric (\ref{metrichd}),  where the $r-t$ sector of the metric is decoupled from the angular parts. In the previous section we generalized the tunneling method to find the corrections to the Hawking temperature for the black hole solution of Einstein-Maxwell theory (Kerr-Newman) in (3+1) dimensions by using the method of complex path. Now we shall use the method, as outlined in the previous section, to find the modified radiation spectrum, by going beyond the semi-classical approximation. This analysis shows that the higher order terms in the WKB ansatz, when included in the theory, do not affect the thermal nature of the spectrum. Grey-body factors do not appear in the radiation spectrum, rather the temperature of the radiation undergoes some higher order corrections, with a perfect blackbody spectrum.

In the beginning of section (\ref{qtbs}) we only considered the semi-classical action (${\cal S}_{0}$) corresponding to the scalar field ansatz ($\Phi$) (\ref{2.9ssp}) and found a solution for that in (\ref{soln}). In order to carryout the density matrix analysis beyond semi-classical limit we need to find a general solution for $\Phi$ considering the higher order terms in $\hbar$. The mechanism for this is already discussed when we find corrected temperature in section (\ref{qtbeyht}). Following the identical steps one can find the expression for ${\cal S}$ as provided in (\ref{2.14}). Finally the cherished solution for the scalar field in presence of the higher order corrections to the semi-classical action, follows from (\ref{2.10ssa}) and (\ref{2.14}) and (\ref{action2}),
\begin{eqnarray}
\Phi=\exp[-{\frac{i}{\hbar}}\omega\left(1+\displaystyle\sum_{i=1}^{\infty}{\gamma_i \hbar^i}\right)(t\pm r^{*})],
\label{highsoln}
\end{eqnarray}
where $r^*$ is the tortoise coordinate.

The left and right moving modes inside and outside the black hole event horizon, following the convention of appendix (\ref{Appendix3C}), now becomes
\begin{eqnarray}
&&\Phi_{in}^{(R)} = e^{-\frac{i}{\hbar}\omega\left(1+\displaystyle\sum_{i=1}^{\infty}{\gamma_i \hbar^i}\right) u_{in}};\,\,\ \Phi_{in}^{(L)}=e^{-\frac{i}{\hbar}\omega\left(1+\displaystyle\sum_{i=1}^{\infty}{\gamma_i \hbar^i}\right) v_{in}},
\nonumber
\\
&&\Phi_{out}^{(R)}=e^{-\frac{i}{\hbar}\omega\left(1+\displaystyle\sum_{i=1}^{\infty}{\gamma_i \hbar^i}\right)u_{out}};\,\,\,\Phi_{out}^{(L)}=e^{-\frac{i}{\hbar}\omega\left(1+\displaystyle\sum_{i=1}^{\infty}{\gamma_i \hbar^i}\right) v_{out}}.
\label{highmodes}
\end{eqnarray}
These inside and outside modes, moving in a particular direction (right or left), can also be connected by the set of transformations (\ref{connect1}) or (\ref{connect2}). This yields
\begin{eqnarray}
\Phi_{in}^{(R)} &=& {e^{-\frac{\pi\omega}{\hbar\kappa}\left(1+\displaystyle\sum_{i=1}^{\infty}{\gamma_i \hbar^i}\right)}}\Phi_{out}^{(R)}\nonumber\\
 &=& {e^{-\frac{\pi\omega}{\hbar{\kappa}'}}}\Phi_{out}^{(R)}\nonumber\\
\Phi_{in}^{(L)} &=& \Phi_{out}^{(L)},
\label{highconnect}
\end{eqnarray}
where we have substituted 
\begin{eqnarray}
\kappa '=\left(1+\displaystyle\sum_{i=1}^{\infty}{\gamma_i \hbar^i}\right)^{-1}\kappa.
\label{highsurface}
\end{eqnarray}
This can be considered as the modified surface gravity in presence of higher $\hbar$ order corrections to the WKB ansatz. 

The physical state representing $n$ number of virtual pairs inside the horizon, when observed from the outside, is now given by
\begin{eqnarray}
|\Psi\rangle = N \displaystyle\sum_n |n^{(L)}_{\textrm{in}}\rangle\otimes|n^{(R)}_{\textrm{in}}\rangle
 = N \displaystyle\sum_n e^{-\frac{\pi n\Omega}{\hbar\kappa '}}|n^{(L)}_{\textrm{out}}\rangle\otimes|n^{(R)}_{\textrm{out}}\rangle.
\label{highstate}
\end{eqnarray}
Now from here on, it is trivial to check that the whole methodology developed for the semiclassical case can be repeated to find the new radiation spectrum. The only difference is the redefinition of the surface gravity ($\kappa$) by $\kappa '$. The final result for the radiation spectrum, for bosons, is now given by
\begin{eqnarray}
\langle n\rangle_{(\textrm{boson})}= \frac{1}{e^{\frac{2\pi\Omega}{\hbar\kappa '}}-1}.
\label{highexpect}
\end{eqnarray}     
Now one can see that the spectrum is still given by the blackbody spectrum with the new corrected Hawking temperature  
\begin{eqnarray}
T_{\textrm{bh}} &=& \frac{\hbar\kappa '}{2\pi}=\frac{\hbar\kappa}{2\pi}\left(1+\displaystyle\sum_{i=1}^{\infty}{\gamma_i \hbar^i}\right)^{-1}\nonumber\\
                &=&\left(1+\displaystyle\sum_{i=1}^{\infty}{\gamma_i \hbar^i}\right)^{-1}T_{\textrm{H}},
\label{correcttemp}
\end{eqnarray}
where $T_{\textrm{H}}$ is the usual semiclassical Hawking temperature, given by (\ref{semihawk}). Note that (\ref{correcttemp}) gives the corrected Hawking temperature for any general static, chargeless black hole solutions with an appropriate choice of metric.

For the cosmological horizon, in presence of other higher order terms in $\hbar$ in the action, the relation between right and left moving modes at two sides are given by 
\begin{eqnarray}
\Phi_{out}^{(R)} &=& \Phi_{{obs}}^{{(R)}} \nonumber\\
\Phi_{out}^{(L)} &=& {e^{\frac{\pi\Omega}{\hbar\kappa '}}}{\Phi_{{obs}}^{{(L)}}},
\label{cosconnect3}
\end{eqnarray}
where $\kappa '$ is defined in (\ref{highsurface}). Subsequently the new radiation spectrum for the bosons turns out to be same as (\ref{expectation}) with $\kappa '$ replacing $\kappa$. Therefore the modified Hawking temperature for the cosmological horizon, given by 
\begin{eqnarray}
T_{\textrm{ch}} &=& -\frac{\hbar\kappa '}{2\pi}=\frac{\hbar\kappa}{2\pi}\left(1+\displaystyle\sum_{i=1}^{\infty}{\gamma_i \hbar^i}\right)^{-1}\nonumber\\
                &=& \left(1+\displaystyle\sum_{i=1}^{\infty}{\gamma_i \hbar^i}\right)^{-1}T_{\textrm{c}},
\label{corcosttemp}
\end{eqnarray}
where $T_{\textrm c}$ is the semiclassical temperature given by (\ref{semicos}).



\section{\label{qtdisc}Discussions}

Let us now summarize the work presented in this chapter. We adopted the quantum tunneling method to study various features of the Hawking effect. Two approaches were followed; one that exploit the principle of detailed balance while the other is inspired from a density matrix type analysis. In the latter the blackbody spectrum was reproduced. 

We exploited the complex path (Hamilton-Jacobi type) approach to study scalar and fermion tunneling. Ingoing and outgoing modes were calculated which essentially gave the respective  absorption/transmission probabilities. Then, by using the principle of detailed balance, the Hawking temperature was identified. Going beyond the semi-classical limit we considered higher order terms (in $\hbar$) in WKB ansatz. It generated some higher order corrections to semi-classical temperature with some unknown coefficients.

By adopting a density matrix approach, we also showed that black holes do emit scalar particles and fermions with a perfect blackbody spectrum where temperature given by the semi-classical Hawking temperature. This result was derived for both black hole (event) horizon and cosmological horizon of arbitrary dimensional static black holes. It was also found that in the presence of higher order corrections to the WKB ansatz the blackbody nature of the modified radiation spectrum does not change. Greybody factors were absent, rather the temperature received some corrections. We calculated the corrected Hawking temperature for both the black hole (event) horizon and also for the cosmological horizon. The temperature corresponding to the modified spectrum, as calculated, reproduced the Hawking temperature at the lowest order.

\begin{subappendices}
\chapter*{Appendix}
\section{\label{appendix3A}Kruskal-like extension for the black hole (event) horizon}
\renewcommand{\theequation}{3A.\arabic{equation}}
\setcounter{equation}{0}  

To perform a Kruskal-like extension of a general chargeless, static metric (\ref{metric}), we first define the tortoise coordinate as
\begin{eqnarray}
r^*=\int{\frac{dr}{g(r)}}.
\label{tortoise}
\end{eqnarray}
It is known that the spacetime structure of the extended regions are extremely sensitive to the different choices for $g(r)$. In order to study the behaviors of the outgoing and ingoing modes with respect to the black hole event horizon ($r_{h}$) we actually need to see only the behavior of the spacetime in a very narrow region just inside and outside of $r_h$. Therefore, for our purpose, we first take the near horizon limit of the metric coefficient,
\begin{eqnarray}
g(r)=(r-r_h)g'(r_h)+\frac{(r-r_h)^2}{2}g''(r_h)+{\cal O}(r-r_h)^3
\label{nearhor}
\end{eqnarray}

In the following we shall consider two different cases for inside and outside (where an observer is present) the event horizon to show that the same spacetime metric is valid in both the regions.  

{\bf Case I}: When $r=r_{\text{obs}}>r_h$ (Outside the event horizon): At a distance ($\rho$) just outside the horizon, $r=r_h+\rho$, such that $|\rho|<<r_h$, one has  $dr=d\rho$ and
\begin{eqnarray}
g(r)=\rho g'(r_h)+\frac{\rho^2}{2}g''(r_h)+{\cal O}(\rho^3)
\label{nearhor2}
\end{eqnarray}
Therefore (\ref{tortoise}) can be integrated over this narrow region to yield
\begin{eqnarray}
r^*_{obs} &=& \frac{1}{g'(r_h)}\ln[\frac{\rho}{g'(r_h)+\rho g''(r_h)/2}]\nonumber\\
                   &=& \frac{1}{g'(r_h)}\ln[\frac{(r-r_h)}{g'(r_h)+(r-r_h) g''(r_h)/2}]
\label{tortoise2}
\end{eqnarray}
The advanced and retarded time coordinates, in this region, are usually defined by 
\begin{eqnarray}
v_{obs} &=& t_{obs}+r^*_{obs}\nonumber\\
u_{obs} &=& t_{obs}-r^*_{obs}
\label{null}
\end{eqnarray}
respectively. With these definitions (\ref{metric}) becomes
\begin{eqnarray}
ds^2=-g(r)du_{obs}dv_{obs}+r^2d\Omega_{D-2}^2
\label{nullmetric}
\end{eqnarray}
Now we make the following two successive coordinate transformations
\begin{eqnarray}
V_{obs} &=& e^{\kappa v_{obs}}\nonumber\\
U_{obs} &=& -e^{-\kappa u_{obs}}.
\label{capital}
\end{eqnarray}
Also, 
\begin{eqnarray}
T_{obs} &=& \frac{(V_{obs}+U_{obs})}{2}\nonumber\\
X_{obs} &=& \frac{(V_{obs}-U_{obs})}{2},
\label{capital2}
\end{eqnarray}
 where $\kappa$ is a constant that will be identified later with the surface gravity. With these transformations we find the desired form of the $r-t$ sector of the metric (\ref{metric}) in Kruskal coordinates,
\begin{eqnarray}
ds^2=-\frac{g(r)}{\kappa^2}e^{-2\kappa r^*_{obs}}(dT_{obs}^2-dX_{obs}^2).
\label{kruskal}
\end{eqnarray}
Putting the values of $g(r)$ from (\ref{nearhor}) and $r^*$ from (\ref{tortoise2}) into the spacetime interval (\ref{kruskal}) and simplifying the pre-factor by choosing $\kappa$ to be the surface gravity ($\kappa=g'(r_\textrm{h})/2$), we find
\begin{eqnarray}
ds^2=\frac{4\left(g'(r_h)^2+\frac{3g'(r_h)g''(r_h)}{2}(r-r_h)+\frac{g''(r_h)^2}{2}(r-r_h)^2\right)}{g'(r_h)^2}(-dT_{obs}^2+dX_{obs}^2).
\label{kruskal2}
\end{eqnarray}
Now using (\ref{capital}) and (\ref{capital2}) we find the Kruskal-like coordinates which are valid outside the event horizon where the observer is present, are given by 
\begin{eqnarray}
T_{obs} &=& \exp^{\kappa r_{{obs}}^{*}}\sinh{\kappa t_{obs}}\nonumber\\
X_{obs} &=& \exp^{\kappa r^*_{obs}}\cosh{\kappa t_{obs}}.
\label{transformation}
\end{eqnarray}
This clearly shows, irrespective of choosing any particular $g(r)$, at the event horizon ($r=r_h$ or $\rho=0$) one does not have any spacetime singularity in this Kruskal-like extension (\ref{kruskal2}). The same is true if one considers higher order terms in the expansion (\ref{nearhor}). The only finite contribution to the interval (\ref{kruskal2}) comes from the linear term in the expansion (\ref{nearhor}) while others vanish at $r=r_h$. This consistency is essential for the calculation of the blackbody spectrum.

{\bf Case II}: When $r=r_{\textrm{in}}<r_h$ (Inside the event horizon): Let us now consider a situation at a distance $\rho$ inside the event horizon. Here $r=r_{\textrm{in}}=r_h-\rho$ and $dr=-d\rho$. Using the same expansion (\ref{nearhor2}) and integrating (\ref{tortoise}), we get the required expression for the tortoise coordinate inside the event horizon, 
\begin{eqnarray}
r^*_{in} &=& \frac{1}{g'(r_h)}\ln[\frac{\rho}{g'(r_h)-\rho g''(r_h)/2}]\nonumber\\
                   &=& \frac{1}{g'(r_h)}\ln[\frac{(r_h-r)}{g'(r_h)-(r_h-r) g''(r_h)/2}].
\label{tortoise3}
\end{eqnarray}
Consider the following coordinate transformations,
\begin{eqnarray}
v_{in} &=& t_{in}+r^*_{in}\nonumber\\
u_{in} &=& t_{in}-r^*_{in},
\label{null2}
\end{eqnarray} 
and
\begin{eqnarray}
V_{in} &=& e^{\kappa v_{in}}\nonumber\\
U_{in} &=& e^{-\kappa u_{in}}.
\label{capital3}
\end{eqnarray}
Also
\begin{eqnarray}
T_{in} &=& \frac{(V_{in}+U_{in})}{2}\nonumber\\
X_{\textrm{in}} &=& \frac{(V_{in}-U_{in})}{2}.
\label{capital4}
\end{eqnarray}
With these coordinate transformations in (\ref{metric}) we are finally left with a metric whose $r-t$ sector is given by 
\begin{eqnarray}
ds^2=\frac{4\left(g'(r_h)^2+\frac{3g'(r_h)g''(r_h)}{2}(r-r_h)+\frac{g''(r_h)^2}{2}(r-r_h)^2\right)}{g'(r_h)^2}(-dT_{in}^2+dX_{{in}}^2).
\label{kruskal3}
\end{eqnarray}
The new Kruskal coordinates which are valid inside the event horizon are now found from (\ref{null2}), (\ref{capital3}) and (\ref{capital4}),
\begin{eqnarray}
T_{in} &=& e^{\kappa r^*_{in}}\cosh{\kappa t_{in}}\nonumber\\
X_{in} &=& e^{\kappa r^*_{in}}\sinh{\kappa t_{in}}.
\label{transformation2}
\end{eqnarray}
One can now realize that all the coordinate transformations used here are identical to the previous case for $r>r_{\textrm h}$ except for the definition of $U_{in}$. There is a relative sign difference between the functional choice of $U_{in}$ and  $U_{obs}$. This new definition ensures that the inner portion of the extended spacetime metric remains timelike. If one follows the earlier definition he/she will end up with a spacelike metric interval which we want to avoid because there is no coordinate singularity at the event horizon. Therefore one can conclude that  
\begin{eqnarray}
ds^2=\frac{4\left(g'(r_h)^2+\frac{3g'(r_h)g''(r_h)}{2}(r-r_h)+\frac{g''(r_h)^2}{2}(r-r_h)^2\right)}{g'(r_h)^2}(-dT^2+dX^2)
\label{kruskal4}
\end{eqnarray} 
is the only metric which is valid in a narrow region on both sides of the event horizon and the Kruskal coordinates outside (where the observer is present) and inside the event horizon are defined by (\ref{transformation}) and (\ref{transformation2}) respectively.

It can be noted that the set of coordinate transformations, given in (\ref{transformation}) and (\ref{transformation2}) are connected by the following transformations in ($t,r^{*}$) coordinate
\begin{eqnarray}
t_{in} &=& t_{obs}\mp\frac{i\pi}{2\kappa}\nonumber\\
r^*_{in} &=&  r^*_{obs}\pm\frac{i\pi}{2\kappa}
\label{connect1}
\end{eqnarray}
In the null coordinates ($u,v$) (\ref{null},\ref{null2})these two relations are recast as
\begin{eqnarray}
u_{in} &=& u_{obs}\mp\frac{i\pi}{\kappa}\nonumber\\
v_{in} &=& v_{obs}.
\label{connect2}
\end{eqnarray}
These relationships are required for connecting various modes defined inside and outside the event horizon. This analysis is performed in subsection \ref{bseh} for the derivation of blackbody radiation spectrum for black hole event horizon.

\section{\label {Appendix3B}Kruskal-like extension for the cosmological horizon}
\renewcommand{\theequation}{3B.\arabic{equation}}
\setcounter{equation}{0} 

To see the behaviour of the spacetime (\ref{metric}) at the cosmological horizon it is required to expand $g(r)$ near the cosmological horizon ($r_h$), given by 
\begin{eqnarray}
g(r)=(r-r_c) g'(r_c)+\frac{(r-r_c)}{2}g''(r_c)+{\cal O}((r-r_{c})^3).
\label{nearhor3}
\end{eqnarray}
The tortoise coordinates inside (where the observer is present; $r=r_{{obs}}<r_{c}$) and outside ($r=r_{{out}}>r_{c}$) the cosmological horizon are defined by
\begin{eqnarray}
r^*_{obs}= \frac{1}{g'(r_c)}\ln[\frac{(r_c-r)}{g'(r_c)-(r_c-r) g''(r_c)/2}]
\label{torcos2}
\end{eqnarray}
and 
\begin{eqnarray}
r^*_{out}= \frac{1}{g'(r_c)}\ln[\frac{(r-r_c)}{g'(r_c)+(r-r_c) g''(r_c)/2}]
\label{torcos1}
\end{eqnarray}
respectively. The sets of null coordinates inside and outside the cosmological horizon are defined as
\begin{eqnarray}
v_{obs} &=& t_{obs}+r^*_{obs}\nonumber\\
u_{obs} &=& t_{obs}-r^*_{obs}
\label{cosnull}
\end{eqnarray} 
and 
\begin{eqnarray}
v_{out} &=& t_{out}+r^*_{out}\nonumber\\
u_{out} &=& t_{out}-r^*_{out}
\label{cosnull2}
\end{eqnarray} 
respectively. Now by exactly mimicking the methodology developed for the case of black hole horizon it can be shown that the metric 
\begin{eqnarray}
ds^2=\frac{4\left(g'(r_c)^2+\frac{3g'(r_{c})g''(r_{c})}{2}(r-r_c)+\frac{g''(r_{c})^2}{2}(r-r_c)^2\right)}{g'(r_c)^2}(-dT^2+dX^2)
\label{kruskal5}
\end{eqnarray}
is defined both inside and outside the horizon. This is the analogue of (\ref{kruskal4}). This time the Kruskal-like coordinates in the region inside and outside the cosmological horizon are, respectively, given by
\begin{eqnarray}
T_{obs} &=& e^{\kappa r^*_{obs}}\cosh{\kappa t_{obs}}\nonumber\\
X_{obs} &=& e^{\kappa r^*_{obs}}\sinh{\kappa t_{obs}}
\label{transformation4}
\end{eqnarray}
and
\begin{eqnarray}
T_{out} &=& e^{\kappa r^*_{out}}\sinh{\kappa t_{out}}\nonumber\\
X_{out} &=& e^{\kappa r^*_{out}}\sinh{\kappa t_{out}},
\label{transformation3}
\end{eqnarray}
where $\kappa=\frac{g'(r_c)}{2}$ is the surface gravity at the cosmological horizon.

From the Kruskal-like extension it is found that the inside ($T_{obs}, X_{obs}$) and outside ($T_{out}, X_{out}$) coordinates, defined in (\ref{transformation4}) and (\ref{transformation3}) respectively, can be connected with each other by the same set of relations defined in (\ref{connect1}) and (\ref{connect2}).

\section{\label{Appendix3C}Identification of ``in'' and ``out'' modes}
\renewcommand{\theequation}{3C.\arabic{equation}}
\setcounter{equation}{0} 
To identify different modes in different regions we use the following convention. If the eigenvalue of the radial momentum operator $\hat p(r)$ , while acting on a specific solution of the semiclassical (WKB) mode (\ref{soln}), is positive, then the mode is right-moving (outgoing). Similarly a left-moving (incoming) mode corresponds to a negative eigenvalue. In this convention the spacelike nature of $\hat p(r)$ must be kept unchanged in any region (i.e. inside or outside the horizon) of the spacetime. 
 
{\bf Black hole horizon:} For all Lovelock black holes, considered in this paper, $\Omega$ (which is the conserved quantity to the timelike Killing vector) is nothing but the mass ($M$) of the black hole. In a region outside the event horizon (position of the observer) ($r>r_h$), $\hat p(r)=-i\hbar{\frac{\partial}{\partial r}}$ and the left ($L$) or right ($R$) moving modes are found as
\begin{eqnarray}
\Phi^{L}_{obs} &=& e^{-\frac{i}{\hbar}\Omega v_{obs}}\label{conv1}\\
\Phi^{R}_{obs} &=& e^{-\frac{i}{\hbar}\Omega u_{obs}}\label{conv2}.
\end{eqnarray}       
To keep the spacelike nature of the momentum operator inside the black hole event horizon, one must use the definition $\hat p(r)=i\hbar{\frac{\partial}{\partial r}}$. Using this one can identify
\begin{eqnarray}
\Phi^{L}_{in} &=& e^{-\frac{i}{\hbar}\Omega v_{in}}\label{conv3}\\
\Phi^{R}_{in} &=& e^{-\frac{i}{\hbar}\Omega u_{in}}\label{conv4}.
\end{eqnarray}
as the left and right moving modes respectively.

{\bf Cosmological horizon:} For the Gauss-Bonnet dS black hole we have an extra cosmological horizon. Here $\Omega$ or equivalently the gravitational mass is given by the BBM mass \cite{BBM} which is negative. In a region inside the cosmological horizon where the observer is present ($r_h<r<r_c$), $\hat p(r)=-i\hbar\frac{\partial}{\partial r}$, and one can identify the left and right moving modes as
\begin{eqnarray}
\Phi^{L}_{obs} &=& e^{-\frac{i}{\hbar}\Omega u_{obs}}\label{conv5}\\
\Phi^{R}_{obs} &=& e^{-\frac{i}{\hbar}\Omega v_{obs}}\label{conv6}.
\end{eqnarray}  
Similarly outside the cosmological horizon ($r>r_c$), one has $\hat p(r)=i\hbar\frac{\partial}{\partial r}$, therefore one finds 
\begin{eqnarray}
\Phi^{L}_{out} &=& e^{-\frac{i}{\hbar}\Omega u_{out}}\label{conv7}\\
\Phi^{R}_{out} &=& e^{-\frac{i}{\hbar}\Omega v_{out}}\label{conv8},
\end{eqnarray}  
as the left and right moving modes respectively.          
\end{subappendices}

\chapter{\label{chap-entropy}Exact differentials and black hole entropy}
We continue our analysis of quantum tunneling by applying its results to study various aspects of black hole thermodynamics. Within classical GTR black holes are pure absorbers and have absolute zero temperature. This property goes against the intuitive idea that black holes have entropy. In early seventies Bekenstein first argued in favour of black hole entropy based on information theory as well as simple aspects of thermodynamics \cite{Beken1}-\cite{Beken3}. He claimed that  entropy of the universe cannot be decreased due to the capture of any object by black holes. For making the total entropy of the universe at least unchanged, a black hole should gain the same amount of entropy which is lost from the rest of the universe. Bekenstein then gave some heuristic arguments to show that black hole entropy must be proportional to its horizon area. He also fixed the proportionality constant as $\frac{ln2}{8\pi}$. The idea of Bekenstien was given a solid mathematical ground when Hawking incorporated quantum fields moving in a background of classical gravity and showed that black holes do emit particles having a black body spectrum with physical temperature $\frac{\hbar\kappa}{2\pi}$, where $\kappa$ is the surface gravity of a black hole  \cite{hk1}-\cite{hk3}. Knowing this expression of black hole temperature (``{\it Hawking temperature}'') one can make an analogy with the `first law of black hole mechanics' and the `first law of thermodynamics' to identify entropy as $S=\frac{A}{4\hbar}$ (in $c=G=k_B=1$ unit), where $A$ is the horizon area of the black hole. Thus it was proven that Bekenstein's constant of proportionality was incorrect and the new proportionality constant is $\frac{1}{4}$. The work of Bekenstein and Hawking thereby leads to the semi-classical result for black hole entropy encapsuled by the Bekenstein-Hawking area law, given by $S_{\textrm {BH}}=\frac{A}{4\hbar}.$

In this chapter we strengthen these ideas and follow a pure thermodynamical method to calculate black hole entropy without using any analogy with the `first law of black hole mechanics'. It is shown that if one considers black holes as pure thermal objects there is no need to use such an analogy as mentioned in the preceding paragraph for computing black hole entropy. The state function nature of entropy plays a key role in its evaluation for black holes. Furthermore the same property also leads to the higher order corrections to the semi-classical Bekenstein-Hawking area law when one considers corrections to the Hawking temperature beyond semi-classical limit as calculated in Chapter-3.

This chapter is organised as follows. In section-\ref{semiflaw} we derive the first law of black hole thermodynamics without using its analogy with the first law of black hole mechanics. The state function property of entropy is then used in section-\ref{calsemien} to calculate semi-classical black hole entropy considering most general black hole spacetime. A similar approach is used in section-\ref{calcoren} to compute black hole entropy beyond the semi-classical limit. The leading (logarithmic) and higher (inverse horizon area) terms are generated as higher order corrections to the Bekenstein-Hawking area law. These ideas for the example of BTZ black hole in (2+1) dimensional gravity is worked out in section-\ref{btz} where we find identical functional form in various results. In section-\ref{traceanomaly} the leading correction to the semi-classical temperature and entropy for (3+1) dimensional black holes is explicitly calculated. Finally section-\ref{dischentropy} includes summary and discussions.

\section{\label{semiflaw}Derivation of the first law of black hole thermodynamics}
Long time back (1973) within the realm of classical general relativity Bardeen, Carter and Hawking gave the ``first law of black hole mechanics'' which states that for two nearby black hole solutions the difference in mass ($M$), area ($A$) and angular momentum ($J$) must be related by \cite{Bardeen}
\begin{eqnarray}
\delta M= \frac{1}{8\pi}{\kappa\delta A}+ \Omega_{\textrm H} \delta J.
\label{bhmech}
\end{eqnarray}
In addition some more terms can appear on the right hand side due to the presence of other matter fields. They found this analogous to the ``first law of thermodynamics'', which states, the difference in energy ($E$), entropy ($S$) and other state parameters of two nearby thermal equilibrium states of a system is given by
\begin{eqnarray}
dE= T dS+ {\textrm {``work terms''}}.
\label{thermo}
\end{eqnarray}
Therefore even in classical general relativity the result (\ref{bhmech}) is appealing due to the fact that both $E$ and $M$ represent the same physical quantity, namely total energy of the system. Although at that time this result was quite surprising as classically, temperature of black holes was absolute zero. So the identification of temperature with surface gravity, as shown by (\ref{bhmech}) and (\ref{thermo}), was meaningless. Consequently, identification of entropy with horizon area was inconsistent.

 However the picture was changed dramatically when Hawking (1975), incorporating quantum effects, discovered \cite{hk1}-\cite{hk3} that black holes do radiate all kinds of particles with a perfect black body spectrum with temperature $T_{\textrm H}=\frac{\hbar\kappa}{2\pi}$. From this mathematical identification of the Hawking temperature ($T_{\textrm H}$) with the surface gravity ($\kappa$) in (\ref{bhmech}), one is left with some analogy between entropy ($S$) and the area of the event horizon($A$), suggested by (\ref{bhmech}) and (\ref{thermo}). The result $S=\frac{A}{4\hbar}$ follows from this analogy.

For such an identification, the horizon area of a black hole is playing the ``mathematical role'' of entropy and does not have a solid physical ground. Also, this naive identification remains completely silent about the role of ``work terms''. But if one does not use this mathematical analogy, rather tries to {\it calculate} entropy, it may appear that these work terms might have some role to play. Therefore the role of these work terms is not transparent in the process of identifying entropy. Moreover, in this analysis one can obtain the ``first law of black hole thermodynamics'' only by deriving the ``first law of black hole mechanics'' and then identifying this with the ordinary ``first law of thermodynamics''.

Now we want to obtain the ``first law of black hole thermodynamics''  by directly starting from the {\it thermodynamical} viewpoint where one does not require the ``first law of black hole mechanics''. From such a law the entropy will be explicitly calculated and not identified, as usually done, by an analogy between (\ref{bhmech}) and (\ref{thermo}). For this derivation we interpret Hawking's result of black hole radiation as 
\begin{itemize}
\item {{\it black holes are thermodynamical objects having mass ($M$) as total energy ($E$) and they are immersed in a thermal bath in equilibrium with physical temperature ($T_{\textrm H}$)}.} 
\end{itemize}
Therefore following the ordinary ``first law of thermodynamics'' we are allowed to write the ``first law of black hole thermodynamics'' as 
\begin{eqnarray}
dM= T_{\textrm H}dS + {\textrm {``work terms on black hole''}},
\label{newlaw}
\end{eqnarray}
where $M$ is the mass of the black hole and $T_{\textrm H}$ is the Hawking temperature. Usually, without deriving the ``first law of black hole mechanics'' one is not able to find ``work terms on black hole'' exactly. But we can always make a dimensional analysis to construct these two terms as proportional to $\Omega_{\textrm H} dJ$ and $\Phi_{\textrm H} dQ$ where $J$ and $Q$ are the angular momentum and charge of the black hole. This is possible since the form of ``angular velocity ($\Omega_{\textrm H}$)'' and ``potential ($\Phi_{\textrm H}$)'' at the event horizon are known individually from classical gravity. These terms can be brought on the right hand side of (\ref{newlaw}) with some prefactors given by dimensionless constants `$a$' and `$b$', such that (\ref{newlaw}) becomes 
\begin{eqnarray}
dM=T_{\textrm H} dS+a\Omega_{\textrm H} dJ+ b\Phi_{\textrm H} dQ.
\label{4.11}
\end{eqnarray}
To fix the arbitrary constants `$a$' and `$b$' let us first rewrite (\ref{4.11}) in the form 
\begin{eqnarray}
dS=\frac{dM}{T_{\textrm H}}+(-\frac{a\Omega_{\textrm H}}{T_{\textrm H}})dJ+(-\frac{b\Phi_{\textrm H}}{T_{\textrm H}})dQ
\label{4.41}
\end{eqnarray}
From the principle of ordinary first law of thermodynamics one must interpret entropy as a {\it state function}. For the evolution of a system from one equilibrium state to another equilibrium state, entropy does not depend on the details of the evolution process, but only on the two extreme points representing the equilibrium states. This universal property of entropy must be satisfied for black holes as well. In fact the entropy of any stationary black hole should not depend on the precise knowledge of its collapse geometry but only on the final equilibrium state. Hence we can conclude that entropy for a stationary black hole is a state function and consequently $dS$ has to be an {\it exact differential}. As a result the coefficients of the right hand side of (\ref{4.41}) must satisfy the three integrability conditions  
\begin{align}
\frac{\partial}{\partial J}(\frac{1}{T_{\textrm H}})\big|_{M,Q}& =\frac{\partial}{\partial M}(-\frac{a\Omega_{\textrm H}}{T_{\textrm H}})\big|_{J,Q}\notag\\
\frac{\partial}{\partial Q}(-\frac{a\Omega_{\textrm H}}{T_{\textrm H}})\big|_{M,J}& =\frac{\partial}{\partial J}(-\frac{b\Phi_{\textrm H}}{T_{\textrm H}})\big|_{M,Q}\notag\\
\frac{\partial}{\partial M}(-\frac{b\Phi_{\textrm H}}{T_{\textrm H}})\big|_{J,Q}& =\frac{\partial}{\partial Q}(\frac{1}{T_{\textrm H}})\big|_{J,M}.
\label{4.71}
\end{align}
As one can see, these relations are playing a role similar to {\it Maxwell's relations} of ordinary thermodynamics. Like Maxwell's relations these three equations do not refer to a process but provide relationships between certain physical quantities that must hold at equilibrium.

The only known stationary solution of Einstein-Maxwell equation with all three parameters, namely Mass ($M$), Charge($Q$) and Angular momentum ($J$) is given by the Kerr-Newman spacetime. All the necessary information for that metric is provided in Appendix-\ref{appendix2B}  and one can readily check that the first, second and third conditions are satisfied only for $a=1,~~a= b~$ and $~b=1$ respectively, leading to the unique solution $a=b=1$. As a result, (\ref{4.11}) immediately reduces to the standard form 
\begin{eqnarray}
dM=T_{\textrm H} dS+\Omega_{\textrm H} dJ+\Phi_{\textrm H} dQ,
\label{sthermo}
\end{eqnarray}
This completes the obtention of the ``first law of black hole thermodynamics'',  for a rotating and charged black hole,  without using the ``first law of black hole mechanics''.

One can make an analogy of (\ref{sthermo}) with the standard first law of thermodynamics given by 
\begin{eqnarray}
dE= TdS- PdV +\mu dN
\label{bappa}
\end{eqnarray}
Knowing $E=M$ (since both represent the same quantity which is the energy of the system) one can infer the correspondence $-\Omega_{\textrm H}\rightarrow P,~J\rightarrow V,~\Phi_{\textrm H}\rightarrow \mu,~Q\rightarrow N$ between the above two cases. Indeed $\Omega_{\textrm H} dJ$ is the work done on the black hole due to rotation and is the exact analogue of the $-pdV$ term. Likewise the electrostatic potential $\Phi_{\textrm {H}}$ plays the role of the chemical potential $\mu$.   


\section{\label{calsemien}Black hole entropy as a state function}
In this section we calculate entropy of the Kerr-Newman black hole from the first law given in (\ref{sthermo}). We solve this first order partial differential equation where state function property of entropy plays a crucial role.

The first step is to rewrite (\ref{sthermo}) as
\begin{eqnarray}
dS=\frac{dM}{T_{\textrm H}}+(\frac{-\Omega_{\textrm H}}{T_{\textrm H}})dJ+(\frac{-\Phi_{\textrm H}}{T_{\textrm H}})dQ,
\label{sthermo1}
\end{eqnarray}
where $dS$ is now an exact differential.

 Note that any first order partial differential equation   
\begin{eqnarray}
df(x,y,z)=U(x,y,z)dx+V(x,y,z)dy+W(x,y,z)dz
\label{4.8}
\end{eqnarray}
is exact if it fulfills these integrability conditions
\begin{eqnarray}
\frac{\partial U}{\partial y}\big|_{x,z}=\frac{\partial V}{\partial x}\big|_{y,z};~~~\frac{\partial V}{\partial z}\big|_{x,y}=\frac{\partial W}{\partial y}\big|_{x,z};~~~\frac{\partial W}{\partial x}\big|_{y,z}=\frac{\partial U}{\partial z}\big|_{x,y}.
\label{4.9}
\end{eqnarray}
If these three conditions hold then the solution of (\ref{4.8}) is given by
\begin{eqnarray}
f(x,y,z)=\int{ Udx} +\int{Xdy}+\int{Ydz},
\label{4.12}
\end{eqnarray}
where
\begin{eqnarray}
X=V-\frac{\partial}{\partial y}{\int{Udx}}
\label{4.13}
\end{eqnarray}
and
\begin{eqnarray}
Y=W-\frac{\partial}{\partial z}[\int{Udx}+{\int Xdy}].
\label{4.14}
\end{eqnarray}

Now comparing (\ref{sthermo1}) and (\ref{4.8}) we find the following dictionary
\begin{align}
(f\rightarrow S,~~x\rightarrow M,~~y\rightarrow J,~~z\rightarrow Q)\nonumber\\
(U\rightarrow\frac{1}{T_{\textrm H}},~~V\rightarrow\frac{-\Omega_{\textrm H}}{T_{\textrm H}},~~W\rightarrow\frac{-\Phi_{\textrm H}}{T_{\textrm H}}).
\label{4.15}
\end{align}
Using this dictionary and (\ref{4.12}), (\ref{4.13}) and (\ref{4.14}) one finds,
\begin{eqnarray}
S=\int{\frac{dM}{T_{\textrm H}}}+\int{XdJ}+\int{YdQ},
\label{4.19}
\end{eqnarray}
where
\begin{eqnarray}
X=(-\frac{\Omega_{\textrm H}}{T_{\textrm H}})-\frac{\partial}{\partial J}{\int\frac{dM}{T_{\textrm H}}}
\label{4.20}
\end{eqnarray}
and
\begin{eqnarray}
Y=(-\frac{\Phi_{\textrm H}}{T_{\textrm H}})-\frac{\partial}{\partial Q}[{\int\frac{dM}{T_{\textrm H}}}+{\int XdJ}].
\label{4.21}
\end{eqnarray}
In order to calculate the semiclassical entropy we need to solve (\ref{4.19}), (\ref{4.20}) and (\ref{4.21}). Note that all the ``work terms'' are appearing in the general expression of the semiclassical entropy of a black hole (\ref{4.19}). Let us first perform the mass integral to get
\begin{eqnarray}
\int\frac{dM}{T_{\textrm H}}=\frac{\pi}{\hbar}\left(2M{[M+(M^2-\frac{J^2}{M^2}-Q^2)^{1/2}}]-Q^2\right),
\label{4.22}
\end{eqnarray}
 where the expression (\ref{4.5}) has been substituted for $T_{\textrm H}^{-1}$. With this result one can check the following equality 
\begin{eqnarray}
\frac{\partial}{\partial J}\int\frac{dM}{T_{\textrm H}}=-\frac{\Omega_{\textrm H}}{T_{\textrm H}}
\label{4.23}
\end{eqnarray}
holds, where $\Omega_{\textrm H}$ is defined in (\ref{angv}). Putting this in (\ref{4.20}) it follows that $X=0$. Using (\ref{4.22}) one can next calculate, 
\begin{eqnarray}
\frac{\partial}{\partial Q}\int\frac{dM}{T_{\textrm H}}=-\frac{\Phi_{\textrm H}}{T_{\textrm H}}.
\label{4.24}
\end{eqnarray}
With this equality and the fact that $X=0$, we find, using (\ref{4.21}), $Y=0$. Exploiting all of the above results, the semiclassical entropy for Kerr-Newman black hole is found to be, 
\begin{eqnarray}
S=\int{\frac{dM}{T_{\textrm H}}}=\frac{\pi}{\hbar}\left(2M{[M+(M^2-\frac{J^2}{M^2}-Q^2)^{1/2}}]-Q^2\right)=\frac{A}{4\hbar}=S_{{\textrm {BH}}},
\label{4.25}
\end{eqnarray}
 which is the standard semi-classical Bekenstein-Hawking area law for Kerr-Newman black hole. The expression for the area ($A$) of the event horizon follows from (\ref{area}). Now it is trivial, as one can check, that all other stationary spacetime solutions, for example Kerr or Reissner-Nordstrom, also fit into the general framework to give the semi-classical Bekenstein-Hawking area law. Thus the universality of the approach is justified.



\section{\label{calcoren}Exact differentials and corrections to the semi-classical black hole entropy} 
We have found that in the semi-classical limit when temperature of a black hole is given by standard Hawking temperature the solution for entropy as followed from the first law matches with the Bekenstein-Hawking value. It has also been discussed in Chapter-3 that if one consider higher order terms (in $\hbar$) in the WKB ansatz for tunneling particle the resulting temperature includes the Hawking temperature only at the lowest order, in addition, some higher order corrections in $\hbar$ appear (\ref{2.26}). The motivation of this section is to consider this corrected temperature in the first law and solve this to find corresponding correctional terms in entropy.  

The modified form of first law of thermodynamics for Kerr-Newman black hole in the presence of corrections to Hawking temperature is 
\begin{eqnarray}
dS_{\textrm {bh}}=\frac{dM}{T_{\textrm {bh}}}+(-\frac{\Omega_{\textrm H}}{T_{\textrm {bh}}})dJ+(-\frac{\Phi_{\textrm H}}{T_{\textrm {bh}}})dQ.
\label{5.1}
\end{eqnarray}
 In this context we further assume that, 
\begin{itemize}
\item{{\it Entropy must be a state function for all stationary spacetimes even in the presence of the quantum corrections to the semi-classical value.}}
\end{itemize}
 This implies that $dS_{\textrm {bh}}$ has to be an exact differential. In the expression for $T_{\textrm {bh}}$ in (\ref{2.26}) there are six undetermined coefficients ($a_1$ to $a_6$) present in $H_{{\textrm {KN}}}$ (\ref{2.15}). The first step in the analysis is to fix these coefficients in such a way that $dS_{\textrm {bh}}$ in (\ref{5.1}) remains an exact deferential. By this restriction we make the corrected black hole entropy independent of any collapse process. For (\ref{5.1}) to be an exact differential the following relations must hold:
\begin{eqnarray}
\frac{\partial}{\partial J}(\frac{1}{T_{\textrm {bh}}})\big|_{M,Q}=\frac{\partial}{\partial M}(-\frac{\Omega_{\textrm H}}{T_{\textrm {bh}}})\big|_{J,Q}
\label{5.2}
\end{eqnarray}
\begin{eqnarray}
\frac{\partial}{\partial Q}(-\frac{\Omega_{\textrm H}}{T_{\textrm {bh}}})\big|_{M,J}=\frac{\partial}{\partial J}(-\frac{\Phi_{\textrm H}}{T_{\textrm {bh}}})\big|_{M,Q}
\label{5.3}
\end{eqnarray}
\begin{eqnarray}
\frac{\partial}{\partial M}(-\frac{\Phi_{\textrm H}}{T_{\textrm {bh}}})\big|_{J,Q}=\frac{\partial}{\partial Q}(\frac{1}{T_{\textrm {bh}}})\big|_{J,M}.
\label{5.4}
\end{eqnarray}
Using the expression of $T_{\textrm {bh}}$ from (\ref{2.26}) and the semiclassical result from (\ref{4.71}), the first condition (\ref{5.2}) reduces to 
\begin{eqnarray}
\frac{\partial}{\partial J}\displaystyle\sum_i \frac{\beta_i\hbar^i}{H^i_{\textrm {KN}}}\big|_{M,Q}=-\Omega_{\textrm H}\frac{\partial}{\partial M}\displaystyle\sum_i\frac{\beta_i\hbar^i}{H^i_{\textrm {KN}}}\big|_{J,Q}
\label{5.41}
\end{eqnarray}
Expanding this equation in powers of $\hbar$, one has the following equality
\begin{eqnarray}
\frac{\partial H_{{\textrm {KN}}}}{\partial J}\big|_{M,Q}=-\Omega_{\textrm H}\frac{\partial H_{{\textrm {KN}}}}{\partial M}\big|_{J,Q}.
\label{5.5}   
\end{eqnarray}
Similarly the other two integrability conditions (\ref{5.3}) and (\ref{5.4}) lead to other conditions on $H_{\textrm {KN}}$,
\begin{eqnarray}
\frac{\partial H_{{\textrm {KN}}}}{\partial Q}\big|_{M,J}=\left(\frac{\Phi_{\textrm H}}{\Omega_{\textrm H}}\right)\frac{\partial{H_{\textrm {KN}}}}{\partial J}\big|_{M,Q}
\label{5.6}   
\end{eqnarray}

\begin{eqnarray}
\frac{\partial H_{{\textrm {KN}}}}{\partial M}\big|_{J,Q}=-\frac{1}{\Phi_{\textrm H}}\frac{\partial H_{\textrm {KN}}}{\partial Q}\big|_{J,M}
\label{5.61}   
\end{eqnarray}
respectively. The number of unknown coefficients present in $H_{{\textrm {KN}}}$ is six and we have only three equations involving them, so the problem is under determined.

 As a remedy to this problem let us first carry out the dimensional analysis for Kerr spacetime and then use the result to reduce the arbitrariness in $H_{\textrm {KN}}$. For $Q=0$ the Kerr-Newman metric reduces to the rotating Kerr spacetime and one can carry the same analysis to find the corrections to Hawking temperature for both scalar particle and fermion tunneling from Kerr spacetime. An identical calculation will be repeated with $Q=0$. The only difference will appear in the dimensional analysis (\ref{2.15}). Since Kerr metric is chargeless the most general expression for corrected Hawking temperature will come out as
\begin{eqnarray}
T_{\textrm {bh}}=T\Big(1+\sum_i\beta_i\frac{\hbar^i}{H_{\textrm K}^i}\Big)^{-1},
\label{5.7}
\end{eqnarray}
where $H_{\textrm K}$ is now given by
\begin{eqnarray}
H_{{\textrm K}}=H_{\textrm {KN}}(Q=0)= a_1r^2_+ + a_2Mr_++a_3M^2.
\label{5.8}
\end{eqnarray}
The first law of thermodynamics for Kerr black hole is
\begin{eqnarray}
dS=\frac{dM}{T_{\textrm H}}+(-\frac{\Omega_{\textrm H}}{T_{\textrm H}})dJ,
\label{5.9}
\end{eqnarray}
where $T_{\textrm H}$ and $\Omega_{\textrm H}$ for Kerr black hole are obtained from their corresponding expressions for the Kerr-Newman case, for $Q=0$, as given in Appendix \ref{appendix2B}. With these expressions one can easily check that $dS$ is an exact differential for Kerr black hole as well since the only integrability condition 
\begin{eqnarray}
\frac{\partial}{\partial J}(\frac{1}{T_{\textrm H}})\big|_{M}=\frac{\partial}{\partial M}(-\frac{\Omega_{\textrm H}}{T_{\textrm H}})\big|_{J}
\label{5.10}
\end{eqnarray} 
is satisfied. As stated earlier the idea behind introducing the Kerr spacetime is to carry out the dimensional analysis for Kerr spacetime first, then demanding that for $Q=0$ the dimensional parameter $H_{{\textrm {KN}}}$ will be same as $H_{{\textrm K}}$. The form of first law for Kerr black hole in presence of corrections to the Hawking temperature, is given by  
\begin{eqnarray}
dS_{\textrm {bh}}=\frac{dM}{T_{\textrm {bh}}}+(-\frac{\Omega_{\textrm H}}{T_{\textrm {bh}}})dJ,
\label{5.11}
\end{eqnarray}
 where the general form of $T_{\textrm {bh}}$ is given in (\ref{5.7}). Now demanding that the corrected entropy of Kerr black hole must be a state function, the following integrability condition  
\begin{eqnarray}
\frac{\partial}{\partial J}(\frac{1}{T_{\textrm {bh}}})\big|_{M}=\frac{\partial}{\partial M}(-\frac{\Omega_{\textrm H}}{T_{\textrm {bh}}})\big|_{J}
\label{5.12}
\end{eqnarray} 
 must hold. Using the semi-classical result from (\ref{5.10}) and considering corrections to all orders in $\hbar$ to the Hawking temperature in (\ref{5.7}) it follows that the above integrability condition is satisfied if the following relation holds
\begin{eqnarray}
\frac{\partial H_{{\textrm K}}}{\partial J}\big|_{M}=-\Omega_{\textrm H}\frac{\partial H_{{\textrm K}}}{\partial M}\big|_{J}.
\label{5.13}
\end{eqnarray} 
 From (\ref{5.8}) it follows that this equality holds only for
\begin{eqnarray}
a_1= 0= a_3
\label{5.131}
\end{eqnarray}
 and the form of $H_{\textrm K}$ is given by 
\begin{eqnarray}
H_{{\textrm K}}=a_2Mr_+.
\label{5.14}
\end{eqnarray} 
Therefore, the corrected form for the Hawking temperature obeying the integrability condition (\ref{5.12}) for the Kerr black hole is given by
\begin{eqnarray}
T_{\textrm {bh}}=T_{\textrm H}\left(1+\displaystyle\sum_i\frac{\beta_i\hbar^i}{(a_2Mr_+)^i}\right)^{-1}=T_{\textrm H}\left(1+\displaystyle\sum_i\frac{\tilde\beta_i\hbar^i}{(Mr_+)^i}\right)^{-1}.
\label{5.141}
\end{eqnarray}

The natural expectation from the dimensional term ($H_{\textrm {KN}}$) in (\ref{2.15}) is that for $Q=0$ it gives the correct dimensional term ($H_{\textrm K}$) in (\ref{5.14}). To fulfil this criterion we must have $a_1= 0= a_3$ in (\ref{2.15}) and this leads to
\begin{align}
H_{{\textrm {KN}}}=a_2Mr_+ + a_4{r_+ Q}+a_5{MQ}+a_6{Q^2}\notag\\
=a_2(Mr_+ + \tilde a_4{r_+ Q}+\tilde a_5{MQ}+\tilde a_6{Q^2}),
\label{5.15}
\end{align}
where $\tilde a_j=\frac{a_j}{a_2}$. Now we are in a position to find the precise form of the dimensional term ($H_{\textrm {KN}}$) satisfying the integrability conditions given in (\ref{5.5}), (\ref{5.6}) and (\ref{5.61}). Note that with the modified expression (\ref{5.15}) the problem of under determination of six coefficients by only three integrability conditions for Kerr-Newman spacetime has been removed. With this expression of $H_{\textrm {KN}}$ one has effectively three undetermined coefficients with three equations and it is straightforward to calculate those coefficients. Putting the new expression of $H_{\text {KN}}$ in (\ref{5.5}), (\ref{5.6}) and (\ref{5.61}) one obtains,      
\begin{eqnarray}
\tilde a_5 - \tilde a_4\frac{r_+}{M}=0 
\label{5.16}
\end{eqnarray}
\begin{eqnarray}
2\tilde a_6 Q + \tilde a_4 \left(\frac{Mr_+ +Q^2}{M}\right)+ \tilde a_5M=-Q 
\label{5.17}
\end{eqnarray}
\begin{eqnarray}
2\tilde a_6 Q + \tilde a_4 \left(r_+ + \frac{J^2Q^2}{M^3(r^2_+ + J^2/M^2)}\right)+ \tilde a_5\left(M+\frac{Q^2r_+}{r^2_+ +J^2/M^2}\right)=-Q. 
\label{5.171} 
\end{eqnarray}
The simultaneous solution of these three equations yields,
\begin{eqnarray}
\tilde a_4= 0= \tilde a_5\notag\\
\tilde a_6= -\frac{1}{2}.
\label{5.172}
\end{eqnarray}
As a result the final form of $H_{\text {KN}}$ derived by the requirements:\\ (i) $H_{\text {KN}}$  must satisfy the integrability conditions (\ref{5.5}, \ref{5.6}, \ref{5.61}), \\ (ii)   $H_{\text {KN}}= H_{\text K}$ for $Q=0$,\\  is given by 
\begin{eqnarray}
H_{{\textrm {KN}}}=a_2(Mr_+-\frac{1}{2}Q^2).
\label{5.18}
\end{eqnarray}
Hence the corrected Hawking temperature for Kerr-Newman black hole is found to be
\begin{eqnarray}
T_{\textrm {bh}}=T\left(1+\displaystyle\sum_i\frac{\beta_i\hbar^i}{a^i_2(Mr_+-\frac{Q^2}{2})^i}\right)^{-1}=T\left(1+\displaystyle\sum_i\frac{\tilde\beta_i\hbar^i}{(Mr_+-\frac{Q^2}{2})^i}\right)^{-1},
\label{5.19}
\end{eqnarray} 
where $\tilde\beta_i=\frac{\beta_i}{a^i_2}$.

We are now in a position to compute the corrected entropy and find the deviations from the semi-classical area law. Comparing (\ref{4.8}) and (\ref{5.1}) with $T_{\textrm {bh}}$ given above we find a similar dictionary as (\ref{4.15}) by modifying semi-classical terms with corrected versions, where necessary, as 
\begin{align}
(f\rightarrow S_{\textrm {bh}},~~x\rightarrow M,~~y\rightarrow J,~~z\rightarrow Q)\nonumber\\
(U\rightarrow\frac{1}{T_{\textrm {bh}}},~~V\rightarrow\frac{-\Omega_{\textrm H}}{T_{\textrm {bh}}},~~W\rightarrow\frac{-\Phi_{\textrm H}}{T_{\textrm {bh}}}).
\label{5.20}
\end{align}
Following this dictionary and (\ref{4.12}), (\ref{4.13}) and (\ref{4.14}) the corrected entropy for Kerr-Newman black hole has the form
\begin{eqnarray}
S_{\textrm {bh}}=\int{\frac{dM}{T_{\textrm {bh}}}}+\int{XdJ}+\int{YdQ},
\label{5.21}
\end{eqnarray}
where
\begin{eqnarray}
X=(-\frac{\Omega_{\textrm H}}{T_{\textrm {bh}}})-\frac{\partial}{\partial J}{\int\frac{dM}{T_{\textrm {bh}}}}
\label{5.22}
\end{eqnarray}
and
\begin{eqnarray}
Y=(-\frac{\Phi_{\textrm H}}{T_{\textrm {bh}}})-\frac{\partial}{\partial Q}[{\int\frac{dM}{T_{\textrm {bh}}}}+{\int XdJ}].
\label{5.23}
\end{eqnarray}
It is possible to calculate $S_{\textrm {bh}}$ analytically up to all orders of $\hbar$. However we shall restrict ourselves up to second order correction to the Hawking temperature. Integration over $M$ yields,   
\begin{eqnarray}
\int\frac{dM}{T}&=&\frac{\pi}{\hbar}(2Mr_+-Q^2)+ 2\pi\tilde\beta_1\hbar\log(2Mr_+-Q^2)-\frac{4\pi\tilde\beta_2\hbar^2}{(2Mr_+-Q^2)^2}+{\textrm {const.}}\nonumber\\
&& +{\textrm {higher order terms}}.
\label{5.24}
\end{eqnarray}
With this result of integration one can check the following relation, 
\begin{eqnarray}
\frac{\partial}{\partial J}\int\frac{dM}{T_{\textrm {bh}}}=-\frac{\Omega_{\textrm H}}{T_{\textrm {bh}}}.
\label{5.25}
\end{eqnarray}
 Therefore $X=0$. Furthermore we get
\begin{eqnarray}
\frac{\partial}{\partial Q}\int\frac{dM}{T_{\textrm {bh}}}=-\frac{\Phi_{\textrm H}}{T_{\textrm {bh}}},
\label{5.26}
\end{eqnarray}
 and using this equality together with $X=0$ we find $Y=0$. The fact that both $X$ and $Y$ pick the most trivial solution as zero in the black hole context, both with or without quantum corrections, is quite unique. The final result for the entropy of the Kerr-Newman black hole in presence of quantum corrections is now given by
\begin{eqnarray}
S_{\textrm {bh}}&=&\frac{\pi}{\hbar}(2Mr_+-Q^2)+ 2\pi\tilde\beta_1\log(2Mr_+-Q^2)-\frac{4\pi\tilde\beta_2\hbar}{(2Mr_+-Q^2)}+{\textrm {const.}}\nonumber\\
&&+{\textrm {higher order terms}}.
\label{5.27}
\end{eqnarray}  
In terms of the semi-classical black hole entropy and horizon area this can be expressed, respectively, as
\begin{eqnarray}
S_{\textrm {bh}}=S_{\textrm {BH}}+ 2\pi\tilde\beta_1\log S_{\textrm {BH}}-\frac{4\pi^2\tilde\beta_2}{S_{\textrm {BH}}}+{\textrm {const.}}+{\textrm {higher order terms}}.
\label{5.28}
\end{eqnarray}  
and
\begin{eqnarray}
S_{\textrm {bh}}=\frac{A}{4}+ 2\pi\tilde\beta_1\log A-\frac{16\pi^2\tilde\beta_2}{A}+{\textrm {const.}}+{\textrm {higher order terms}}.
\label{5.29}
\end{eqnarray}  
The first term in the expression (\ref{5.28}) is the usual semi-classical Bekenstein-Hawking entropy and the other terms are due to higher order corrections. The logarithmic and inverse area terms have appeared as the leading and non leading corrections to the entropy and area law. It is to be noted that although this derivation is performed for the Kerr-Newman black hole the functional form for corrected entropy is also valid for the less general black hole spacetimes, {\it i.e.} Kerr, Reissner-Nordstrom and Schwarzschild examples. However in each case one has different values of horizon area ($A$) and arbitrary coefficients $\beta_1$ and $\beta_2$. In the next chapter we shall demonstrate the computation for the coefficient of the leading (logarithmic) correction ($\beta_1$) separately for all cases.

\section{\label{btz}Corrected entropy of (2+1) dimensional BTZ black hole}   

So far our studies on black hole thermodynamics are contained within (3+1) dimensions. However the approach discussed above is also applicable for black holes in other dimensions (both higher and lower). In the present section we are addressing a particular case of (2+1) dimensions for which black hole solutions were found by Banados, Teitelboim and Zanelli (BTZ) \cite{Banados}. 

The metric for the (2+1) dimensional BTZ black hole with negative cosmological constant $\Lambda= -\frac{1}{l^2}$ and in the unit of $8G_3=1$, where $G_3$ is the three dimensional Newton's constant, is given by \cite{Banados},   
\begin{eqnarray}
ds^2 = -N^2dt^2+N^{-2}dr^2+r^2(N^\phi dt+d\phi)^2,
\label{1.01btz}
\end{eqnarray}
with the lapse function
\begin{eqnarray}
N^2(r) = -M+\frac{r^2}{l^2}+\frac{J^2}{4r^2},
\label{1.02}
\end{eqnarray}
and
\begin{eqnarray} 
N^\phi (r) = -\frac{J}{2r^2}. 
\label{1.03}
\end{eqnarray}
Here $M$ and $J$ are respectively the mass and angular momentum of the BTZ black hole. Outer (event) and inner horizons are obtained by setting $g^{rr}=N^2= 0$. This yields, 
\begin{eqnarray} 
r_{\pm} = \frac{l}{\sqrt 2}\left(M\pm\sqrt{M^2-\frac{J^2}{l^2}}\right)^{1/2}.
\label{1.04h}
\end{eqnarray}
Therefore the horizon radius is a function of both $M$ and $J$. The area of the event (outer) horizon is given by, 
\begin{equation}
A=2 \pi r_+.
\label{1.041}
\end{equation}
The angular velocity at the event horizon is given by 
\begin{eqnarray}
\Omega= -\frac{g_{\phi t}}{g_{\phi \phi}}\Big|_{r=r_+}=\frac{J}{2r^2_+}.
\label{1.06om}
\end{eqnarray}

There exist some important works based on Cardy formula \cite{Carlip} and statistical method \cite{Govind} which show that under certain circumstances the semi-classical entropy of BTZ black hole undergoes some higher order corrections. In fact both these studies reported the leading order correction to be logarithmic. Our aim here is to use the ideas of tunneling mechanism and exact differentials and look for any higher order corrections to BTZ black hole entropy. These ideas have already been worked out for (3+1) dimensional black holes. A detailed description of the same for BTZ black hole would be a repeating task. We thus prefer only to outline major steps that leads to some important results.

To compute the semi-classical Hawking temperature let us first isolate the $r-t$ sector of the metric (\ref{1.01btz}) by making the following coordinate transformation near the event horizon,
\begin{eqnarray} 
d\chi=d\phi-\Omega dt
\label{1.05}
\end{eqnarray}
i.e, 
\begin{eqnarray}
\chi=\phi-\Omega t.
\label{new11}
\end{eqnarray}
In the near horizon approximation, using the transformation (\ref{new11}) the metric(\ref{1.01btz}) can be written in the desired form,
\begin{eqnarray}
ds^2 = -N^2dt^2+N^{-2}dr^2+r_+^2d\chi^2,
\label{new1}
\end{eqnarray}
 where the $r-t$ sector is isolated from the angular part ($d\chi^2$).

Now considering the radial trajectory, {\it i.e.,} the $r-t$ sector of the metric (\ref{new1}), the expression of semi-classical Hawking temperature follows from the analysis presented in Chapter-3 (section-\ref{qthj}), in the following form, 
\begin{eqnarray}
T_{\textrm{H}}=\frac{\hbar}{4}{\left(Im\int_C{\frac{dr}{\sqrt {g_{tt}(r)g^{rr}(r)}}}\right)}^{-1}
\label{new3}
\end{eqnarray} 
Considering the metric (\ref{new1}), the Hawking temperature for the BTZ black hole is found to be,
\begin{eqnarray}
T_{\textrm{H}}=\frac{\hbar}{4}{\left(Im\int_C{\frac{dr}{N^2(r)}}\right)}^{-1}=\frac{\hbar}{4}{\left(l^2Im\int_C{\frac{r^2dr}{(r^2-r^2_+)(r^2-r^2_-)}}\right)}^{-1}.      
\label{new4}
\end{eqnarray}
The integration is to be performed remembering that the particle tunnels from just behind the event horizon to the outer region. So the integrand has a simple pole at $r=r_+.$ Choosing the contour as a half loop going around this pole from left to right and integrating, we obtain the desired result for the semi-classical Hawking temperature, 
\begin{eqnarray}
T_{\textrm{H}}=\frac{\hbar}{2\pi l^2}\left(\frac{r^2_+-r^2_-}{r_+}\right).
\label{1.07}
\end{eqnarray}

Similarly it is trivial to carry out the steps outlined in Chapter-3 (for beyond semi-classical case) to find the corrected temperature of BTZ black hole, given by,
\begin{eqnarray}
T_{\textrm{bh}}=T_{\textrm{H}}\Big(1+\sum_i\beta_i\frac{\hbar^i}{r^i_+}\Big)^{-1},
\label{rev19}
\end{eqnarray}
where $T_{\textrm{H}}$ is given by (\ref{1.07}) and also we have made a dimensional analysis by introducing the horizon radius in higher order terms. This choice is obvious for BTZ black hole since in $(2+1)$ dimensions in the units $8G_3=1$ and $c=k_B=1$ the Planck constant ($\hbar$) is of the order of Planck length ($l_P$) and independent of Planck mass ($m_{P}$).

With the well known expression of the semi-classical Hawking temperature  (\ref{1.07}) we now proceed with the calculation of the semi-classical Bekenstein-Hawking entropy. The corrections to these semi-classical results will be considered later. Consider the first law of thermodynamics for a chargeless rotating black hole (also follows from (\ref{sthermo}) for $Q=0$),
\begin{eqnarray}
dM= T_{\textrm{H}}dS+\Omega dJ,
\label{1.08}
\end{eqnarray}
which can be written in the form
\begin{eqnarray}
dS(M,J)= \frac{dM}{T_{\textrm{H}}}-\frac{\Omega}{T_{\textrm{H}}}dJ.
\label{1.09}
\end{eqnarray}
Inverting the Hawking temperature found from (\ref{1.07}) and (\ref{1.04h}) of the BTZ black hole, we obtain,
\begin{eqnarray}
\frac{1}{T_{\textrm{H}}}(M,J)= \frac{\sqrt 2\pi l \left(M+\sqrt{M^2-\frac{J^2}{l^2}}\right)^{1/2}}{\hbar\sqrt{M^2-\frac{J^2}{l^2}}}.
\label{1.10bt}
\end{eqnarray} 
The angular velocity (\ref{1.06om}) simplifies to,
\begin{eqnarray}
\Omega=-\frac{g_{\phi t}}{g_{\phi\phi}}|_{r=r_+}=\frac{J}{l^2(M+\sqrt{M^2-\frac{J^2}{l^2}})}.
\label{1.101}
\end{eqnarray}
Using (\ref{1.10bt}) and (\ref{1.101}) yields,
\begin{eqnarray}
\frac{\Omega}{T_{\textrm{H}}}(M,J)= \frac{\sqrt 2 \pi J}{l\hbar\left(M+\sqrt{M^2-\frac{J^2}{l^2}}\right)^{1/2}\sqrt{M^2-\frac{J^2}{l^2}}}.
\label{1.11}
\end{eqnarray}
Note that $dS$ in (\ref{1.09}) is an exact differential since it satisfies the relation,
\begin{eqnarray}
\frac{\partial}{\partial J}\left(\frac{1}{T_{\textrm{H}}}\right)=\frac{\partial}{\partial M}\left(\frac{-\Omega}{T_{\textrm{H}}}\right),
\label{new5}
\end{eqnarray}
 as may be checked on exploiting the equations (\ref{1.10bt}), (\ref{1.11}). Now, in order to calculate the entropy ($S$), (\ref{1.09}) has to be solved. Here it is easy to follow the method used for the Kerr-Newman case in section-\ref{calsemien}. In fact the present case is much simpler version of (\ref{sthermo1}). Looking into the solution (\ref{4.19}) of (\ref{sthermo1}), it is now trivial to write down the solution of (\ref{1.09}) for the semi-classical result of entropy for the spinning BTZ black hole, as
\begin{eqnarray}
S= \int\frac{1}{T_{\textrm{H}}}dM + \int\frac{-\Omega}{T_{\textrm{H}}}dJ -\int\frac{\partial}{\partial J}\left(\int\frac{1}{T_{\textrm{H}}}dM\right)dJ.
\label{1.16}
\end{eqnarray}
The solution of the integral over $dM$ gives
\begin{eqnarray}
\int\frac{1}{T_{\textrm{H}}}dM= 4\pi \frac{l}{\sqrt 2 \hbar}(M+\sqrt{M^2-\frac{J^2}{l^2}})^{1/2}=\frac{4\pi r_+}{\hbar}. 
\label{1.17}
\end{eqnarray}
Having this result, it can be easily checked that the following relation also holds,
\begin{eqnarray} 
\frac{\partial}{\partial J}\left(\int\frac{dM}{T_{\textrm{H}}}\right)=\frac{-\Omega}{T_{\textrm{H}}}.
\label{1.18}
\end{eqnarray}
Substituting the above relation in (\ref{1.16}) and using (\ref{1.17}), immediately leads to the entropy of the spinning BTZ black hole,
\begin{eqnarray}
S=\int\frac{1}{T_{\textrm{H}}}dM=\frac{4\pi r_+}{\hbar}. 
\label{1.19}
\end{eqnarray}
If we write this in terms of horizon area (\ref{1.041}) by reinstating the unit $8G_3=1$ , we get
\begin{eqnarray}
S=\frac{A}{4\hbar G_3}=S_{{\textrm {BH}}}, 
\label{1.1911}
\end{eqnarray}
 which is the well known semi-classical Bekenstein-Hawking area law.

Now it is also possible to calculate higher order corrections to (\ref{1.1911}) in the following manner. First consider the expression of modified Hawking temperature ($T_{\textrm{bh}}$) and plug it into the first law, so that,
\begin{equation}
dS_{{\textrm {bh}}}= \frac{dM}{T_{\textrm{bh}}}-\frac{\Omega}{T_{\textrm{bh}}}dJ.
\label{1.21}
\end{equation}
Here also, one can check, $dS_{{\textrm {bh}}}$ is a perfect differential since the following relationship,
\begin{eqnarray}
\frac{\partial}{\partial J}\sum_i\frac{1}{T_{\textrm{H}}}\left(1+\frac{\beta_i\hbar^i}{r^i_+}\right)=\frac{\partial}{\partial M}\sum_i\frac{-\Omega}{T_{\textrm{H}}}\left(1+\frac{\beta_i\hbar^i}{r^i_+}\right)
\label{1.22}
\end{eqnarray}
 holds. This can be proved order by order by expanding the summation and substituting $r_+$, $\frac{1}{T}$ and $\frac{\Omega}{T}$ from (\ref{1.04h}), (\ref{1.10bt}) and (\ref{1.11}) respectively.

Following the analysis of section-\ref{calcoren} the expression of corrected entropy for BTZ black hole is given by,
$$S_{\textrm {bh}}(M,J)= \int\frac{1}{T_{\textrm{H}}}\sum_i\left(1+\frac{\beta_i \hbar^i}{r^i_+}\right)dM + \int{\sum_i\frac{-\Omega}{T_{\textrm{H}}}\left(1+\frac{\beta_i \hbar^i}{r^i_+}\right)}dJ -$$
\begin{eqnarray}\int\frac{\partial}{\partial J}\left(\int\frac{1}{T}\sum_i\left(1+\frac{\beta_i \hbar^i}{r^i_+}\right)dM\right)dJ,
\label{1.23}
\end{eqnarray}
which is the modified version of (\ref{1.16}). It is possible to solve (\ref{1.23}) analytically for all orders . Let us now restrict upto second order corrections. In that case we expand the summation over the integrands in (\ref{1.23}) upto the second order, which yields,
$$S_{{\textrm {bh}}}(M,J)= \int\frac{1}{T}\left(1+\frac{\beta_1 \hbar}{r_+}+\frac{\beta_2 \hbar^2}{r^2_+}+{\cal O}(\hbar^3)\right)dM + \int\frac{-\Omega}{T}\left(1+\frac{\beta_1 \hbar}{r_+}+\frac{\beta_2 \hbar^2}{r^2_+}+{\cal O}(\hbar^3)\right)dJ -$$
\begin{eqnarray}\int\frac{\partial}{\partial J}\left(\int\frac{1}{T_{\textrm{H}}}\left(1+\frac{\beta_2 \hbar}{r_+}+\frac{\beta_2 \hbar^2}{r^2_+}+{\cal O}(\hbar^3)\right)dM\right)dJ.
\label{1.24}
\end{eqnarray}
Solving the integral over $dM$ in (\ref{1.24}) by using the value of $r_+$ from (\ref{1.04h}) yields,
$$\int\frac{1}{T_{\textrm{H}}}\left(1+\frac{\beta_1\hbar}{r_+}+\frac{\beta_2\hbar^2}{r^2_+}+{\cal O}(\hbar^3)\right)dM= \frac{4\pi l}{\sqrt 2\hbar}(M+\sqrt{M^2-\frac{J^2}{l^2}})^{1/2}+ 2\pi\beta_1\log(M+\sqrt{M^2-\frac{J^2}{l^2}})$$
\begin{equation}
-\frac{4\sqrt2\pi\beta_2\hbar}{l^2(M+\sqrt{M^2-\frac{J^2}{l^2}})^{1/2}}+{\cal O}(\hbar^3).
\label{1.25}
\end{equation}
Using this result one can establish the following relationship between the integrands of the second and third integral of (\ref{1.24}), 
\begin{eqnarray}
\frac{\partial}{\partial J}\int\frac{1}{T_{\textrm{H}}}\left(1+\frac{\beta_1 \hbar}{r_+}+\frac{\beta_2 \hbar^2}{r^2_+}+{\cal O}(\hbar^3)\right)dM={\frac{-\Omega}{T_{\textrm{H}}}\left(1+\frac{\beta_1 \hbar}{r_+}+\frac{\beta_2 \hbar^2}{r^2_+}+{\cal O}(\hbar^3)\right)}~
\label{1.26}
\end{eqnarray}
Substituting this in (\ref{1.24}) and using (\ref{1.25}) we find the entropy of the BTZ black hole including higher order corrections,
$$ S_{{\textrm {bh}}}=\frac{4\pi l}{\sqrt 2\hbar}(M+\sqrt{M^2-\frac{J^2}{l^2}})^{1/2}+ 2\pi\beta_1\log(M+\sqrt{M^2-\frac{J^2}{l^2}})-$$
\begin{equation}
\frac{4\sqrt2\pi\beta_2\hbar}{l^2(M+\sqrt{M^2-\frac{J^2}{l^2}})^{1/2}}+{\cal O}(\hbar^3)+ const..
\label{1.27}
\end{equation}
Equation (\ref{1.27}) can be expressed in terms of horizon (outer) radius to yield,
\begin{equation}
S_{\textrm {bh}}= \frac{4\pi r_+}{\hbar}+4\pi\beta_1\log {r_+} -\frac{4\pi\beta_2\hbar}{l}(\frac{1}{r_+}) +{\cal O}(\hbar^3)+const..
\label{1.28}
\end{equation}
Substituting $r_+$ from (\ref{1.041}) and reinstating the unit $8G_3=1$, we can write this expression for entropy in terms of horizon area of the BTZ black hole as given by,
\begin{equation}
S_{\textrm {bh}}= \frac{A}{4\hbar G_3}+4\pi\beta_1\log A-\frac{64\pi^2\beta_2\hbar G_3}{l}(\frac{1}{A}) +{\cal O}(\hbar^3)+const..
\label{1.29}
\end{equation}
This can also be written in terms of semi-classical Bekenstein-Hawking entropy (\ref{1.1911}),
\begin{equation}
S_{\textrm {bh}}= S_{\textrm {BH}}+4\pi\beta_1\log(S_ {\textrm {BH}})-\frac{16\pi^2\beta_2}{l}(\frac{1}{{S_\textrm {BH}}}) +{\cal O}(\hbar^3)+const..
\label{1.30}
\end{equation} 
The first term in (\ref{1.30}) is the usual semi-classical Bekenstein-Hawking entropy (\ref{1.1911}) while the other terms are corrections due to quantum effects. We see that a logarithmic correction appears in the leading order. Incidentally, we also find inverse of area term in sub-leading order (\ref{1.29}) which was not discussed in other approaches \cite{Carlip,Govind}.

Thus as an important observation we find that for both (3+1) and (2+1) dimensions the functional form of the corrected entropy (\ref{5.28}) and (\ref{1.30}) are identical. The reason behind such a behavior might have some deeper implications. Especially, these correctional terms play dominant role near the Planck scale, the physics of which is not very well understood.

\section{\label{traceanomaly}Trace anomaly and leading correction to semi-classical expressions} 

Some recent works based on field theory \cite{Fursaev,fur2}, quantum geometry \cite{Partha,partha2}, statistical mechanics \cite{Das, das2, das3}, Cardy formula \cite{Carlip,carlip2}, brick wall method \cite{Hooft,hooft2} and tunneling method \cite{Majhitrace} have confirmed that there exist corrections to the semi-classical results in black hole thermodynamics.  Despite of diversity among several approaches it has been found that most general form of black hole entropy includes logarithmic and inverse horizon area corrections to the semiclassical value. But they in general differ in fixing the coefficient of the leading/logarithmic correction. At the present status no approach has been able to tell us anything about other coefficients coming with inverse area terms. 

So far we have found that higher order corrections to the Hawking temperature (\ref{5.19}) as well as entropy (\ref{5.28}) of various black holes have quite generic structures. These results also matches with the works mentioned above. We also recall that there is one unknown coefficient coming with each of these higher order corrections to the area law. In this section we fix the normalisation ($\tilde\beta_1$) of the leading correction to the semi-classical black hole entropy and temperature. Unlike the existing works which fail to include all spacetime metrics, here we fix $\tilde\beta_1$ for all spacetime metrics in (3+1) dimensional Einstein gravity. This coefficient is found to be related with the trace anomaly of the stress tensor for the scalar field.

We start by considering the scalar particle action for the Kerr-Newman spacetime is given by (\ref{2.16})
\begin{eqnarray}
{\cal S}(r,t)=\left({\cal S}_0(r,t)+\sum_i\hbar^i{\cal S}_i(r,t)\right)=\left({\cal S}_0(r,t)+\sum_i\frac{\tilde\beta_i\hbar^i}{(Mr_+ -\frac{Q^2}{2})^i}{\cal S}_0(r,t)\right),
\label{6.1}
\end{eqnarray}
 where the appropriate form for $H_{KN}$ from (\ref{5.18}) is considered. Taking the first order ($\hbar$) correction in this equation we can write the following relation for the imaginary part of the outgoing particle action   
\begin{eqnarray}
{\textrm {Im}{\cal S}_1^{\textrm {out}}}(r,t)=\frac{\tilde\beta_1}{(Mr_+ -\frac{Q^2}{2})}{\textrm {Im}}{\cal S}_0^{\textrm {out}}(r,t).
\label{6.2}
\end{eqnarray}
The imaginary part for the semi-classical action for an outgoing particle can be found from (\ref{2.13}), (\ref{2.17}) and (\ref{2.23}) as 
\begin{eqnarray}
{\textrm {Im}}{\cal S}_0^{\textrm {out}}(r,t)=-2\omega {\textrm {Im}}\int_C{\frac{dr}{\sqrt{f(r)g(r)}}}.
\label{6.3}
\end{eqnarray}

Let us make an infinitesimal scale transformation of the metric coefficients in (\ref{2.7ht}) parametrized by the constant factor `$k$' such that $\bar f(r)=k f(r)\simeq (1+\delta k)f(r)$ and $\bar g(r)=k^{-1} g(r)\simeq (1+\delta k)^{-1}g(r)$. From the scale invariance of the Klein-Gordon equation in (\ref{2.8sp}) it follows that the Klein-Gordon field ($\Phi$) should transform as $\Phi=k^{-1}\Phi$. Since $\Phi$ has a dimension of mass, one interprets that the black hole mass ($M$) should transform as $M=k^{-1}M\simeq (1+\delta k)^{-1}M$ under the infinitesimal scale transformation. Therefore the other two black hole parameters ($Q$, $a$) and the particle energy $\omega$ should also transform as $M$ does. Using these it is straightforward to calculate the transformed form of (\ref{6.2}) and (\ref{6.3}) to get

\begin{eqnarray}
{\textrm {Im}\overline{\cal S}_1^{\textrm {out}}}(r,t)=\frac{\tilde\beta_1}{(\overline M\overline r_+ -\frac{\overline Q^2}{2})}{\textrm {Im}}\overline{\cal S}_0^{\textrm {out}}(r,t)=\frac{\tilde\beta_1}{(Mr_+ -\frac{Q^2}{2})}(1+\delta k){\textrm {Im}}{\cal S}_0^{\textrm {out}}(r,t),
\label{6.4}
\end{eqnarray}
and
\begin{eqnarray}
\frac{\delta {\textrm {Im}}{{\cal S}}^{{\textrm {out}}}_1(r,t)}{\delta k}=\frac{\tilde\beta_1}{(Mr_+ -\frac{Q^2}{2})}{\textrm {Im}}{\cal S}_0^{\textrm {out}}(r,t).
\label{6.5}
\end{eqnarray}

Now consider the massless scalar field action
\begin{eqnarray}
{\cal S}=\frac{1}{2}\int{\sqrt{-g}\nabla_{\mu}\Phi\nabla^{\mu}\Phi}
\label{sca}
\end{eqnarray}
 Under a constant scale transformation of the metric coefficients ($g_{\mu\nu}\rightarrow\overline{g}_{\mu\nu}$) this action is not invariant in the presence of trace anomaly. This lack of conformal invariance is given by the following relation 
\begin{eqnarray}
\frac{\delta {{\cal S}}}{\delta k}=\frac{1}{2}\int{d^4x\sqrt{-g}(<T^{\mu}_{~\mu}>^{(1)}+<T^{\mu}_{~\mu}>^{(2)}+...)},
\label{6.6}
\end{eqnarray} 
 where $<T^{\mu}_{~\mu}>^{(i)}$' s  are the trace of the regularised stress energy tensor calculated for the $i$-th loop. However, in the literature \cite{Hawkzeta, Dewitt}, only the first order loop calculation has been carried out and this gives
\begin{eqnarray}
\frac{\delta {\textrm {Im}}{{\cal S}}^{{\textrm {out}}}_1(r,t)}{\delta k}=\frac{1}{2}{\textrm {Im}}\int{d^4x\sqrt{-g}(<T^{\mu}_{~\mu}>^{(1)})},
\label{6.7}
\end{eqnarray} 
where, for a scalar background, the form of trace anomaly is given by \cite{Hawkzeta, Dewitt}
\begin{eqnarray}
<T^{\mu}_{~\mu}>^{(1)}=\frac{1}{2880\pi^2}\left(R_{\mu\nu\rho\sigma}R^{\mu\nu\rho\sigma}-R_{\mu\nu}R^{\mu\nu}+\nabla_{\mu}\nabla^{\mu}R\right)
\label{6.9}
\end{eqnarray} 
Now integrating (\ref{6.3}) around the pole at $r=r_+$ we get
\begin{eqnarray}
{\textrm {Im}}{\cal S}_0^{\textrm {out}}(r,t)=-2\pi\omega\frac{(r_+M-\frac{Q^2}{2})}{(r_+-M)}
\label{6.9a}
\end{eqnarray}
Putting this in (\ref{6.5}) and comparing with (\ref{6.7}) we find
\begin{eqnarray}
\tilde\beta_1=-\frac{(M^2-Q^2-\frac{J^2}{M^2})^{1/2}}{4\pi\omega}{\textrm {Im}}\int{d^4x\sqrt{-g}<T^{\mu}_{~\mu}>^{(1)}}.
\label{6.8}
\end{eqnarray} 
Equation (\ref{6.8}) gives the general form of the coefficient associated with the leading correction to the semi-classical entropy for any stationary black hole. To get $\tilde\beta_1$ for a particular black hole in (3+1) dimensions one needs to solve both (\ref{6.8}) and (\ref{6.9}) for that black hole. In the remaining sections we shall take different spacetime metrics and explicitly calculate $\tilde\beta_1$ for them.

\subsection{\label{sch}Schwarzschild black hole} 
For $Q=0=J$ the Kerr-Newman spacetime metric reduces to the Schwarzschild spacetime and from (\ref{6.8}) it follows that
\begin{eqnarray}
\tilde\beta_1=-\frac{M}{4\pi\omega}{\textrm {Im}}\int{d^4x\sqrt{-g}<T^{\mu}_{~\mu}>^{(1)}}.
\label{6.10}
\end{eqnarray} 
The effective energy of the tunneling particle ($\omega$) for the Schwarzschild spacetime is given by the Komar conserved quantity corresponding the timelike Killing vector ($K_{\xi^{\mu}_{(t)}}$) evaluated at the event horizon. An exact calculation of the Komar integral (\ref{eeff}) gives $\omega=M$ (also follows from (\ref{ms}) for $Q=0=M$), where $M$ is the mass of Schwarzschild black hole. Therefore we get 
\begin{eqnarray}
\tilde\beta_1=-\frac{1}{4\pi}{\textrm {Im}}\int{d^4x\sqrt{-g}<T^{\mu}_{~\mu}>^{(1)}}.
\label{6.11}
\end{eqnarray} 
A similar result was found by Hawking \cite{Hawkzeta}, where the path integral approach based on zeta function regularization was adopted. The path integral for standard Einstein-Hilbert gravity  was modified due to the fluctuations coming from the scalar field in the black hole spacetime. 

To find the trace anomaly of the stress tensor (\ref{6.9}) we calculate the following invariant scalars for Schwarzschild black hole, given by 
\begin{align}
R_{\mu\nu\rho\sigma}R^{\mu\nu\rho\sigma} &=\frac{48M^2}{r^6},\notag\\
R_{\mu\nu}R^{\mu\nu} &=0\label{6.11a}\\
R=0.\notag
\end{align}
Using these we can find $<T^{\mu}_{~\mu}>^{(1)}$ from (\ref{6.9}) and inserting it in (\ref{6.11}) yields,
\begin{align}
\tilde\beta_1^{({\textrm {Sch}})} &=-\frac{1}{4\pi}\frac{1}{2880\pi^2}{\textrm {Im}}\int_{r=2M}^{\infty}\int_{\theta=0}^{\pi}\int_{\phi=0}^{2\pi}\int_{t=0}^{-8\pi i M}{\frac{48M^2}{r^6}r^2 \sin\theta dr d\theta d\phi dt}\notag\\
&=\frac{1}{180\pi}.
\label{6.12}
\end{align} 
The corrected entropy/area law (\ref{5.28}, \ref{5.29}) is now given by,
\begin{align}
S_{\textrm {bh}}^{\textrm {(Sch)}}=S_{\textrm {BH}}+ \frac{1}{90}\log S_{\textrm {BH}}+{\textrm {higher order terms}},
\nonumber\\
=\frac{A}{4}+\frac{1}{90}\log A + {\textrm {higher order terms.}}
\label{6.13}
\end{align}   

The expression for corrected temperature follows from (\ref{5.19}) and (\ref{6.12}), given by, 
\begin{eqnarray}
T_{\textrm{bh}}^{\textrm{Sch}}=T_{\textrm{H}}\left(1+\frac{\hbar}{180\pi Mr_+}+\textrm{higher order terms}\right).
\label{corsc}
\end{eqnarray}

\subsection{\label{rn}Reissner-Nordstrom black hole} 
For the Reissner-Nordstrom black hole, putting $J=0$ in (\ref{6.8}), we get
\begin{eqnarray}
\tilde\beta_1=-\frac{(M^2-Q^2)^{1/2}}{4\pi\omega}{\textrm {Im}}\int{d^4x\sqrt{-g}<T^{\mu}_{~\mu}>^{(1)}},
\label{6.14}
\end{eqnarray} 
 where the particle energy is again given by the Komar energy integral corresponding to the timelike Killing field $\xi^{\mu}_{(t)}$. Unlike the Schwarzschild case, however,  the effective energy for Reissner-Nordstrom black hole observed at a distance $r$, is now given by $a=0$ limit in (\ref{ms}),  
\begin{eqnarray} 
\omega=(M-\frac{Q^2}{r}).
\label{6.15} 
\end{eqnarray}   
For a particle undergoing tunneling $r=r_+=(M+\sqrt{M^2-Q^2})$, we get $\omega=(M^2-Q^2)^{1/2}$ and therefore (\ref{6.14}) gives
\begin{eqnarray}
\tilde\beta_1=-\frac{1}{4\pi}{\textrm {Im}}\int{d^4x\sqrt{-g}<T^{\mu}_{~\mu}>^{(1)}}.
\label{6.16}
\end{eqnarray}
This has exactly the same functional form as (\ref{6.11}). To calculate this integral, we first simplify the integrand given in (\ref{6.9}), for a Reissner-Nordstrom black hole,
\begin{align}
R_{\mu\nu\rho\sigma}R^{\mu\nu\rho\sigma} &=\frac{8(7Q^4-12MQ^2r+6M^2r^2)}{r^8},\notag\\
R_{\mu\nu}R^{\mu\nu} &=\frac{4Q^4}{r^8},\label{6.16a}\\
R=0.\notag
\end{align}
With these results $<T^{\mu}_{~\mu}>^{(1)}$ is obtained and, finally,
\begin{align}
\tilde\beta_1^{(\textrm {RN})} &=-\frac{1}{4\pi}\frac{1}{2880\pi^2}{\textrm {Im}}\int_{r=r_+}^{\infty}\int_{\theta=0}^{\pi}\int_{\phi=0}^{2\pi}\int_{t=0}^{-i\beta}{<T^{\mu}_{~\mu}>^{(1)}r^2 \sin\theta dr d\theta d\phi dt}\notag\\
&=\frac{1}{180\pi}(1+\frac{3}{5}\frac{r_-^2}{r_+^2-r_+r_-}).
\label{6.16b}
\end{align} 
Therefore the corrected entropy/area law for a Reissner-Nordstrom black hole is now given by (\ref{5.28}, \ref{5.29})
\begin{eqnarray}
S_{\textrm {bh}}^{\textrm {(RN)}}=S_{\textrm {BH}}+ \frac{1}{90}(1+\frac{3}{5}\frac{r_-^2}{r_+^2-r_+r_-})\log S_{\textrm {BH}}+{\textrm {higher order terms}}.
\nonumber\\
=\frac{A}{4}+ \frac{1}{90}(1+\frac{3}{5}\frac{r_-^2}{r_+^2-r_+r_-})\log{A}+{\textrm {higher order terms.}}
\label{6.16c}
\end{eqnarray}
Unlike the Schwarzschild black hole here the prefactor of the logarithmic term is not a pure number. This is because the presence of charge on the outer region of the event horizon includes a contribution to the matter sector. Therefore the charge ($Q$) directly affects the dynamics of the system which in turn is related to entropy. It is interesting to see that in the extremal limit the prefactor of the logarithmic term blows up, suggesting that there cannot be a smooth limit from non-extremal to the extremal case. This is in agreement with a recent work \cite{Carrolpaper} where it is argued that the {\it extremal limit} of the Reissner-Nordstrom black hole is different from the extremal case itself. For the extremal case the region between inner and outer horizons disappears but in the {\it extremal limit} this region no longer disappears, rather it approaches a patch of $AdS_{2}\times S^{2}$. As a result the non-extremal to extremal limit is not continuous. 

Finally the corrected temperature for the this black hole is found from (\ref{5.19}), given by
\begin{eqnarray}
T_{\textrm{bh}}^{\textrm{RN}}=T_{\textrm{H}}\left(1+\frac{\tilde\beta_1\hbar}{(Mr_+-\frac{Q^2}{2})}+\textrm{higher order terms}\right),
\end{eqnarray}
where $\tilde\beta_1$ is expressed in  (\ref{6.16b}).

\subsection{\label{kerr}Kerr black hole}
The Kerr black hole is the chargeless limit of the Kerr-Newman black hole. This is an axially symmetric solution of Einstein's equation and has two Killing vectors $\xi^{\mu}_{(t)}$ and $\xi^{\mu}_{(\phi)}$ as already discussed in Chapter-2. 

 For a Kerr black hole (\ref{6.8}) reduces to
\begin{eqnarray}
\tilde\beta_1=-\frac{(M^2-\frac{J^2}{M^2})^{1/2}}{4\pi\omega}{\textrm {Im}}\int{d^4x\sqrt{-g}<T^{\mu}_{~\mu}>^{(1)}}
\label{6.17}
\end{eqnarray}
where $\omega$ is represented by (\ref{2.13a}). In $Q=0$ limit the results for the effective conserved charges (\ref{ms},\ref{jeff1}) yield $\omega=(M^2-\frac{J^2}{M^2})^{1/2}$ and therefore (\ref{6.17}) becomes
\begin{eqnarray}
\tilde\beta_1=-\frac{1}{4\pi}{\textrm {Im}}\int{d^4x\sqrt{-g}<T^{\mu}_{~\mu}>^{(1)}}
\label{6.18}
\end{eqnarray}
which is exactly same as the two previous cases. The invariant scalars for Kerr spacetime are given by
\begin{align}
R_{\mu\nu\rho\sigma}R^{\mu\nu\rho\sigma} &=-\frac{96M^2\left(\alpha_1+15\alpha_2\cos{2\theta}+6a^4(a^2-10r^2)\cos{4\theta}+a^6\cos{6\theta}\right)}{(a^2+2r^2+a^2\cos{2\theta})^6},\notag\\
\alpha_1 &=(10a^6-180a^4r^2+240a^2r^4-32r^6), \notag\\
\alpha_2 &=(a^4-16a^2r^2+16r^4)\notag\\
R_{\mu\nu}R^{\mu\nu} &=0\label{6.18a}\\
R=0,\notag
\end{align}
from which the trace $<T^{\mu}_{~\mu}>^{(1)}$ in (\ref{6.9})is obtained. Now performing the integration we get  
\begin{align}
\tilde\beta_1^{(\textrm {K})} &=-\frac{1}{4\pi}\frac{1}{2880\pi^2}{\textrm {Im}}\int_{r=r_+}^{\infty}\int_{\theta=0}^{\pi}\int_{\phi=0}^{2\pi}\int_{t=0}^{-i\beta}{<T^{\mu}_{~\mu}>^{(1)}r^2 \sin\theta dr d\theta d\phi dt}\notag\\
&=\frac{1}{180\pi}.
\label{6.18b}
\end{align} 
Therefore the corrected entropy/area law for a Kerr black hole that follows from (\ref{5.28}, \ref{5.29}) is given by, 
\begin{eqnarray}
S_{\textrm {bh}}^{\textrm {(K)}}=S_{\textrm {BH}}+ \frac{1}{90}\log S_{\textrm {BH}}+{\textrm {higher order terms}},
\nonumber\\
=\frac{A}{4}+\frac{1}{90}\log A + {\textrm {higher order terms.}}
\label{6.18c}
\end{eqnarray} 
This result is identical to the Schwarzschild black hole. This can be physically explained by the following argument. The difference between the Schwarzschild and Kerr spacetimes is due to spin($a$). Unlike charge ($Q$), which has a contribution to the matter part, spin is arising in Kerr spacetime because of one extra Killing direction. This difference is purely geometrical and has nothing to do with the dynamics of the system and as a result there is no difference between the structure of corrected entropy in these two cases. 

Likewise the corrected temperature has the identical expression given in (\ref{corsc})  where $T_{\textrm{H}}$ now represents the Hawking temperature for the Kerr black hole.

\subsection{\label{kn}Kerr-Newman black hole}
The general expression for $\tilde\beta_1$ in (\ref{6.8}) involves the total energy of the tunneling particle, given by (\ref{2.13a}). Unlike the Kerr black hole, in this case the effective energy faced by a particle at a finite distance from the horizon is not the same as felt at infinity. Because of the presence of electric charge ($Q$) it is modified. This was also the case for Reissner-Nordstrom black hole where one extra term ($-\frac{Q^2}{r}$) arose in (\ref{6.15}) due to the charge of the black hole. Using the effective expressions (\ref{ms}) and (\ref{jeff1}) in (\ref{2.13a}) we obtain $\omega=(M^2-Q^2-\frac{J^2}{M^2})^{1/2}$. Therefore $\tilde\beta_1$ for the Kerr-Newman spacetime reads as, 
\begin{eqnarray}
\tilde\beta_1=-\frac{1}{4\pi}{\textrm {Im}}\int{d^4x\sqrt{-g}<T^{\mu}_{~\mu}>^{(1)}},
\label{6.20}
\end{eqnarray}
 which is identical to the previous expressions. The invariant scalars for Kerr-Newman black holes are given by
\begin{align}
R_{\mu \nu \rho \sigma } R^{\mu \nu \rho \sigma } &= \frac{128 }{(a^2+2 r^2+a^2 \cos{2 \theta })^6}
\nonumber\\ 
 &[192 r^4 (Q^2-2m r)^2-96 r^2 (Q^2-3 m r) (Q^2-2 m r)(a^2+2 r^2+a^2 \cos{2 \theta})) \nonumber \\ 
 &+ (7 Q^2-18 m r)(Q^2-6 m r) (a^2+2 r^2+a^2 cos{2 \theta})^2\nonumber\\&-3 m^2 (a^2+2 r^2+a^2 \cos{2 \theta})^3)],\notag\\
R_{\mu\nu}R^{\mu\nu} &= \frac{64 Q^4}{\left(a^2+2 r^2+a^2 \cos{2 \theta}\right)^4},\label{6.20a}\\
R=0.\notag
\end{align}
Simplifying $<T^{\mu}_{\mu}>^{(1)}$ in (\ref{6.9}) and performing the integration in (\ref{6.20}) one finds
\begin{align}
\tilde\beta _1^{\textrm {KN}} &= \frac{r_+^2 + r_+ r_- -Q^2}{5760 \pi  r_+^4 (r_+ -r_-) (r_+ r_- -Q^2)^{5/2}}\left(\alpha_1 + \frac{r_+\sqrt{r_+r_- -Q^2}}{r_+^2 + r_+r_- - Q^2} (9Q^8-\alpha_2r_+ + \alpha_3r_+^2r_-^2)\right)\label{6.20b}\\
 &{\textrm {with,}}\notag\\
\alpha_1 &=9 Q^4 [ r_+^4\tan^{-1}{(\frac{r_+}{\sqrt{-Q^2+r_- r_+}})} +  (Q^2-r_-r_+)^2\cot^{-1}{\frac{r_+}{\sqrt{-Q^2+r_- r_+}}}]\notag\\
\alpha_2 &=6Q^6r_+-41Q^4r_+^3+32r_+^4r_-^3+2Q^2r_-(9Q^4+13q^2r_+^2+32r_+^4)\notag\\
\alpha_3 &= 9Q^4+64Q^2r_+^2+32r_+^4.\notag
\end{align}
The corrected entropy/area law now follows from (\ref{5.28}) and (\ref{5.29}),
\begin{align}
S_{\textrm {bh}}^{\textrm {(KN)}}&=S_{\textrm {BH}}+ \frac{r_+^2 + r_+ r_- -Q^2}{2880  r_+^4 (r_+ -r_-) (r_+ r_- -Q^2)^{5/2}}\times\nonumber\\&\left(\alpha_1 + \frac{r_+\sqrt{r_+r_- -Q^2}}{r_+^2 + r_+r_- - Q^2} (9Q^8-\alpha_2r_+ + \alpha_3r_+^2r_-^2)\right)\times\nonumber\\
&\log S_{\textrm {BH}} +{\textrm {h. o. terms}}.\nonumber\\
&=\frac{A}{4}+\frac{r_+^2 + r_+ r_- -Q^2}{2880  r_+^4 (r_+ -r_-) (r_+ r_- -Q^2)^{5/2}}\times\nonumber\\&\left(\alpha_1 + \frac{r_+\sqrt{r_+r_- -Q^2}}{r_+^2 + r_+r_- - Q^2} (9Q^8-\alpha_2r_+ + \alpha_3r_+^2r_-^2)\right)\times\nonumber\\
&\log A+{\textrm {h. o. terms}}.
\label{6.20c}
\end{align} 
For $Q=0$ the above prefactor of the logarithmic term reduces to $\frac{1}{90}$, the coeffecient for the Kerr spacetime. This proves the robustness of the result.

Finally plugging the result for $\tilde\beta_1$ from (\ref{6.20b}) into (\ref{5.19}) gives the desired result of corrected temperature of the Kerr-Newman black hole.


\section{\label{dischentropy}Discussions} 
Let us now summarise the content of this chapter. We have given a new and simple approach to derive the ``first law of black hole thermodynamics'' from the thermodynamical perspective where one does not require the ``first law of black hole mechanics''. The key point of this derivation was the observation that ``black hole entropy'' is a {\it {state function}}. In the process we obtained some relations involving black hole entities, playing a role analogous to {\it {Maxwell's relations}}, which must hold for any stationary black hole. Based on these relations, we presented a systematic calculation of the semi-classical Bekenstein-Hawking entropy taking into account all the ``work terms on a black hole''. This approach is applicable to any stationary black hole solution. The standard semi-classical area law was reproduced. An interesting observation that has been come out of the calculation was that the work terms did not contribute to the final result of the semi-classical entropy.

To extend our method for calculating entropy in the presence of higher order corrections we used the result for the corrected Hawking temperature as derived in Chapter-3. However this result involved a number of arbitrary constants. Demanding that the corrected entropy be a state function it was possible to find the appropriate form of the corrected Hawking temperature. By using this result we explicitly calculated the entropy with higher order corrections. In the process we again found that work terms on black hole did not contribute to the final result of the corrected entropy. This analysis was done for the Kerr-Newman spacetime and it was trivial to find the results for other stationary spacetimes like (i) Kerr, (ii) Reissner-Nordstrom and (iii) Schwarzschild by taking appropriate limits. Identical results were found for (2+1) dimensional BTZ black hole. It is important to note that the functional form for the corrected entropy is same for all these stationary black holes. The logarithmic and inverse area terms as leading and next to leading corrections were quite generic up-to a dimensionless prefactor.   
      
The coefficient of the leading order correction was calculated and it was found to be proportional with the trace anomaly evaluated in the 1st order loop approximation. In particular the coefficient of the leading correction was computed exactly for all known black hole solutions of Einstein gravity in $(3+1)$ dimensions. Since the result of trace anomaly is known only in the first loop approximation we could not tell anything about the coefficients coming with higher (non-leading) corrections to temperature and entropy.

\chapter{\label{chap-leading-correction}Thermodynamics of noncommutativity-inspired Schwarzschild black hole }


We have, so far, found modifications to the Bekenstein-Hawking area law and Hawking temperature of various black holes. These modifications were due to the higher order corrections in the WKB ansatz for the tunneling quantum field. However, this is not the only way that expressions for thermodynamic variables may be modified from their standard semi-classical values. Indeed, similar modifications may be generated by the noncommutative corrections to gravity, which also have black hole solutions. These are usually referred as noncommutative inspired black holes \cite{nico5, Smail}, \cite{Spall}-\cite{spa}. This approach has drawn significant attention in recent times (for an extensive list of references we refer review articles \cite{ncreview,Banerjee:2009gr}). Unlike the usual NC geometry where noncommutativity is introduced in the coordinate sector, here, it appears in the matter sector. In the following section we elaborate on this point and outline the method for obtaining the simplest (Schwarzschild) type black hole solution.
  
\section{{\label{ncsch}}Noncommutative geometry inspired Schwarzschild black hole}  
The usual definition of mass density in terms of the Dirac delta function in commutative space does not hold good in noncommutative space because of the position-position uncertainty relation. In noncommutative space mass density is defined by replacing the Dirac delta function by a Gaussian distribution of minimal width $\sqrt\theta$  in the following way \cite{Smail} {\footnote{This definition of mass density has been related to the Voros product between two coherent states defined in noncommutative space \cite{mod5}.}}
\begin{eqnarray}    
\rho_{\theta}(r) = \dfrac{M}{{(4\pi\theta)}^{3/2}}e^{-{\frac{r^2}{4\theta}}}
\label{1.01}
\end{eqnarray}
where the noncommutative parameter $\theta$ is a small ($\sim$ Plank length$^2$) positive number. Using this expression one can write the mass of the black hole of radius $r$ in the following way
\begin{eqnarray}
m_\theta(r) = \int_0^r{4\pi r'^{2}\rho_{\theta}(r')dr' }=\frac{2M}{\sqrt\pi}\gamma(3/2 , r^2/4\theta) 
\label{1.02}
\end{eqnarray}
where $\gamma(3/2 , r^2/4\theta)$ is the lower incomplete gamma function defined as
\begin{eqnarray}
\gamma(a,x)=\int_0^x t^{a-1} e^{-t} dt.
\label{1.021}
\end{eqnarray}
In the limit $\theta\rightarrow 0$ it becomes the usual gamma function $(\Gamma_{{\textrm{total}}})$. Therefore $m_\theta(r)\rightarrow M$ is the commutative limit of the noncommutative mass $m_\theta(r)$.

   To find a solution of Einstein equation with the noncommutative mass density of the type (\ref{1.01}), the temporal component of the energy momentum tensor ${(T_{\theta})}_\mu^\nu$ is identified as, ${(T_{\theta})}_t^t=-\rho_\theta$. Now demanding the condition on the metric coefficients ${(g_\theta)}_{tt}=-{(g_\theta)}^{rr}$ for the noncommutative Schwarzschild metric and using the covariant conservation of energy momentum tensor ${(T_{\theta})}_\mu^\nu~_{;\nu}=0$, the energy momentum tensor can be fixed to the form,
\begin{eqnarray}
{(T_\theta)}_\mu^\nu={\textrm {diag}}{[-\rho_\theta,p_r,p',p']},
\label{1.022}
\end{eqnarray}
 where, $p_r=-\rho_\theta$ and $p'=p_r-\frac{r}{2}\partial_r\rho_\theta=-(1+\frac{r}{2}\partial_r)\rho_{\theta}$. This form of energy momentum tensor is different from the perfect fluid because here $p_r$ and $p'$ are not same. Using (\ref{1.01}) we obtain,
\begin{eqnarray}
p'=\Big[\frac{r^2}{4\theta}-1\Big]\frac{M}{(4\pi\theta)^{\frac{3}{2}}}e^{-\frac{r^2}{4\theta}}
\label{ref2}
\end{eqnarray}
i.e. the pressure is anisotopic. But for $r<<\sqrt\theta$, the first term in (\ref{ref2}) drops out and $p'=-\rho_\theta=p_r$, i.e. the energy-momentum tensor takes the isotropic form. When $r\rightarrow 0$ the energy density tends to a constant value $-\frac{M}{(4\pi\theta)^{\frac{3}{2}}}$. On the other hand, at the large values of $r$ ($r>>\sqrt\theta$) all the components of the energy-momentum tensor very quickly tend to zero and so the pressure is again isotropic and the Schwarzschild vacuum solution is well applicable. 

     The solution of Einstein equation (in $c=G=1$ unit) ${(G_\theta)}^{\mu\nu}=8\pi{{(T_\theta)}^{\mu\nu}}$, using (\ref{1.022}) as the matter source, is given by the line element \cite{Smail},   
\begin{eqnarray}
ds^2=-f_\theta(r)dt^2+\frac{dr^2}{f_\theta(r)}+r^2d\Omega^2;\,\,\, f_\theta(r)=-{(g_\theta)}_{tt}=\left(1-\frac{4M}{r\sqrt\pi}\gamma(\frac{3}{2},\frac{r^2}{4\theta})\right)
\label{1.04}
\end{eqnarray}
 Incidentally, this is same if one just replaces the mass term in the usual commutative Schwarzschild space-time by the noncommutative mass $m_\theta(r)$ from (\ref{1.02}). Also observe that for $r>>\sqrt\theta$ the above noncommutative metric reduces to the standard Schwarzschild form.

     The metric (\ref{1.04}) represents a self-gravitating, anisotropic fluid type matter. The existence of the radial pressure in the small length scale ($r<<\sqrt\theta$) is due to the quantum vacuum fluctuation and it balances the inward gravitational pull to prevent the collapse of the matter to a point. This is reminiscent of earlier works \cite{Frolov}-\cite{shanki} where such a phenomenon is associated with the occurrence of a de-Sitter metric inside the black hole $(f_\theta(r)<0)$. As we now show the introduction of noncommutativity naturally induces a de-Sitter metric for $r<<\sqrt\theta$. In this limit the metric coefficient $f_\theta(r)$ in (\ref{1.04}) reduces to,
\begin{eqnarray}
f_\theta(r)\simeq 1-\frac{Mr^2}{3\sqrt{\pi}\theta^{\frac{3}{2}}}.
\label{ref3}
\end{eqnarray}
Therefore in this limit the metric (\ref{1.04}) reduces to a de-Sitter metric with cosmological constant 
\begin{eqnarray}
\Lambda_\theta=\frac{M}{3\sqrt\pi\theta^{3/2}}
\label{ref4}
\end{eqnarray}
which has a constant scalar curvature, given by, 
\begin{eqnarray}
R_\theta=\frac{4M}{\sqrt\pi\theta^{3/2}}.
\label{ref1}
\end{eqnarray}
Consequently there is no curvature singularity present any more, instead one finds a de-Sitter core of constant positive curvature surrounding the close vicinity of the singularity at $r=0$. This is in agreement with \cite{Frolov}-\cite{shanki} where the existence of the inner de-Sitter core was mentioned. Taking the commutative limit $\theta\rightarrow 0$ in (\ref{ref1}) immediately manifests the singularity.

It is interesting to note that the noncommutative metric (\ref{1.04}) is still stationary, static and spherically symmetric as in the commutative case. One or more of these properties is usually violated for other approaches \cite{Obre}-\cite{Kobak} of introducing noncommutativity, particularly those based on Seiberg-Witten maps that relate commutative spaces with noncommutative ones.

      The event horizon of the black hole can be found by setting ${(g_\theta)}_{tt}\Big|_{r=r_h}=0$ in (\ref{1.04}), which yields,
\begin{eqnarray}
r_h=\frac{4M}{\sqrt\pi}\gamma(\frac{3}{2},\frac{r^2_h}{4\theta}).
\label{1.05}
\end{eqnarray}
Keeping upto the leading order $\frac{1}{\sqrt{\theta}}e^{-{M^2}/{\theta}}$, we find
\begin{eqnarray}
r_h \simeq 2M\left(1-\frac{2M}{\sqrt{\pi\theta}}e^{{-M^2}/{\theta}}\right)  
\label{1.06}
\end{eqnarray}

       Now for a general stationary, static and spherically symmetric space time the Hawking temperature ($T_h$) is related to the surface gravity ($\kappa$) by the following relation \cite{Majhi1}
\begin{eqnarray}
T_h=\frac{\kappa}{2\pi};\,\,\,\kappa = [\frac{1}{2}\frac{d{(g_\theta)}_{tt}}{dr}]_{r=r_h}. 
\label{1.061}
\end{eqnarray}
Therefore the Hawking temperature for the noncommutative Schwarzschild black hole is given by,
\begin{eqnarray}
T_h = {\frac{1}{4\pi}}\left[{\frac{1}{r_h}}- {\frac{r_h^2}{4\theta^{3/2}}}\frac{e^-{\frac{{r_h}^2}{4\theta}}}{\gamma({\frac{3}{2}},{\frac{r^2_h}{4\theta}})}\right].
\label{1.08}
\end{eqnarray}
 To write the Hawking temperature in the regime $\frac{r^2_h}{4\theta}>>1$ as a function of $M$ we will use (\ref{1.06}). Keeping upto the leading order in $\theta$ we get
\begin{eqnarray}
T_h\simeq\frac{1}{8{\pi}M}\left[1-\frac{4M^3}{{\sqrt\pi} {\theta}^{3/2}}{e^{-M^2/\theta}}\right].
\label{1.10}
\end{eqnarray}
We will now use the integrated form of the first law of thermodynamics,  
\begin{eqnarray}
\int dS_{{\textrm{bh}}}=\int \frac{dM}{T_h}.
\label{1.1}
\end{eqnarray}
to calculate the Bekenstein-Hawking entropy. Upto the leading order in $\theta$ this is given by,
\begin{eqnarray}
S_{\textrm{bh}}=\frac{A_{\theta}}{4}; \,\,\ A_{\theta}=4\pi r_h^2\simeq 16\pi M^2-64\sqrt{\frac{\pi}{\theta}}M^3e^{-\frac{M^2}{\theta}}.
\label{1.13}
\end{eqnarray}
Although this is functionally identical to the Benkenstein-Hawking area law in the commutative space, note that, there is a NC correction to the semiclassical value. In $\theta\rightarrow 0$ limit this extra contribution vanishes and we recover the known result. In the next section we consider a new graphical analysis and find all such NC modifications to all orders in $\theta$.


\section{{\label{graphnc}}Graphical analysis and noncommutative corrections to the semiclassical area law}

We have analytically seen above that in the limit $\frac{r^2_h}{4\theta}>>1$ the noncommutative version of the semi-classical Bekenstein-Hawking area law holds upto the leading order in $\theta$. This motivates us to see whether this law holds for all orders in $\theta$, irrespective of the limit we have mentioned. Since analytically it seems very difficult, this issue will be discussed by a graphical analysis.

It will be always useful for us to write the right hand side of (\ref{1.1}) in terms of the horizon $r_h$ of the black hole. Using (\ref{1.05}) we have
\begin{eqnarray}
dM= \frac{\sqrt{\pi}}{4\gamma(\frac{3}{2},\frac{r^2_h}{4\theta})}\Big[1-\frac{r^3_h}{4{\theta}^{\frac{3}{2}}}\frac{e^{-\frac{r^2_h}{4\theta}}}{\gamma(\frac{3}{2},\frac{r^2_h}{4\theta})}\Big]dr_h.
\label{1.2}
\end{eqnarray}
Substituting this and (\ref{1.08}) in (\ref{1.1}), we get a closed form relation,
\begin{eqnarray}
\frac{dS_{{\textrm{bh}}}}{dr_h}=\frac{\pi^{\frac{3}{2}}r_h}{\gamma(\frac{3}{2},\frac{r^2_h}{4\theta})}.
\label{1.3}
\end{eqnarray}
This will be compared graphically with the quantity $\frac{dS_{{\textrm{bh}}}}{dr_h}$ calculated from the semi-classical Bekenstein-Hawking area law (\ref{1.13}). This yields, using (\ref{1.13}),
\begin{eqnarray}
\frac{dS_{{\textrm{bh}}}}{dr_h}\Big|_{\textrm{semiclassical}}=2\pi r_h.
\label{1.4}
\end{eqnarray}
Now $\frac{dS_{{\textrm{bh}}}}{dr_h}$ is plotted as a function of $r_h$ (for both equations (\ref{1.3}) and (\ref{1.4})) in figure (\ref{fig1}). 
\begin{figure}[h] 
\centering
\includegraphics[angle=0,width=7cm,keepaspectratio]{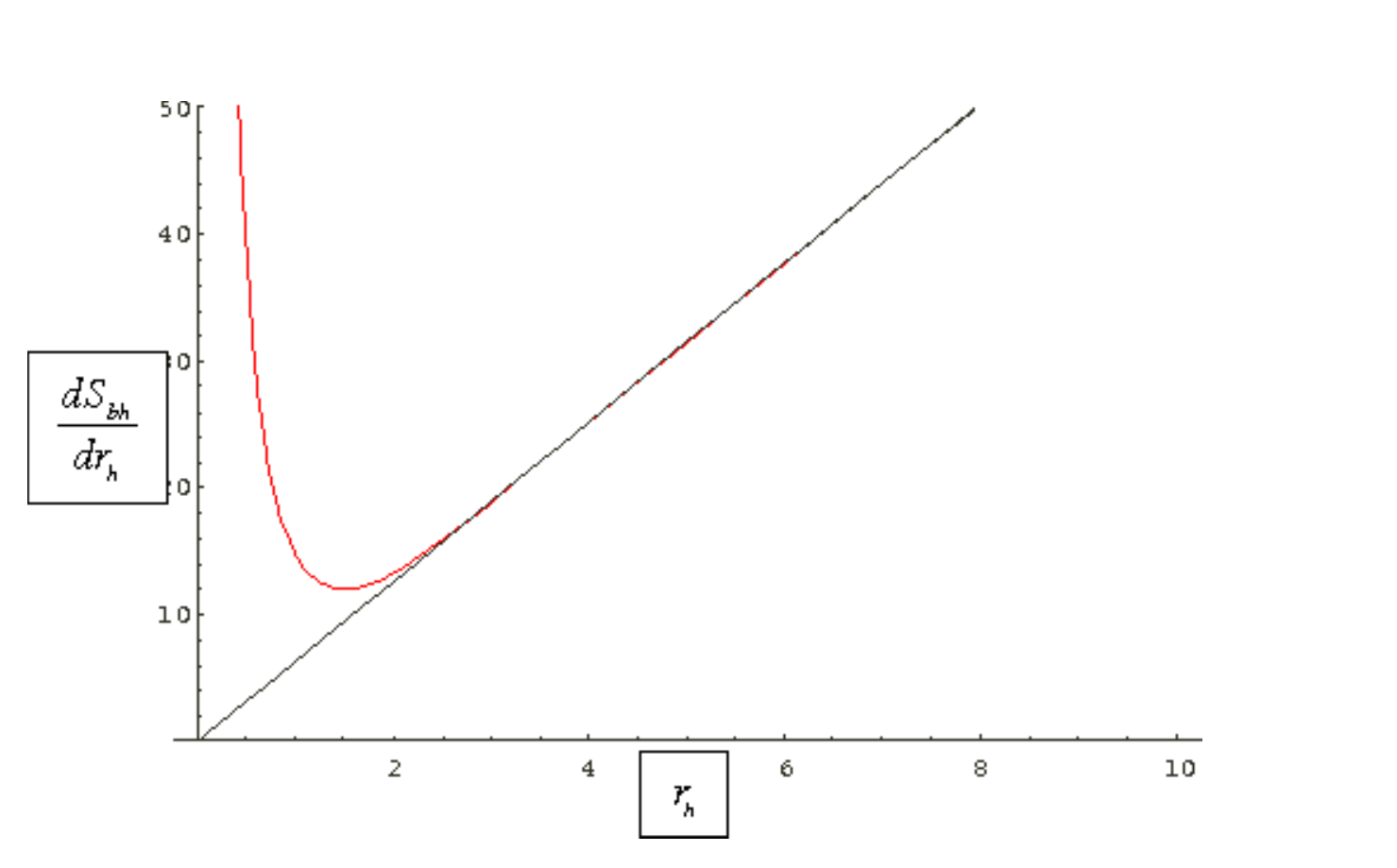}
\caption[]{\it $\frac{dS_{\textrm{bh}}}{dr_h}$ Vs. $r_h$ plot. $\frac{dS_{\textrm{bh}}}{dr_h}$ is plotted in units of $4\theta$ and $r_h$ is plotted in units of $2\sqrt\theta$. Red/upper curve: for eq. (\ref{1.3}), Black/lower curve: for eq. (\ref{1.4}).}
\label{fig1}
\end{figure}
It is interesting to see that semiclassical area law still holds for $r_h\gtrsim4.8\sqrt\theta$ since the two curves exactly coincide. To further understand this issue we solve for $r_h$ by equating (\ref{1.3}) and (\ref{1.4}) to obtain,
\begin{eqnarray}
\gamma\Big(\frac{3}{2},\frac{r_h^2}{4\theta}\Big)=\frac{\sqrt{\pi}}{2}
\label{gamma}
\end{eqnarray}
which is put in the form,
\begin{eqnarray}
\int_0^{\frac{r_h^2}{4\theta}}dt \sqrt{t} e^{-t}=\frac{\sqrt{\pi}}{2}
\label{int}
\end{eqnarray}
A numerical analysis yields the saturated bound for $\frac{r_h^2}{\theta}$ as $(4.8)^{2}$. This shows that the linear area law holds for values of $r_h\gtrsim 4.8\sqrt\theta$.

     Now in the region $r_h<4.8\sqrt\theta$ the two curves do not coincide. So there is a deviation from the usual area law. Also one can see from figure (\ref{fig1}) that the curve for (\ref{1.3}) (red/upper curve) attains a minimum value at $r_h=3.0\sqrt\theta$ and then sharply diverges for $r_h<3.0\sqrt\theta$ which is physically unreasonable since the change of Bekenstein-Hawking entropy with the horizon is expected to be unidirectional. This point will be cleared in the following analysis.

\begin{figure}[h] 
\centering
\includegraphics[angle=0,width=7cm,keepaspectratio]{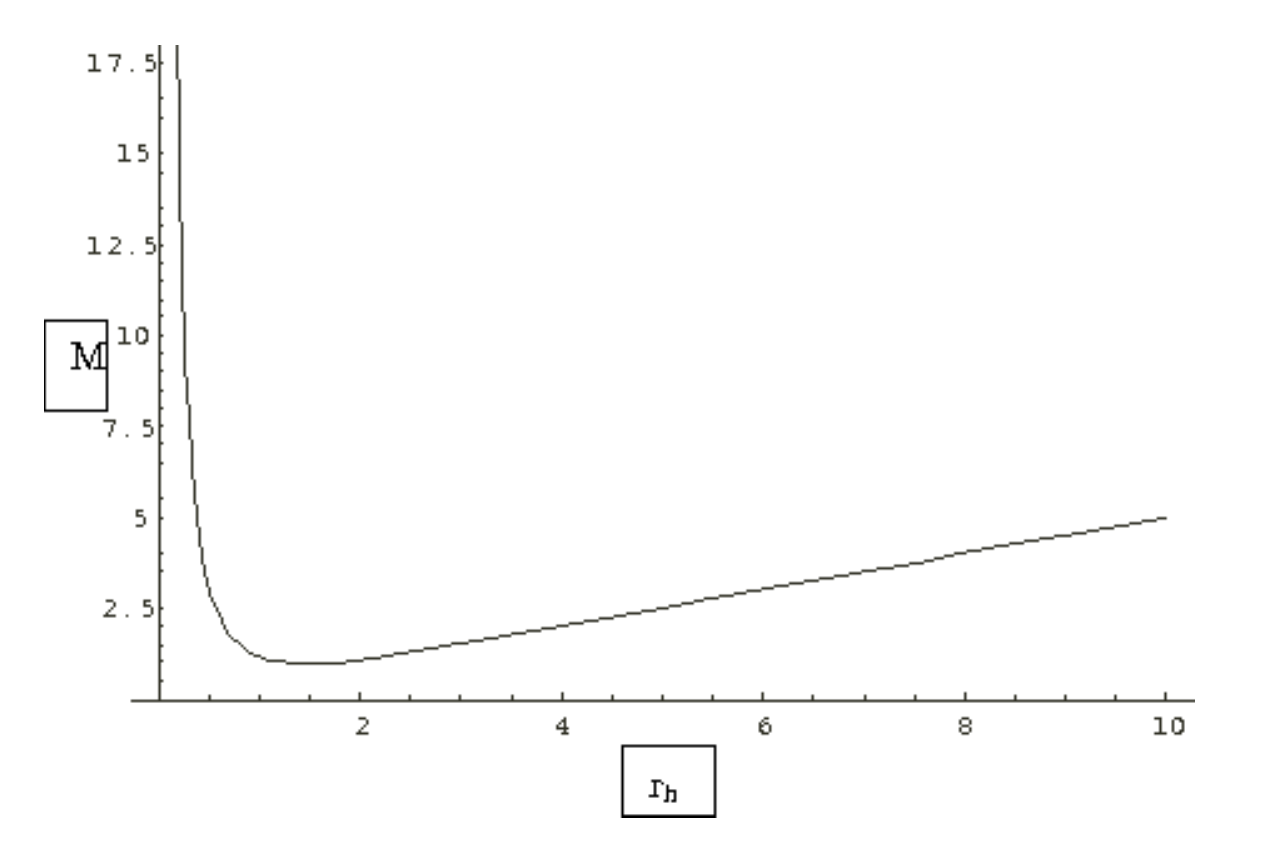}
\caption[]{{\it $M$ Vs. $r_h$ plot. $M$ is plotted in units of $2\sqrt\theta$ and $r_h$ is plotted in units of $2\sqrt\theta$ for eq. (\ref{1.05}).}}
\label{fig2}
\end{figure}

       In figure (\ref{fig2}) we plot the black hole mass $M$ as a function of $r_h$ (for equation (\ref{1.05})). It shows there is a minimum value ($M_0=1.9\sqrt\theta$) of $M$ at $r_h=3.0\sqrt\theta$ and noncommutativity introduces new behavior with respect to the standard Schwarzschild black hole \cite{Smail,Park}:\\
(i) Two distinct horizons occur for $M>M_0$: one inner (Cauchy) horizon and one outer (event) horizon. \\
(ii) One degenerate horizon occurs at $r_h=3.0\sqrt\theta$ for $M=M_0$.\\
(iii) No horizon occurs for $M<M_0$.\\  
In the case of $M>>M_0$, the inner horizon shrinks to zero while the outer horizon approaches the Schwarzschild radius $2M$. These features are also explained in \cite{Smail,Park}. Now we plot $T_h$ as a function of $r_h$ in figure (\ref{fig3}) (for equation (\ref{1.08})). It is observed that for $r_h<3.0\sqrt\theta$ there is no black hole because physically $T_h$ cannot be negative. Therefore the black hole only exists in the region $r_h\geq3.0\sqrt\theta$ for which there is only one horizon.
\begin{figure}[h] 
\centering
\includegraphics[angle=0,width=7cm,keepaspectratio]{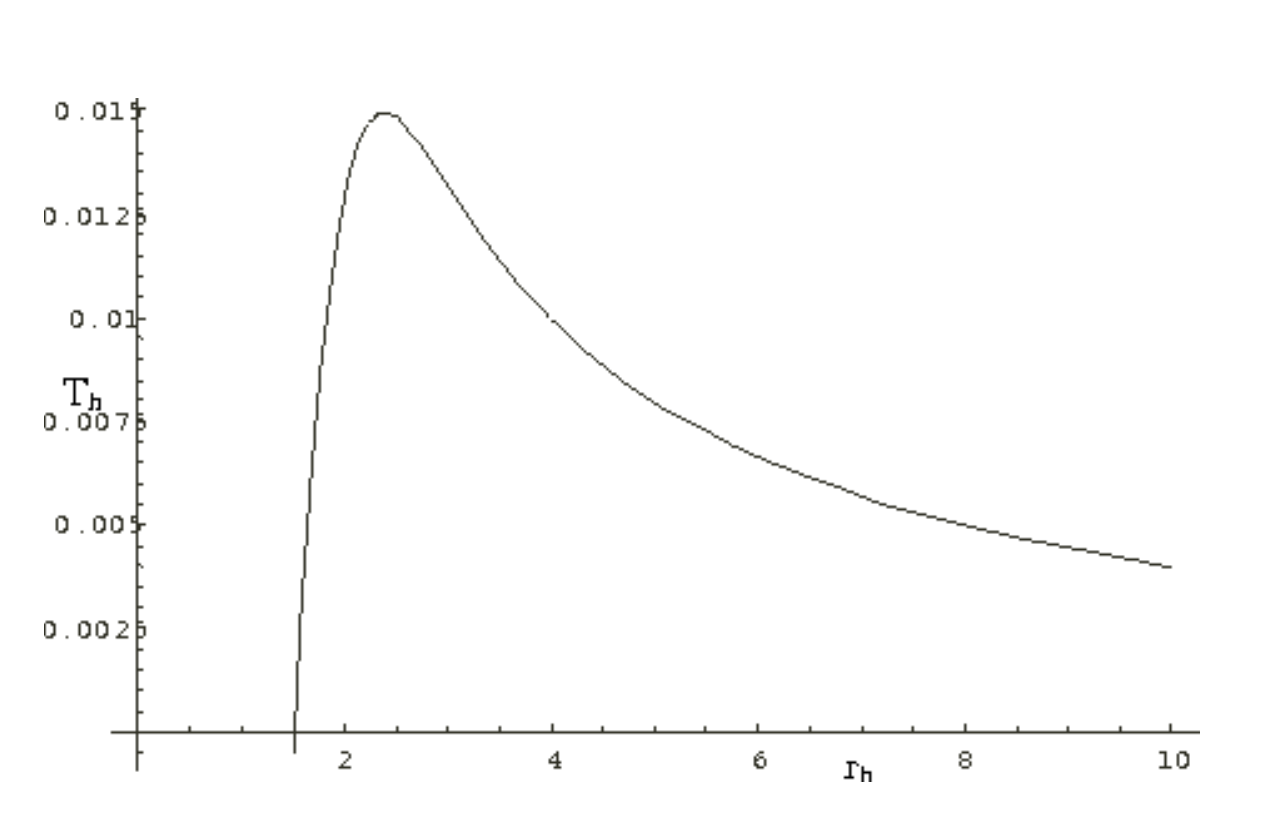}
\caption[]{{\it $T_h$ Vs. $r_h$ plot. $T_h$ is plotted in units of $\frac{1}{\sqrt\theta}$ and $r_h$ is plotted in units of $2\sqrt\theta$ for eq. (\ref{1.08}).}}
\label{fig3}
\end{figure}

       Now the minimum of $\frac{dS_{\textrm{bh}}}{dr_h}$ (see figure (\ref{fig1})) occurs for $r_h=3\sqrt\theta$ which just saturates the limit of physical validity of the black hole. Thus the sharp increase of $\frac{dS_{\textrm{bh}}}{dr_h}$ for $r_h<3\sqrt\theta$ is in the unphysical domain and hence ignored.

        So it is clear from figure (\ref{fig1}) that the semi-classical area law is not satisfied in the region $3.0\sqrt\theta\leq r_h<4.8\sqrt\theta$ while for $r_h\gtrsim4.8\sqrt\theta$ the Bekenstein-Hawking area law holds perfectly. We now find the correction to this area law such that it will describe the entropy for the complete physical region.

       To do this we proceed as follows. The first step is to expand (\ref{1.3}) in powers of the upper incomplete gamma function $\Gamma(\frac{3}{2},\frac{r^2_h}{4\theta})$,
\begin{eqnarray}
\Gamma(a,x)=\int_x^\infty t^{a-1}e^{-t}dt
\end{eqnarray}
so that, 
\begin{eqnarray}
\frac{dS_{{\textrm{bh}}}}{dr_h}&=&\frac{\pi^{\frac{3}{2}}r_h}{\frac{\sqrt\pi}{2}-\Gamma(\frac{3}{2},\frac{r^2_h}{4\theta})}
\nonumber
\\
&=&2\pi r_h\Big[1-\frac{2}{\sqrt\pi}\Gamma(\frac{3}{2},\frac{r^2_h}{4\theta})\Big]^{-1}
\nonumber
\\
&=&2\pi r_h\Big[1+\frac{2}{\sqrt\pi}\Gamma(\frac{3}{2},\frac{r^2_h}{4\theta})+\frac{4}{\pi}\Gamma^2(\frac{3}{2},\frac{r^2_h}{4\theta})+\frac{8}{\pi^{\frac{3}{2}}}\Gamma^3(\frac{3}{2},\frac{r^2_h}{4\theta})+........\Big].
\label{2.2n}
\end{eqnarray}
The above expansion is valid only when $|\frac{2}{\sqrt\pi}\Gamma(\frac{3}{2},\frac{r_h^2}{4\theta})|<1$. This is proved by a graphical analysis. The plot (\ref{fig9}) shows that $|\frac{2}{\sqrt\pi}\Gamma(\frac{3}{2},\frac{r_h^2}{4\theta})|$ is always less than $1$ for the entire black hole region.
\begin{figure}[h] 
\centering
\includegraphics[angle=0,width=7cm,keepaspectratio]{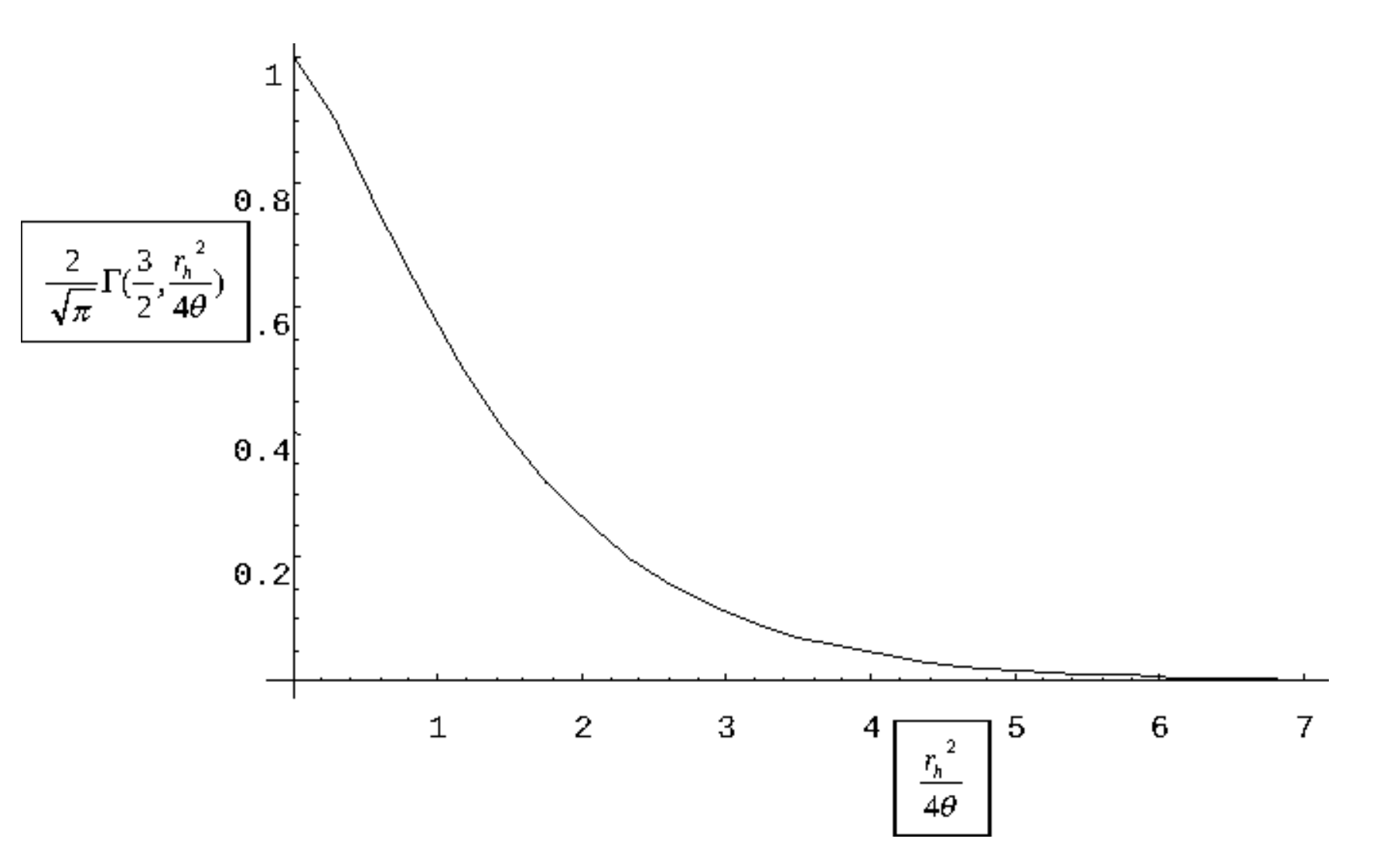}
\caption[]{{\it{$\frac{2}{\sqrt\pi}\Gamma(\frac{3}{2},\frac{r_h^2}{4\theta})$ Vs. $\frac{r_h^2}{4\theta}$ plot.}}}
\label{fig9}
\end{figure}

      The first term in (\ref{2.2n}) corresponds to the usual area law. The other terms are therefore interpreted as corrections to the area law. To justify this we will take the help of graphical analysis. Taking only the first order correction, $\frac{dS_{{\textrm{bh}}}}{dr_h}$ is written as 
\begin{eqnarray}
{\frac{dS_{{\textrm{bh}}}}{dr_h}}^{(1)}= 2\pi r_h\Big[1+\frac{2}{\sqrt\pi}\Gamma(\frac{3}{2},\frac{r^2_h}{4\theta})\Big].
\label{2.3n}
\end{eqnarray}
\begin{figure}[h] 
\centering
\includegraphics[angle=0,width=7cm,keepaspectratio]{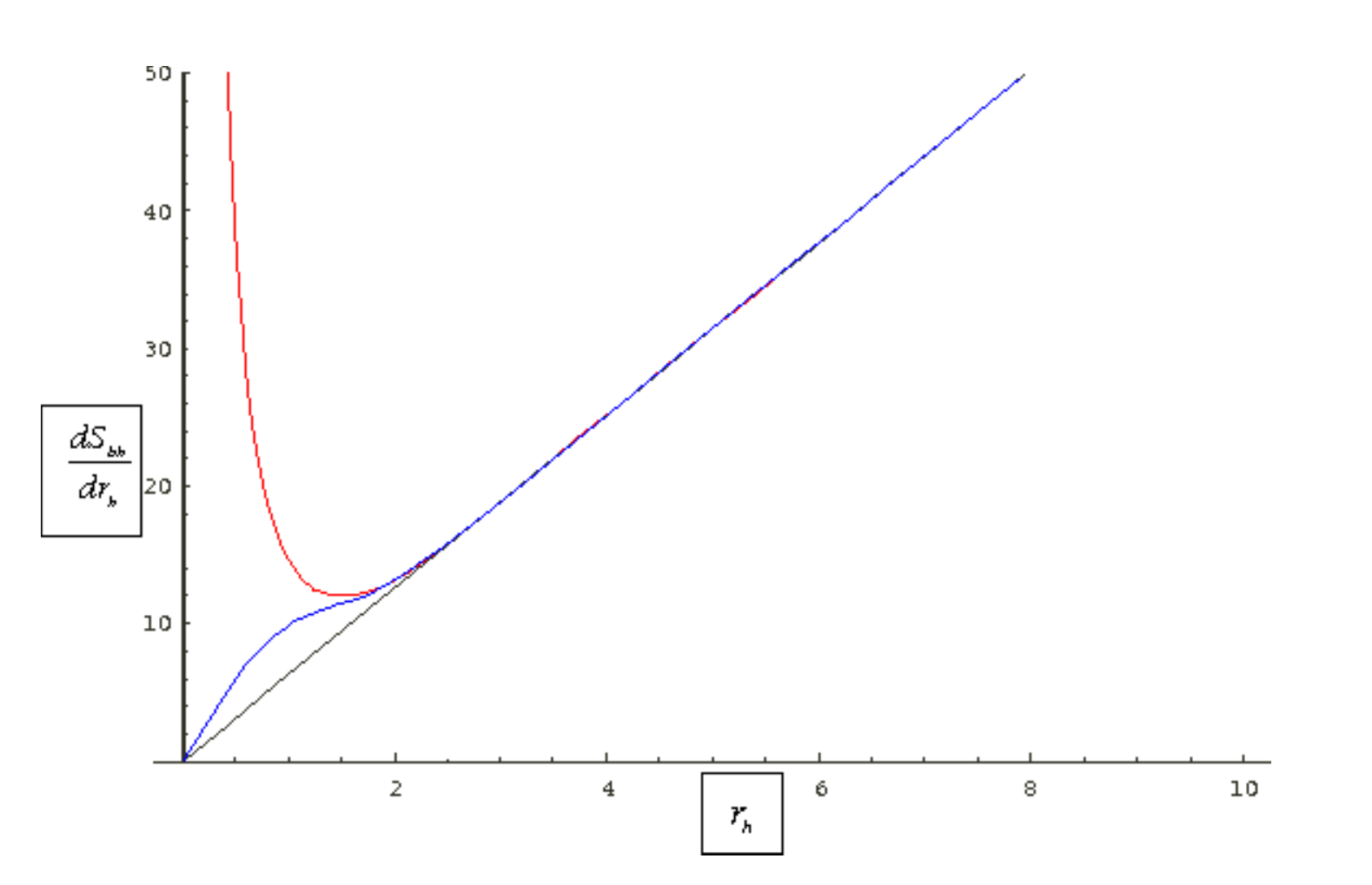}
\caption[]{{\it $\frac{dS_{\textrm{bh}}}{dr_h}$ Vs. $r_h$ plot. {$\frac{dS_{\textrm{bh}}}{dr_h}$ is plotted in units of $4\theta$ and $r_h$ is plotted in units of $2\sqrt\theta$. Red/upper curve: for eq. (\ref{1.3}), black/lower curve: for eq. (\ref{1.4}) and blue/middle curve: for eq. (\ref{2.3n})}}.}
\label{fig5}
\end{figure}
The variation of $\frac{dS_{{\textrm{bh}}}}{dr_h}$ versus $r_h$ for equations (\ref{1.3}), (\ref{1.4}) and (\ref{2.3n}) is shown in figure (\ref{fig5}). It is observed that the blue curve (corresponding to (\ref{2.3n})) has the correct linear behaviour for $r_h\gtrsim4.8\sqrt\theta$. Below this it agrees with the red curve almost till the extremal (physical) limit $r_h=3.0\sqrt\theta$, near which it shows a slight deviation. To improve this situation, the next order correction in (\ref{2.2n}) is included,
\begin{eqnarray}
{\frac{dS_{{\textrm{bh}}}}{dr_h}}^{(2)}= 2\pi r_h\Big[1+\frac{2}{\sqrt\pi}\Gamma(\frac{3}{2},\frac{r^2_h}{4\theta})+\frac{4}{\pi}\Gamma^2(\frac{3}{2},\frac{r^2_h}{4\theta})\Big].
\label{2.4n}
\end{eqnarray}
\begin{figure}[h] 
\centering
\includegraphics[angle=0,width=7cm,keepaspectratio]{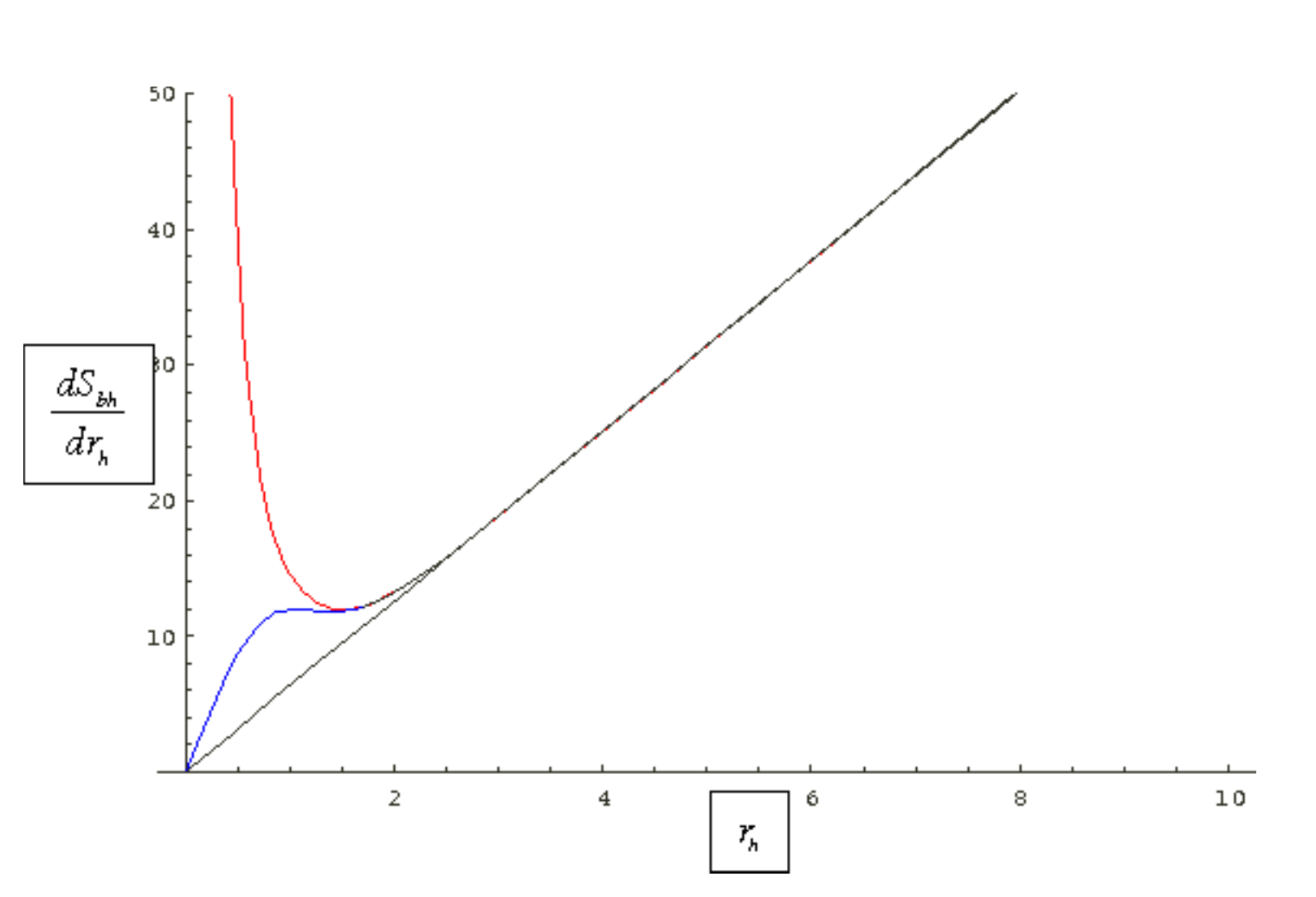}
\caption[]{{\it{$\frac{dS_{\textrm{bh}}}{dr_h}$ Vs. $r_h$ plot. $\frac{dS_{\textrm{bh}}}{dr_h}$ is plotted in units of $4\theta$ and $r_h$ is plotted in units of $2\sqrt\theta$. Red/upper curve: for eq. (\ref{1.3}), black/lower curve: for eq. (\ref{1.4}) and blue/middle curve: for eq. (\ref{2.4n})}}.}
\label{fig6}
\end{figure}
This is now plotted in figure (\ref{fig6}) along with equations (\ref{1.3}) and (\ref{1.4}). It shows that the blue curve coincides with the red curve for the entire physical domain $r_h\geq3.0\sqrt\theta$. Incidentally, if the third order correction had been included (see figure (\ref{fig7})), the matching would extend below the extremal limit $r_h=3.0\sqrt\theta$. In fact the curves begin to coincide from $r_h=2.6\sqrt\theta$ which actually lies in the unphysical domain and hence is inconsequential. Therefore we conclude that it is both necessary and sufficient to take upto the second order correction in the variation of the Bekenstein-Hawking entropy with the horizon $r_h$ of the black hole and (\ref{2.4n}) should eventually lead to the required correction to the area law in the region of our interest. Now integrating over $r_h$, (\ref{2.4n}) yields
\begin{eqnarray}
S_{{\textrm{bh}}}&=& \pi r^2_h-\sqrt{\frac{\pi}{\theta}}~r^3_h e^{-\frac{r^2_h}{4\theta}}-6\sqrt{\pi\theta}~r_h e^{-\frac{r^2_h}{4\theta}}-6\pi\theta\Big(1-{\textrm {Erf}}(\frac{r_h}{2\sqrt\theta})\Big)
\nonumber
\\
&+&2\sqrt\pi ~r^2_h \Gamma(\frac{3}{2},\frac{r^2_h}{4\theta})+8\int r_h\Gamma^2(\frac{3}{2},\frac{r^2_h}{4\theta})dr_h. 
\label{2.5n}
\end{eqnarray}
This is the desired expression for the entropy in the entire physical region of the black hole that is valid to all orders in $\theta$. Taking the large radius limit ($\frac{r_h^2}{4\theta}>>1$) and keeping terms up to the leading order ($\frac{1}{\sqrt\theta}e^{-\frac{1}{\theta}}$) immediately reproduces (\ref{1.13}).

        Expressing (\ref{2.5n}) in terms of the semi-classical noncommutative area $A_{\theta}=4\pi r_h^2$ (\ref{1.13}) the cherished area law is obtained, 
\begin{eqnarray}
S_{{\textrm{bh}}}&=& \frac{A_{\theta}}{4}-\frac{A_{\theta}^{\frac{3}{2}}}{8\pi\sqrt\theta} e^{-\frac{A_{\theta}}{16\pi\theta}}-3\sqrt{\theta A_{\theta}} e^{-\frac{A_{\theta}}{16\pi\theta}}-6\pi\theta\Big(1-{\textrm {Erf}}(\frac{1}{4}\sqrt\frac{A_{\theta}}{\pi\theta})\Big)
\nonumber
\\
&+&\frac{A_{\theta}}{2\sqrt\pi} \Gamma(\frac{3}{2},\frac{A_{\theta}}{16\pi\theta})+\frac{1}{\pi}\int \Gamma^2(\frac{3}{2},\frac{A_{\theta}}{16\pi\theta})dA_{\theta}. 
\label{2.51n}
\end{eqnarray}
\begin{figure}[h] 
\centering
\includegraphics[angle=0,width=7cm,keepaspectratio]{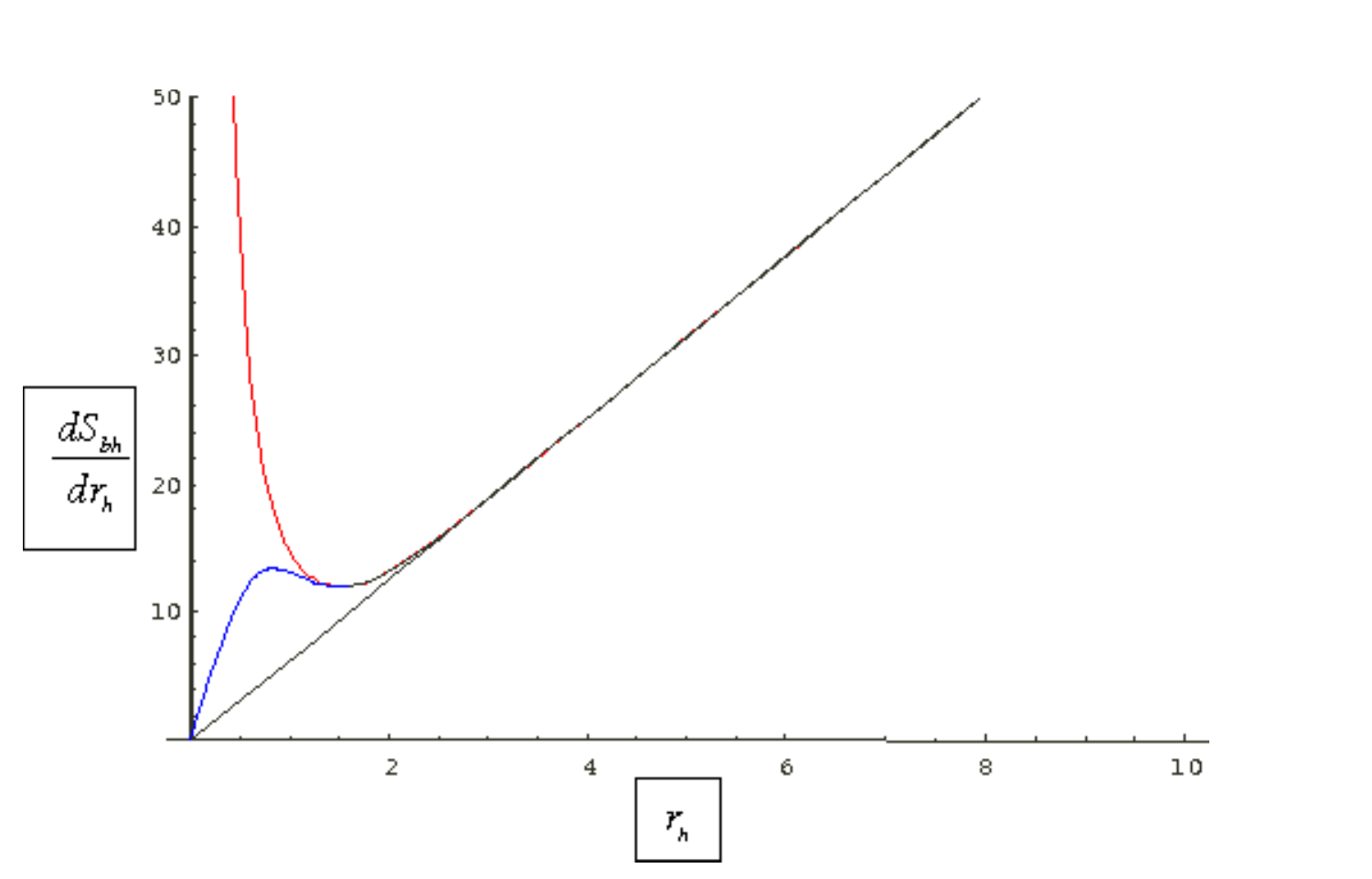}
\caption[]{{\it{$\frac{dS_{\textrm{bh}}}{dr_h}$ Vs. $r_h$ plot. $\frac{dS_{\textrm{bh}}}{dr_h}$ is plotted in units of $4\theta$ and $r_h$ is plotted in units of $2\sqrt\theta$. Red/upper curve: for eq. (\ref{1.3}), black/lower curve: for eq. (\ref{1.4}) and blue/middle curve: for third order correction.}}}
\label{fig7}
\end{figure}
The first term yields the noncommutative version of the famous Bekenstein-Hawking semi-classical area law. The other terms are the corrections to the area law. These corrections, contrary to the higher $\hbar$ corrections, do not involve any logarithmic terms. Rather, they involve exponentials of the noncommutative semi-classical area $A$ as well as the error function. Taking the large area limit $(\frac{A_{\theta}}{16\pi\theta}>>1)$ and retaining terms upto the leading order $(\frac{1}{\sqrt\theta}e^{-\frac{1}{\theta}})$, the general structure in (\ref{2.51n}) reduces to (\ref{1.13}). Finally, in the commutative limit $\theta\rightarrow 0$, all terms except the $\frac{A_{\theta}}{4}$ term separately vanish and the usual semi-classical Bekenstein-Hawking area law is reproduced.

    There is a further point that deserves some attention. From (\ref{1.01}) it is observed that $r_\theta=2\sqrt\theta$ might be interpreted as the radius of some sphere where noncommutative effects cannot be ignored. In that case the particular combination ($\frac{A_{\theta}}{16\pi\theta}$) appearing in (\ref{2.51n}) could be regarded as the ratio between the areas of the black hole and the noncommutative sphere.

We now show that at the extremal limit $r_h=3.0\sqrt\theta$ (which corresponds to the zero temperature degenerate horizon state) the entropy vanishes. To show this the appropriate limit of (\ref{2.5n}) from $r_h=3.0\sqrt\theta$ upto some arbitrary $r_h$ is taken,

\begin{eqnarray}
S_{{\textrm{bh}}}\Big|_{r_h=3.0\sqrt\theta}^{r_h}&=& \Big[\pi r^2_h-\sqrt{\frac{\pi}{\theta}}~r^3_h e^{-\frac{r^2_h}{4\theta}}-6\sqrt{\pi\theta}~r_h e^{-\frac{r^2_h}{4\theta}}-6\pi\theta\Big(1-{\textrm {Erf}}(\frac{r_h}{2\sqrt\theta})\Big)
\nonumber
\\
&+&2\sqrt\pi ~r^2_h \Gamma(\frac{3}{2},\frac{r^2_h}{4\theta})+8\int r_h\Gamma^2(\frac{3}{2},\frac{r^2_h}{4\theta})dr_h\Big]_{r_h=3.0\sqrt\theta}^{r_h}. 
\label{2.52}
\end{eqnarray}
\begin{figure}[h] 
\centering
\includegraphics[angle=0,width=7cm,keepaspectratio]{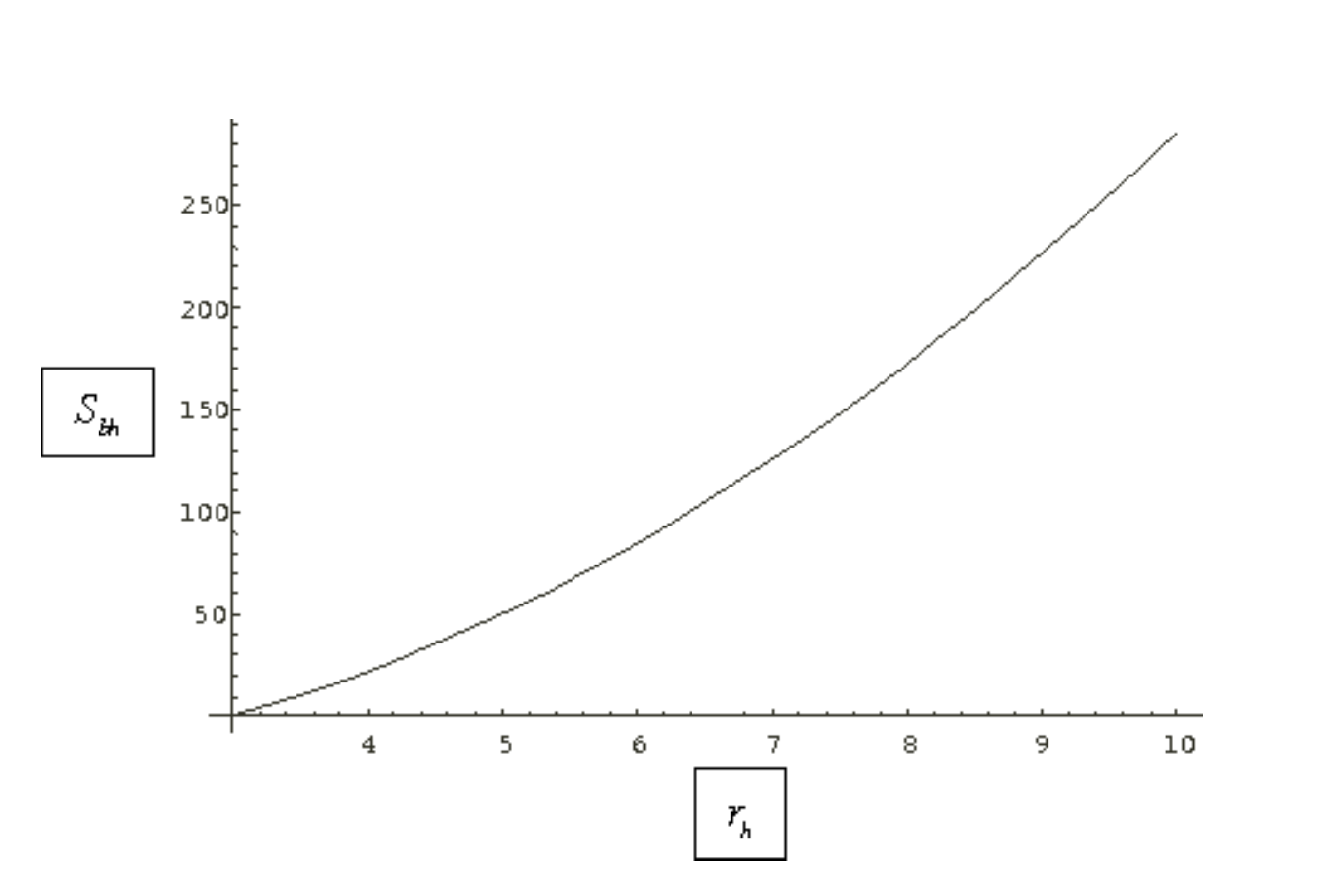}
\caption[]{{\it{$S_{\textrm{bh}}$ Vs. $r_h$ plot. $S_{\textrm{bh}}$ is plotted in units of $\theta$ and $r_h$ is plotted in units of $\sqrt\theta$.}}}
\label{fig8}
\end{figure}
A numerical plot (figure \ref{fig8}) clearly reveals that $S_{\textrm{bh}}=0$ at the extremal point $r_h=3.0\sqrt\theta$. 

With the above analysis we are now sure that noncommutative effects indeed modify the semi-classical results of black holes defined in the commutative spacetime. However these effects are purely semi-classical and do not include any higher order corrections to the WKB ansatz. In the next section we incorporate both NC and higher order (in $\hbar$) effects in the ansatz and find out a general expression for entropy/temperature for this black hole.

\section{\label{ncs} Noncommutative Schwarzschild black hole and corrected area law} 

In the previous section, the modification to the semi-classical expressions (\ref{1.13} and \ref{2.51}) was explicitly dictated by the NC parameter ($\theta$). In this section we contain ourselves to the leading order in $\theta$ (\ref{1.13}) and calculate temperature/entropy by going beyond the known semi-classical limit.

As NC Schwarzschild black hole is also a static and spherically symmetric solution it is trivial to use tunneling method to obtain the corrected Hawking temperature identical to (\ref{2.26}) [with $Q=0=a$]. In fact mimicking the steps carried out for the commutative case, it is trivial to obtain the corrected temperature [again with $Q=0=a$] (\ref{5.19}) and corrected entropy identical with (\ref{5.28}, \ref{5.29}). However a distinction due to the NC parameter ($\theta$) should be manifested in the definition of the entropy or horizon area of these terms. In the following we now calculate the coefficient of the leading order correction ($\tilde\beta_1$), following (\ref{6.8} with $Q=0=a$), for this black hole. Interestingly it is also found to have the NC modifications.

We start by calculating the invariant scalars for the metric (\ref{1.04}). This is found to be 
\begin{eqnarray}
R_{abcd}R^{abcd} &=& \frac{48M^2}{r^6}+\frac{M^2e^{-(r^2/2\theta)}}
{4\pi r^6\theta^5}\times\nonumber\\
&& \left[r^{10}+16\alpha_1+32\theta^3e^{(r^2/4\theta)}\Gamma(\frac{3}{2},\frac{r^2}{4\theta})\alpha_2+768e^{(r^2/4\theta)}\Gamma(\frac{3}{2},\frac{r^2}{4\theta})\right]
\label{Rabcd}\\
&&{\textrm{where,}}~~~ \alpha_1=(r^6\theta^2-\sqrt\frac{\pi}{\theta}r^5\theta^3e^{(r^2/4\theta)}-4\sqrt\frac{\pi}{\theta}r^3\theta^4e^{(r^2/4\theta)})\nonumber\\
&& \alpha_2=\left(\frac{r^5}{\sqrt\theta}-24\sqrt\pi\theta^2e^{(r^2/4\theta)}+4\theta^2(\frac{r^2}{\theta})^{3/2}\right)
\nonumber
\end{eqnarray}
\begin{eqnarray}
R_{ab}R^{ab}=\frac{M^2e^{-(r^2/2\theta)}(r^4-8r^2\theta+32\theta^2)}{8\pi\theta^5}\\
R=-\frac{Me^{-(r^2/4\theta)}(r^2-8\alpha)}{2\sqrt\pi\theta^{5/2}}
\label{Rab} 
\end{eqnarray}
Note that in the commutative limit ($\theta\rightarrow 0$) the above results match with the known results of the standard vacuum Schwarzschild spacetime metric, for which $R_{abcd}R^{abcd}=\frac{48M^2}{r^6},~~R_{ab}R^{ab}=0,~~R=0$. 
To find the trace anomaly (\ref{6.9}), we now evaluate
\begin{eqnarray}
\nabla_a\nabla^aR &= -\frac{Me^{-r^2/2\theta}}{8\pi\theta^5(r^2/\theta)^{1/2}}\left[2M(r^2-12\theta)(r^2/\theta)^{3/2}\theta+\sqrt\pi e^{r^2/4\theta}\alpha_3\right.\nonumber\\&\left.+4Me^{(r^2/4\theta)}\Gamma(3/2,r^2/4\theta)\alpha_4\right]~~~~~\label{nabla}\\
& {\textrm{where,}}~~~\alpha_3 = \left(r^5-22r^3\theta+72r\theta^2-2M(r^4-20r^2\theta+48\theta^2)\right)\nonumber\\
& \alpha_4 = (r^4-20r^2\theta+48\theta^2).\nonumber 
\end{eqnarray}
Exploiting all these results, the trace anomaly is calculated from (\ref{6.9}), upto the leading order in 
${\cal O}(e^{-\frac{r^2}{4\theta}})$, as 
\begin{eqnarray}
\langle{T^{\mu}}_{\mu}\rangle^{(1)}&=&\frac{1}{2880\pi^{2}}
\left[\left(\frac{48M^{2}}{r^{6}}-\frac{4M^2}{\sqrt{\pi}\theta^{5/2}}
\frac{e^{-r^{2}/(4\theta)}}{r}\right)
-\frac{Me^{-r^{2}/(4\theta)}}{8\sqrt{\pi}\theta^{9/2}}[r^4-2Mr^3-22r^{2}
\theta+40Mr\theta]\right]\nonumber\\
&&+\mathcal{O}(e^{-r^{2}/(2\theta)})~.\nonumber
\end{eqnarray}
Substituting this in (\ref{6.8}) and performing the integral yields
\begin{eqnarray}
\tilde{\beta}_{1}=\frac{M}{180\pi\omega}
\left[1+\frac{2M}{\sqrt{\pi\theta}}(1-\frac{2M^2}{\theta})
e^{-M^{2}/(\theta)}\right]+\mathcal{O}(\sqrt{\theta}e^{-M^{2}/(\theta)})~.
\label{coeff_1a}
\end{eqnarray}
To compute $\omega$, we calculate the Komar energy 
integral (\ref{eeff}). For the spacetime metric (\ref{1.04}) we get 
\begin{eqnarray}
\omega=M\left[1-\frac{r}{\sqrt{\pi\theta}}(1+\frac{r^2}{2\theta})
e^{-r^{2}/(4\theta)}\right]+\mathcal{O}(\sqrt{\theta}e^{-r^{2}/(4\theta)})~.
\label{komar1a}
\end{eqnarray}
Note that at spatial infinity ($r\rightarrow\infty$), the Komar energy ($\omega$) is nothing but the commutative mass ($M$) of the spacetime as expected. However for finite distance the effective energy involves NC corrections. Also, for the commutative limit ($\theta\rightarrow 0$), $\omega= M$ holds at any radial distance outside the event horizon which is a well known fact for the standard Schwarzschild black hole. Near the event horizon (\ref{1.06}), the above expression for $\omega$ simplifies to
\begin{eqnarray}
\omega=M\left[1-\frac{2M}{\sqrt{\pi\theta}}(1+\frac{2M^2}{\theta})
e^{-M^{2}/\theta}\right]+\mathcal{O}(\sqrt{\theta}e^{-M^{2}/(\theta)})~.
\label{komar1b}
\end{eqnarray}
Substituting this in the expression for $\tilde{\beta}_{1}$ in (\ref{coeff_1a}), we obtain :
\begin{eqnarray}
\tilde{\beta}_{1}=\frac{1}{180\pi}
\left[1+\frac{4M}{\sqrt{\pi\theta}}e^{-M^{2}/\theta}\right]+\mathcal{O}(\sqrt{\theta}e^{-M^{2}/\theta})~.
\label{coeff_1b}
\end{eqnarray}
Exploiting (\ref{5.28}, \ref{5.29}) and (\ref{coeff_1b}), we find the corrected entropy/area law (upto leading order correction) for the NC Schwarzschild black hole
\begin{eqnarray}
S_{\textrm{bh}}^{\textrm{NSch}}  &=& \frac{A_{\theta}}{4\hbar}+\frac{1}{90}(1+\frac{4M}{\sqrt{\pi\theta}}e^{-M^2/\theta})\ln\frac{A_{\theta}}{\hbar} + \mathcal{O}(\sqrt{\theta}e^{-\frac{M^2}{\theta}})
\nonumber\\        
        &=& S_{BH}+\frac{1}{90}(1+\frac{4M}{\sqrt{\pi\theta}}e^{-M^2/\theta})\ln S_{BH}+\mathcal{O}(\sqrt{\theta}e^{-\frac{M^2}{\theta}})
\label{corr_entr2}
\end{eqnarray}
This expression of entropy for NC Schwarzschild black hole involves both the NC and quantum effects. The first term in this expression is the semi-classical entropy (including $\theta$ modified horizon area) and the next term is the leading correction in $\hbar$. The coefficient of the logarithmic correction is different from the standard Schwarzschild black hole due to the presence of noncommutative parameter ($\theta$). In the commutative limit $\theta\rightarrow 0$, the expression for the corrected entropy exactly matches with the standard Schwarzschild case where the coefficient of the leading correction is $\frac{1}{90}$ (see chapter 4, eq. \ref{6.13}).

The corrected temperature, in presence of the leading correction for both $\theta$ and $\hbar$, is now given by
\begin{eqnarray}
T_{\textrm{bh}}^{\textrm{NSch}}=T_{\textrm{H}}\left(1+\frac{\hbar}{180\pi Mr_+}\left[1+\frac{4M}{\sqrt{\pi\theta}}e^{-M^{2}/\theta}\right]+\textrm{higher order terms}\right).
\label{cornsc}
\end{eqnarray}

\section{\label{disccorrection}Discussions} 
We have made a detailed investigation of the Hawking temperature, entropy and the area law for a Schwarzschild black hole whose metric is modified by effects of noncommutative effect. The noncommutative version of the semi-classical Bekenstein-Hawking area law holds in the region $r_h\geq4.8\sqrt\theta$. The linear relation between entropy and area was violated below this horizon radius till the extremal point $r_h=3.0\sqrt\theta$. From a graphical analysis we found the correctional terms to the area law to cover the complete physical domain ($r_h\geq3.0\sqrt\theta$) of the black hole. The correction terms involved exponentials as well as error functions. We have also considered the higher order (in $\hbar$) effects in the WKB ansatz and calculated the leading order correction to the entropy and temperature where both NC and quantum effects were present.



\chapter{\label{chap-phase-transition}Ehrenfest's scheme, thermodynamic geometry and black hole phase transitions}
The indispensable connection between black holes and thermodynamics often allows us to exchange ideas in these two fields. In this chapter we further build on this fact and use fundamental tools of thermodynamics to black hole systems. Particularly we study phase transition phenomena in black holes by using its analogy with standard fluid systems. Our approach is inspired by Clausius-Clapeyron and Ehrenfest's ideas for the identification and classification of phase transitions \cite{stan, zeman}. Although several works \cite{hp1}-\cite{hp14} on black hole phase transitions are present in literature, the above concepts, in fact, had never been used for black holes. For the first time we introduce such an approach to study black hole phase transitions, which, interestingly is found to be quite powerful.

This chapter is organized in the following manner. In section-\ref{subehr} we derive Ehrenfest relations for rotating black holes. In the next section (\ref{kads}) phase transition in Kerr anti-de Sitter black holes has been discussed. Here we show that there exists a second order phase transition from the lower to higher mass black holes. In section \ref{thergeo} we follow a thermodynamic geometry approach to study phase transition in the Kerr-AdS black hole and compare the results found here with those found by using the Ehrenfest's scheme. The last section is for summary and discussions.

\section{\label{subehr}Derivation of Ehrenfest's relations for black holes} 
In this section we use Ehrenfest's ideas and implement them for black hole systems undergoing phase transitions. It leads us to a set of Ehrenfest like relations for black holes (although we consider the rotating black holes here, it is easy to translate it for charged black holes as well). The key point of this derivation lies on the known analogy between the first law of black hole thermodynamics, given by (\ref{sthermo}) and other fluid systems, given by (\ref{bappa}).  

Using the above mentioned analogy the Gibbs energy for black holes can be constructed in the following form,  
\begin{eqnarray}
G=M-\Omega_{\textrm{H}} J-TS.
\label{equation 19a}
\end{eqnarray}
The differential form of this is written as,
\begin{eqnarray}
dG=-Jd\Omega_H-SdT\label{dg}
\end{eqnarray}
From this equation, black hole entropy and the angular momentum can be written as the derivatives of the Gibbs free energy as,
\begin{eqnarray}
&&S=-\left(\frac{\partial G}{\partial T}\right)_{\Omega_H}\\
&&J= -\left(\frac{\partial G}{\partial \Omega_H}\right)_{T}
\end{eqnarray}
Since, by definition, the Gibbs free energy (\ref{equation 19a}) is a state function, $dG$ is an exact differential and hence we get the following Maxwell relation from (\ref{dg})
\begin{eqnarray}
\left(\frac{\partial J}{\partial T}\right)_{\Omega_H}=\left(\frac{\partial S}{\partial \Omega_H}\right)_{T}\label{maxwellrelation}
\end{eqnarray}

Note that a system, undergoing a first order phase transition, has a continuous Gibbs energy ($G$) but the first order derivatives of $G$ are discontinuous. The infinitesimal variation of $G$ is equal at both phases which yields the Clausius-Clapeyron (CC) equation for such systems. The famous example is the liquid to vapor phase transition which satisfies CC equation. Similarly, in a second order phase transition Gibbs free energy and its first order derivatives {\it i.e.} entropy and angular momentum (analog of volume), are all continuous. However, all the second order derivatives of $G$, namely, heat capacity at constant pressure ($C_P$), volume expansion coefficient ($\alpha$) and compressibility ($k_T$) are discontinuous at the critical point. So at a phase transition point, characterized by some temperature $T_c$ and an angular velocity $\Omega_H$, one has
\begin{eqnarray}
&&G_1=G_2\label{freeenergyequality}\\
&&S_1=S_2 \label{entropyequality}\\
&&J_1=J_2\label{angularmomentumequality}
\end{eqnarray}
where the subscripts 1 and 2 denote the values of the different physical quantities in the two phases.

Let us now consider the equality (\ref{entropyequality}). If the temperature and angular velocity are increased infinitesimally to $T+dT$ and $\Omega_H+d\Omega_H$ then
\begin{eqnarray}
S_1+dS_1=S_2+dS_2\label{2ndentropyequality}
\end{eqnarray}
From (\ref{entropyequality}) and (\ref{2ndentropyequality}) we find
\begin{eqnarray}
dS_1=dS_2\label{entropychange}
\end{eqnarray}
Taking $S$ as a function of $T$ and $\Omega_H$
\begin{eqnarray}
S=S(T,\Omega_H)
\end{eqnarray}
we write the infinitesimal change in entropy as
\begin{eqnarray}
dS=\left(\frac{\partial S}{\partial T}\right)_{\Omega_H}dT+\left(\frac{\partial S}{\partial \Omega_H}\right)_{T}d\Omega_H
\end{eqnarray}
Using (\ref{maxwellrelation}), the above equation takes the form,
\begin{eqnarray}
dS&=\frac{C_{\Omega_H}}{T}dT+J\alpha d\Omega_H
\end{eqnarray}
where 
\begin{equation}
\alpha=\frac{1}{J}\left(\frac{\partial J}{\partial T}\right)_{\Omega_H} 
\label{volexp}
\end{equation}
is the coefficient of change in angular momentum and 
\begin{equation}
C_{\Omega_H}=T\left(\frac{\partial S}{\partial T}\right)_{\Omega_H}. 
\label{corsp}
\end{equation}
Since $J$ is same in both phases,
\begin{eqnarray}
&&dS_1=\frac{C_{\Omega_{H_1}}}{T}dT+J\alpha_1 d\Omega_H \label{ds1}\\
&&dS_2=\frac{C_{\Omega_{H_2}}}{T}dT+J\alpha_2 d\Omega_H \label{ds2}
\end{eqnarray} 
Now use of the condition (\ref{entropychange}) requires the equality of (\ref{ds1}) and (\ref{ds2}),
\begin{eqnarray}
\frac{C_{\Omega_{H_1}}}{T}dT+J\alpha_1 d\Omega_H=\frac{C_{\Omega_{H_2}}}{T}dT+J\alpha_2 d\Omega_H.
\end{eqnarray} 
This gives the first Ehrenfest's equation  
\begin{eqnarray}
-\left(\frac{\partial\Omega_H}{\partial T}\right)_{S}=\frac{C_{\Omega_{H_2}}-C_{\Omega{_{H_1}}}}{TJ(\alpha_2-\alpha_1)}
\label{eren1}
\end{eqnarray} 
We now take the constancy of angular momentum (\ref{angularmomentumequality}) which, under infinitesimal change of temperature and angular velocity, gives
\begin{eqnarray}
dJ_1=dJ_2.\label{angularmomentumchange}
\end{eqnarray}
Taking angular momentum as a function of $T$ and $\Omega_H$,
\begin{eqnarray}
J=J(T,\Omega_H)
\end{eqnarray}
we get the differential relation
\begin{eqnarray}
dJ&=&\left(\frac{\partial J}{\partial T}\right)_{\Omega_H}dT+\left(\frac{\partial J}{\partial \Omega_H}\right)_Td\Omega_H\\
&=&J\alpha dT-Jk_Td\Omega_H
\end{eqnarray}
where 
\begin{equation}
k_T=\frac{1}{J}\left(\frac{\partial J}{\partial {\Omega_H}}\right)_T
\label{compr}
\end{equation}
is the analog of compressibility. Since $dJ$ is same for the two phases, we get the second Ehrenfest's equation by mimicking the previous steps used to derive (\ref{eren1})
\begin{eqnarray}
-\left(\frac{\partial\Omega_{H}}{\partial T}\right)_{J}=\frac{\alpha_{2}-\alpha_{1}}{k_{T_2}-k_{T_1}}.
\label{eren2}
\end{eqnarray} 
This completes our task of deriving Ehrenfest equations (\ref{eren1}) and (\ref{eren2}) for rotating black holes . We are thus in a position to analyze and classify black hole phase transitions using the tool of thermodynamics. In the subsequent part of this chapter we perform such studies.

For future convenience we provide a table showing the analogy between the usual thermodynamic variables and black hole parameters:
\begin{table*}[th]
\centering
\begin{tabular}{|c|c|}
\hline
Standard Fluids & Black Holes
 \\
\hline\hline
   Energy ($E$)  & Mass ($M$) \\ \hline
    Volume ($V$) & Angular momentum ($J$)/ Charge ($Q$) \\ \hline
    Pressure ($P$) & Angular velocity ($-\Omega$)/ Electric potential ($-\Phi$) \\ \hline
    Entropy ($S$) & Entropy ($S$) \\ \hline
    Temperature ($T$) & Temperature ($T$) \\ \hline
    First law: $dE=TdS-PdV$ & First law: $dM=TdS+\Omega dJ / \Phi dQ $ \\ \hline
   
\end{tabular} \caption{Analogy between standard thermodynamical systems and rotating/charged black holes}
\end{table*}

\section{\label{kads}Phase transition in Kerr Anti-de Sitter (AdS) black holes}

\subsection{The Kerr-AdS black hole}

Kerr AdS black hole is a solution of Einstein equation in (3+1) dimensions with a negative cosmological constant $\Lambda=-\frac{3}{l^2}$. It is characterised by two parameters, namely mass $M$ and angular momentum $J$. The spacetime metric of the Kerr AdS black hole in Boyer-Lindquist coordinates ($t,~r,~\theta,~\phi$) is given by,
\begin{equation}
ds^{2} = -\frac{\Delta_r}{\rho^2}\left(dt-\frac{a\sin^2\theta}{\Xi}d\phi\right)^2 + \frac{\rho^2}{\Delta_r}dr^2 + \frac{\rho^2}{\Delta_{\theta}}d\theta^2 + \frac{\Delta_{\theta}\sin^2\theta}{\rho^2}\left(adt-\frac{r^2+a^2}{\Xi}d\phi\right)^2
\label{1}  
\end{equation}
where, 
\begin{align}
\rho^2& =r^2+a^2\cos^2\theta \\
\Xi& =1-\frac{a^2}{l^2}  \notag \\
\Delta_r& =(r^2+a^2)(1+\frac{r^2}{l^2})-2mr  \notag \\
\Delta_{\theta}& =1-\frac{a^2}{l^2}\cos^2\theta  \notag \\
\end{align}
The position of the event horizon, denoted by $r_+$, is given by the largest solution of the polynomial  
\begin{eqnarray}
\Delta_r|_{r=r_+}=(r_+^2+a^2)(1+\frac{r_+^2}{l^2})-2mr_+=0. 
\label{3}
\end{eqnarray}
From the above relation one finds the expression for the black hole mass in terms of its horizon radius as 
\begin{eqnarray}
M=\frac{m}{\Xi}=\frac{1}{2(1-\frac{a^2}{l^2})r_+}(r_+^2+a^2)(1+\frac{r_+^2}{l^2})
\label{4}
\end{eqnarray}
The angular velocities at the event horizon ($\Omega_H$) and at infinity ($\Omega_{\infty}$) are   
\begin{eqnarray}
\Omega_H=\frac{a(1-\frac{a^2}{l^2})}{r_+^2+a^2}\\
\Omega_{\infty}=-\frac{a}{l^2}.
\label{5}
\end{eqnarray}
Note that unlike the asymptotically flat Kerr black hole here the angular velocity at infinity does not vanish, rather it is proportional to the cosmological constant. The effective angular velocity ($\Omega$) that appears in thermodynamics is given by the difference between $\Omega_H$ and $\Omega_{\infty}$\cite{Cald}-\cite{Cardoso}
\begin{eqnarray}
\Omega=\Omega_H-\Omega_{\infty}=\frac{a(1+\frac{r_+^2}{l^2})}{r_+^2+a^2}
\label{6}
\end{eqnarray}
Semiclassical Hawking temperature and entropy for Kerr AdS black hole are given by \cite{Cald}-\cite{Cardoso}
\begin{eqnarray}
T=\frac{\kappa}{2\pi}=\frac{r_+\left(1+\frac{a^2}{l^2}+3\frac{r_+^2}{l^2}-\frac{a^2}{r_+^2}\right)}{4\pi(r_+^2+a^2)}
\label{7}
\end{eqnarray}
and
\begin{eqnarray}
S=\frac{A}{4}=\frac{\pi(r_+^2+a^2)}{(1-\frac{a^2}{l^2})},
\label{8}
\end{eqnarray}
where $\kappa$ is the surface gravity at the event horizon and $A$ is the horizon area of the black hole.

To perform a study based on the tools of thermodynamics, let us first re-express various thermodynamic quantities in terms of appropriate variables. The generalised Smarr formula for Kerr AdS black hole is given by \cite{Cald}
\begin{eqnarray}
M=M(S,J)=\left[\frac{S}{4\pi}+\frac{\pi J^2}{S} + \frac{J^2}{l^2}+\frac{S}{2\pi l^2}\left(\frac{S}{\pi}+\frac{S^2}{2\pi^2 l^2}\right)\right]^{\frac{1}{2}}.
\label{10}
\end{eqnarray}
From this relation one can find the expression for the Hawking temperature ($T=\frac{\partial M}{\partial S}$) and angular velocity ($\Omega=\frac{\partial M}{\partial J}$), as given by 
\begin{eqnarray}
T=T(M,S,J)=\frac{1}{8\pi M}\left(1-\frac{4\pi^2J^2}{S^2}+\frac{4S}{\pi l^2}+\frac{3S^2}{\pi^2l^4}\right),
\label{9}
\end{eqnarray}
and
\begin{eqnarray}
\Omega=\frac{\pi J}{MS}\left(1+\frac{S}{\pi l^2}\right).
\label{11}
\end{eqnarray}
The above equation may also be interpreted as an expression for the angular momentum in terms of mass, entropy and angular velocity. Since this equation will be useful in later analysis we write it as, 
\begin{eqnarray}
a=\frac{J}{M}=\frac{S\Omega}{\pi}\left(1+\frac{S}{\pi l^2}\right)^{-1}
\label{12}
\end{eqnarray}
where the new variable $a$ is the angular momentum per unit mass.

Instead of keeping $l$ throughout the analysis, it is convenient to absorb it by scaling the thermodynamic variables. We define $Tl,\frac{M}{l},\frac{J}{l^2}, \frac{S}{l^2}$, $\Omega l$ as the temperature ($T$), Mass ($M$), angular momentum ($J$), entropy ($S$) and angular velocity ($\Omega$) of the black hole. In terms of these newly defined variables (\ref{9}) is written as,
\begin{eqnarray}
T=\frac{1}{8\pi M}\left(1-\frac{4\pi^2J^2}{S^2}+\frac{4S}{\pi}+\frac{3S^2}{\pi^2}\right)
\label{temp}
\end{eqnarray}
Similarly, (\ref{10}),(\ref{11}) and \ref{12}) are written as,
\begin{eqnarray}
M^2&=&\frac{S}{4\pi}+\frac{\pi J^2}{S} + J^2+\frac{S}{2\pi}\left(\frac{S}{\pi}+\frac{S^2}{2\pi^2}\right)\label{M2}\\
\Omega&=&\frac{\pi J}{MS}\left(1+\frac{S}{\pi}\right)\\
\frac{J}{S}&=&\frac{M\Omega}{\pi+S}\label{JS}
\end{eqnarray}
Now using (\ref{JS}) we substitute $J$ in (\ref{M2}) to express $M^2$ in terms of $S$ and $\Omega$. This yields
\begin{eqnarray}
M^2 =\frac{S}{4\pi}\frac{1+\frac{2S}{\pi}(1+\frac{S}{2\pi})}{1-\frac{\Omega^2 S}{\pi(1+\frac{S}{\pi})}}.
\label{Msquare}
\end{eqnarray}
Likewise, replacing $M$ (\ref{Msquare}) in (\ref{temp}), the semi-classical temperature is found to be,
\begin{eqnarray}
T=\sqrt{\frac{S(\pi+S)^3}{(\pi+S-S\Omega^2)}}\left[\frac{\pi^2-2\pi S(\Omega^2-2)-3S^2(\Omega^2-1)}{4\pi^{\frac{3}{2}}S(\pi+S)^2}\right].
\label{14}
\end{eqnarray}
From the above equation we see that $T$ is real only when
\begin{eqnarray}
\pi+S-S\Omega^2>0
\label{cond}
\end{eqnarray}
which means
\begin{eqnarray}
\Omega^2<1+\frac{\pi}{S}\label{sir}
\end{eqnarray}
This imposes a restriction on $\Omega$ for a fixed value of the entropy. Note that when $\Omega^2=1+\frac{\pi}{S}$, the temperature diverges. The first law of thermodynamics for the Kerr-AdS black hole is given by \cite{gib}
\begin{equation}
dM=TdS+\Omega dJ
\label{fla}
\end{equation}
where $T$ is the Hawking temperature and $\Omega$ is the difference between the angular velocities at the event horizon ($\Omega_H$) and at infinity ($\Omega_{\infty}$). To get further insight into the thermodynamical behaviour of the black hole we calculate the semi-classical specific heat at constant angular velocity (which is the analogue of $C_{P}$) from (\ref{14}) by using the relation $C_{\Omega}=T\left(\frac{\partial S}{\partial T}\right)_{\Omega}$. This is found to be 
\begin{eqnarray}
C_{\Omega}=\frac{2S(\pi+S)(\pi+S-S\Omega^2)\left(\pi^2-2\pi S(\Omega^2-2)-3S^2(\Omega^2-1)\right)}{(\pi+S)^3(3S-\pi)-6S^2(\pi+S)^2\Omega^2+S^3(4\pi+3S)\Omega^4}
\label{15}
\end{eqnarray}

Let us first plot (\ref{14}) and (\ref{15}) with respect to $S$ for a fixed value of $\Omega$ ($\Omega=0.3$). These plots are depicted in figure 6.1. From figure 6.1(a) we find that the semiclassical Hawking temperature ($T$) is continuous when plotted with the semi-classical entropy ($S$). For a first order phase transition (like liquid to vapour) the first order derivative of Gibbs free energy, i.e. volume and entropy are discontinuous. The continuity of entropy ($S$) (and also $J$) suggests that for fixed $\Omega$ there is no first order phase transition taking place in the Kerr AdS black hole. Nevertheless there is a minimum temperature for a certain value of $S$ ($S=1.09761$). Figure 6.1(b) shows a discontinuity in $C_{\Omega}$ at this critical value of entropy ($S=S_c=1.09761$). However in this plot the discontinuity of specific heat is infinite which is completely different from the ordinary thermodynamical systems  (for example feromagnetic to paramagnetic transformation) where finite discontinuity occurs. Note that this transition between two phases of Kerr-AdS black holes have different entropies. As entropy is proportional to the square of the black hole mass the critical point $S_{c}$ is essentially separating two branches of Kerr-AdS black holes with ``smaller'' and ``larger'' masses. 
\begin{figure}[ht]
\centering
\includegraphics[angle=0,width=14cm,keepaspectratio]{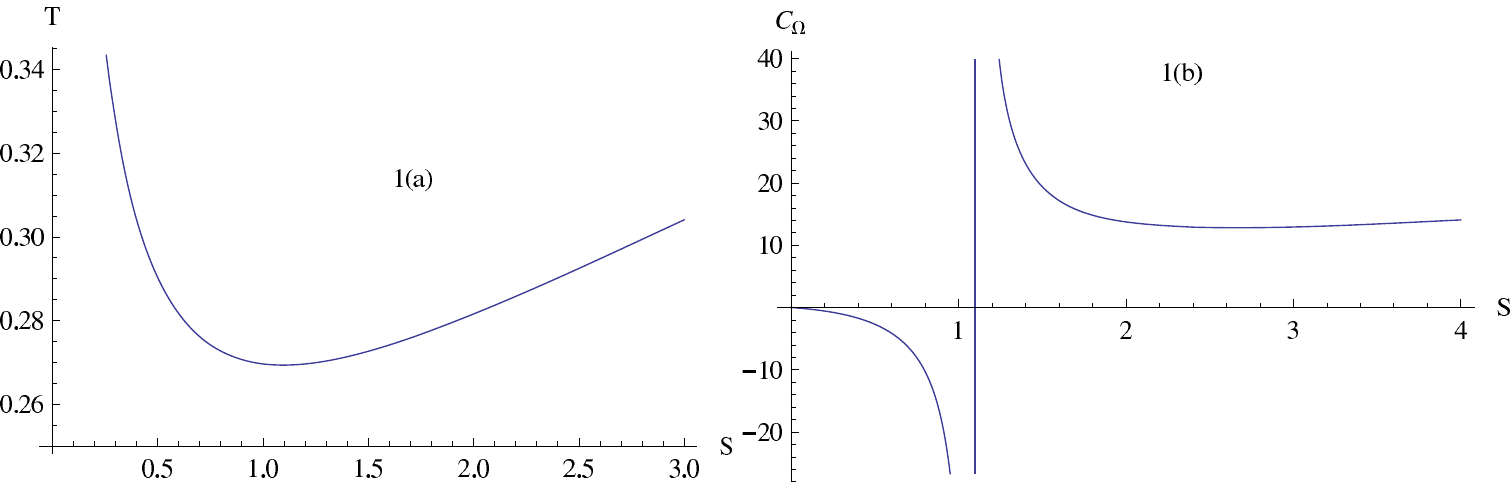}
\caption[]{{Semi-classical Hawking temperature ($T$) and Specific heat ($C_{\Omega}$) vs entropy ($S$) plot for fixed $\Omega=0.3$.}}
\label{figure 1}
\end{figure}

Now we move to the next section where we implement the Ehrenfest scheme to classify this phase transition.

\subsection{Ehrenfest relations and Kerr-AdS black hole}

We noted earlier that there is a discontinuity in $C_{\Omega}$ as shown in figure 6.1(b). Nevertheless, presence of discontinuities in $\alpha$ and $k_T$ are also necessary to show that a phase transition is taking place. As mentioned earlier, since $C_{\Omega}$, $\alpha$ and $k_T$ are all second order derivatives of the Gibbs free energy, they must be discontinuous in a second order phase transition. In order to check the validity of Ehrenfest's relations (\ref{eren1}, \ref{eren2}), we first calculate these physical entities and plot them. 

First using (\ref{JS}) and (\ref{Msquare}) we can express $J$ as a function of $S$ and $\Omega$, as given by
\begin{eqnarray}
J(S,\Omega)=\frac{S\Omega\sqrt{\frac{S(\pi+S)^3}{(\pi+S-S\Omega^2)}}}{2\pi^{\frac{3}{2}}(\pi+S)}.
\label{26}
\end{eqnarray}
Using the definition of $a$ (\ref{12}) we write (\ref{volexp}) as,
\begin{eqnarray}
J\alpha=M \left(\frac{\partial a}{\partial T}\right)_{\Omega}+a\left(\frac{\partial M}{\partial T}\right)_{\Omega}
\label{20}
\end{eqnarray}
Now to calculate the first term of the right hand side we need a functional relationship between $a, T$ and $\Omega$. To do so we rewrite the semiclassical entropy of the Kerr AdS black hole (\ref{JS}) in terms of $\Omega$ and $a$ as
\begin{eqnarray}
S=\frac{\pi a}{\Omega-a}.
\label{21}
\end{eqnarray}
In (\ref{14}) $T$ was expressed as $T=T(S,\Omega)$. Substituting (\ref{21}) in (\ref{14}) we write the temperature in terms of $a$ and $\Omega$ as,
\begin{eqnarray}
T=T(\Omega,a)=\frac{2a\Omega^2+\Omega a^2-2a-\Omega}{4\pi\sqrt{a(a-\Omega)(a\Omega-1)}}.
\label{22}
\end{eqnarray}
Now it is straightforward to calculate $\left(\frac{\partial a}{\partial T}\right)_{\Omega}$. To calculate the second term of the right hand side of (\ref{20}), we write $\left(\frac{\partial M}{\partial T}\right)_{\Omega}=\left(\frac{\partial M}{\partial S}\right)_{\Omega}\left(\frac{\partial S}{\partial T}\right)_{\Omega}$. Now using (\ref{Msquare}) and (\ref{14}) we find, $\left(\frac{\partial M}{\partial T}\right)_{\Omega}$. Making use of these results, the value of $a$ from (\ref{21}) in (\ref{20}) and substituting $J$ from (\ref{26}) we finally obtain
the analog of volume expansion coefficient,
\begin{eqnarray}
\alpha=\frac{\sqrt{4\pi^3 S(\pi+S)(\pi +S-S\Omega^2)}[6\Omega(\pi+S)^2-2S\Omega^2(2\pi+3S)]}{(\pi+S)^3(3S-\pi)-6S^2(\pi+S)^2\Omega^2+S^3(4\pi+3S)\Omega^4}
\label{25alp}
\end{eqnarray}
  
Now to find an expression for the analog of compressibility $k_T=\frac{1}{J}\left(\frac{\partial J}{\partial {\Omega}}\right)_T$, we first write $J$ in terms of $a$ and $\Omega$ in the following manner,
\begin{eqnarray}
J=\frac{1}{2}\sqrt{\frac{a(\Omega-a)}{(1-a\Omega)}}\left[\frac{a\Omega}{(\Omega-a)^2}\right].
\label{26aa}
\end{eqnarray}
Since $J$ cannot be expressed in terms of $T$ and $\Omega$ we shall use the rules of partial differentiation to find $k_T$ from (\ref{22}) and (\ref{26aa}). From the theorem $dJ=\left(\frac{\partial J}{\partial a}\right)_{\Omega}da+\left(\frac{\partial J}{\partial \Omega}\right)_{a}d\Omega$ we can write
\begin{eqnarray}
\left(\frac{\partial J}{\partial\Omega}\right)_{T}=\left(\frac{\partial J}{\partial a}\right)_{\Omega}\left(\frac{\partial a}{\partial \Omega}\right)_T+\left(\frac{\partial J}{\partial \Omega}\right)_{a}
\label{301}
\end{eqnarray}
The above equation is written in a more useful form by substituting $\left(\frac{\partial a}{\partial \Omega}\right)_T=-\frac{\left(\frac{\partial T}{\partial \Omega}\right)_a}{\left(\frac{\partial T}{\partial a}\right)_{\Omega}}$. This gives
\begin{eqnarray}
\left(\frac{\partial J}{\partial\Omega}\right)_{T}=\frac{\left(\frac{\partial J}{\partial \Omega}\right)_{a}\left(\frac{\partial T}{\partial a}\right)_{\Omega}-\left(\frac{\partial T}{\partial \Omega}\right)_{a}\left(\frac{\partial J}{\partial a}\right)_{\Omega}}{\left(\frac{\partial T}{\partial a}\right)_{\Omega}}.
\end{eqnarray}
Using (\ref{22}) and (\ref{26aa}) in the above equation we obtain $k_T$ in terms of $a$ and $\Omega$. Finally eliminating $a$ in favour of $S$ and $\Omega$ by using (\ref{21}) and then substituting $J$ from (\ref{26}), we find,
\begin{eqnarray}
k_T &=& \frac{(3S-\pi)(\pi+S)^3+2S(\pi+S)^2(4\pi+3S)\Omega^2-S^2(2\pi+3S)^2\Omega^4}{(3S-\pi)\Omega(\pi+S)^3-6S^2(\pi+S)^2\Omega^3+S^3(4\pi+3S)\Omega^5}\nonumber\\
\label{JKT}
\end{eqnarray}

\begin{figure}[ht]
\centering
\includegraphics[angle=0,width=14cm,keepaspectratio]{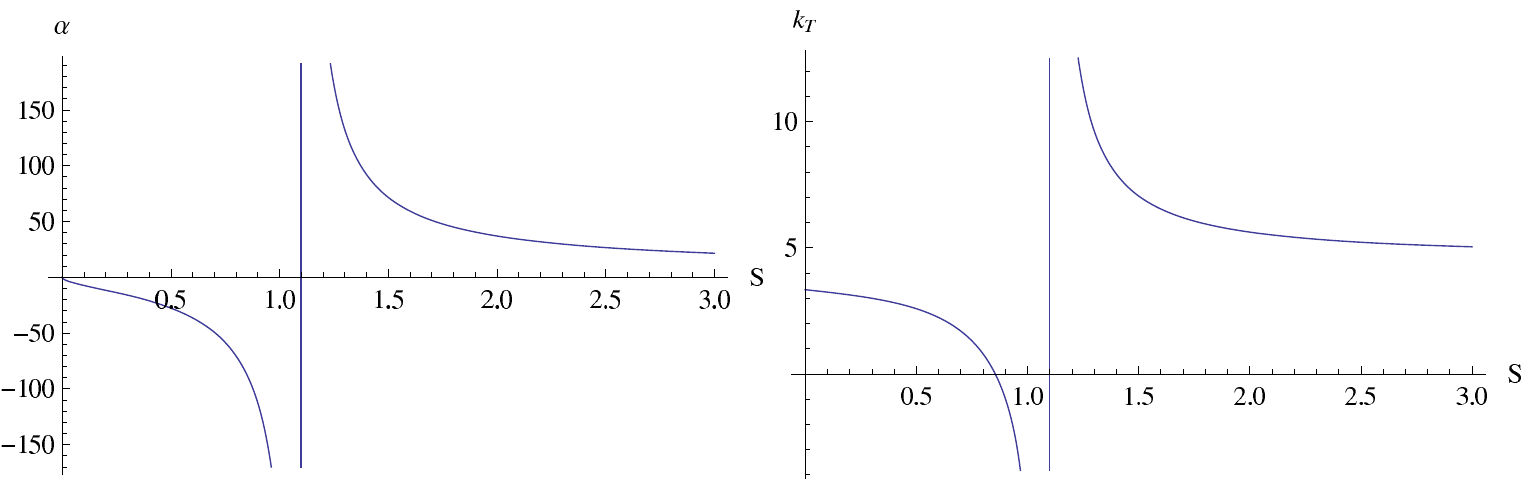}
\caption[]{{$\alpha$ and $k_T$ vs entropy ($S$) plot for fixed $\Omega=0.3$.}}
\label{figure 2}
\end{figure}

Let us now plot (\ref{25alp}) and (\ref{JKT}) with respect to $S$ for a fixed value of $\Omega=0.3$. These two plots are shown in (\ref{figure 2}) and show a discontinuity in both these quantities at the same critical value of $S_c=1.09761$ (see figure 6.1(b)). With these results we are now convinced about a genuine phase transition in Kerr-AdS black hole.


To check the validity of Ehrenfest relations, we start by considering the left hand sides of both Ehrenfest's equation (\ref{eren1}) and (\ref{eren2}). The left hand side of (\ref{eren1}) can be calculated easily from the relation (\ref{14}). This is found to be 
\begin{eqnarray}
-\left(\frac{\partial\Omega}{\partial T}\right)_S=\frac{4\pi^{\frac{3}{2}}(\pi+S-S\Omega^2)^2\sqrt\frac{S(\pi+S)^3}{(\pi+S-S\Omega^2)}}{S\Omega\left(S(\pi+S)(2\pi+3S)\Omega^2-3(\pi+S)^3\right)}.
\label{31}
\end{eqnarray}
To calculate the left hand side of (\ref{eren2}), equation (\ref{22}) is written as
\begin{eqnarray}
dT=\left(\frac{\partial T}{\partial a}\right)_{\Omega}da+\left(\frac{\partial T}{\partial \Omega}\right)_{a}d\Omega~.
\label{dt}
\end{eqnarray} 
For a process where $J$ is constant, from (\ref{26aa}), one has  
\begin{eqnarray}
\left(\frac{\partial J}{\partial a}\right)_{\Omega}da+\left(\frac{\partial J}{\partial \Omega}\right)_{a}d\Omega=0~.
\label{j0}
\end{eqnarray}
Now eliminating terms involving $da$ from the above two relations and then rearranging terms, yields,
\begin{eqnarray}
-\left(\frac{\partial T}{\partial\Omega}\right)_{J}=\frac{\left(\frac{\partial T}{\partial a}\right)_{\Omega}\left(\frac{\partial J}{\partial \Omega}\right)_{a}-\left(\frac{\partial T}{\partial \Omega}\right)_{a}\left(\frac{\partial J}{\partial a}\right)_{\Omega}}{\left(\frac{\partial J}{\partial a}\right)_{\Omega}}.
\label{29}
\end{eqnarray}
Finally using (\ref{22}) and (\ref{26aa}) and substituting $a$ in terms of $\Omega$ and $S$ from (\ref{21}) we find, 
\begin{eqnarray}
-\left(\frac{\partial\Omega}{\partial T}\right)_J&=&\frac{4\pi^{\frac{3}{2}}(\pi+S-S\Omega^2)^2\sqrt\frac{S(\pi+S)^3}{(\pi+S-S\Omega^2)}}{S\Omega\left(S(\pi+S)(2\pi+3S)\Omega^2-3(\pi+S)^3\right)}.
\label{2ndlh}
\end{eqnarray}
which gives the left hand side of the second Ehrenfest's equation (\ref{eren2}) and matches with (\ref{31}).

In order to calculate the right hand sides, $\Omega$ must be treated as a constant (this is analogous to fixing pressure while performing an experiment). This would help us to re-express $C_\Omega$ (\ref{15}), $\alpha$ (\ref{25alp}), and $k_T$ (\ref{JKT}), which have functional forms $\frac{f(S)}{g(S)}$,  $\frac{h(S)}{g(S)}$  and  $\frac{k(S)}{g(S)}$ respectively,  infinitesimally close to the critical point ($S_0$). Note that they all have the same denominator which satisfies the relation $g(S_0)= 0$. This observation is crucial in the ensuing analysis.  

The expressions of $C_{\Omega}$, $\alpha$ and $k_T$ in the two phases ($ i=1,2 $) are respectively given by ${C_{\Phi}}\big|_{S_i}=C_{\Phi_i}$, ${\alpha}\big|_{S_i}=\alpha_i$ and $k_T\big|_{S_i}=k_{T_i}$. To obtain the R.H.S. of (\ref{eren1}) we first simplify it's numerator:
\begin{eqnarray}
C_{\Phi_{2}}-C_{\Phi_{1}}=\frac{f(S_2)}{g(S_2)}-\frac{f(S_1)}{g(S_1)}
\end{eqnarray} 

Taking the points close to the critical point we may set  $f(S_2)=f(S_1)=f(S_0)$ since $f(S)$ is well behaved. However since $ g(S_0)=0 $ we do not set $g(S_2)=g(S_1)=g(S_0)$. Thus 
\begin{eqnarray}
C_{\Phi_{2}}-C_{\Phi_{1}}=f(S_0)\left(\frac{1}{g(S_2)}-\frac{1}{g(S_1)}\right).
\end{eqnarray} 
Following this logic one derives,
\begin{eqnarray}
\frac{C_{\Phi_2}-C_{\Phi_1}}{T_0 Q(\alpha_2-\alpha_1)}=\frac{f(S_0)}{T_0 h(S_0)}=\frac{4\pi^{\frac{3}{2}}(\pi+S_0-S_0\Omega^2)^{\frac{3}{2}}(\pi+S_0)^{\frac{1}{2}}}{\sqrt{S_0}\Omega[3(\pi+S_0)^2-S_0\Omega^2(2\pi+3S_0)]}
\label{eh2}
\end{eqnarray}
and, similarly,
\begin{eqnarray}
\frac{Q(\alpha_{2}-\alpha_{1})}{Q(k_{T_2}-k_{T_1})}=\frac{h(S_0)}{k(S_0)}=\frac{4\pi^{\frac{3}{2}}(\pi+S_0-S_0\Omega^2)^{\frac{3}{2}}(\pi+S_0)^{\frac{1}{2}}}{\sqrt{S_0}\Omega[3(\pi+S_0)^2-S_0\Omega^2(2\pi+3S_0)]}.
\label{eh3} 
\end{eqnarray}

Remarkably we find that the divergence in $C_{\Omega}$ is canceled with that of $\alpha$ in the first equation and the same is true for the case of $\alpha$ and $k_T$ in the second equation. From (\ref{31}, \ref{2ndlh}, \ref{eh2}, \ref{eh3}) the validity of the Ehrenfest's equations is established. Hence this phase transition in Kerr-AdS black hole is a genuine second order transition.

\section{\label{thergeo}Thermodynamic geometry in Kerr-AdS black hole}

Recently, analysis of black hole phase transitions from the point of view of thermodynamic state space (Ruppeiner) geometry has drawn some attention. In this approach Ruppeiner metric is defined by the Hessian of the entropy and this gives the pair correlation function \cite{Ruppeiner:1995zz, Ruppeiner:1983zz}. Also, the invariant length on the thermodynamic state space, defined by the Ruppeiner metric, when exponentiated, gives the probability distribution of fluctuations around the maximum entropy state \cite{Aman:2003ug}. Using the tools of Riemannian geometry, Ruppeiner curvature scalar ($R$) is defined and this is interpreted as the correlation volume multiplied by some proportionality constant \cite{Ruppeiner:2008kd}. Normally it is believed that the idea of correlation length has its root in the microscopic details of the system. But, remarkably, in the Ruppeiner geometry, a purely thermodynamic quantity ($R$) is claimed to serve the same purpose as the correlation length. This interpretation has been found quite successful in describing many second order phase transitions where this curvature scalar diverges at the critical point \cite{Shen:2005nu,Sahay:2010wi}. Now we examine whether such an interpretation holds for phase transition in Kerr AdS black hole.

The Ruppeiner metric is defined as \cite{Ruppeiner:1995zz}
\begin{equation}
dS^2=g_{ij}^{R}dX^idX^j
\label{rupp}
\end{equation} 
where, $g_{ij}=-\frac{\partial^2{S(X^k)}}{\partial X^i\partial X^j},~~{\textrm{and}}~~X^i\equiv X^i(M,N^a)$. Here $N^a$'s are all other extensive variables of the system. For Kerr AdS black hole $N^a=J$. In order to find $g_{ij}$ it is desirable to express $S$ in terms of $M$ and $J$. However from (\ref{M2}) we see that M is expressed as a function of $S$ and $J$ which is not invertible. In fact in this situation we can calculate the Weinhold metric which is defined in the following way \cite{Weinhold}, 
\begin{equation}
dS_{W}^2=g_{ij}^{W}dX^i dX^j
\label{wein}
\end{equation}
where $g_{ij}^{W}=\frac{\partial^2{M(X^k)}}{\partial X^i\partial X^j},~~{\textrm{and}}~~X^i\equiv X^i(S,J)$. It is well known that Ruppeiner metric and Weinhold metric are related by a conformal factor \cite{Mru, Ferrara}
\begin{equation}
dS_{R}^2=\frac{1}{T}dS_{W}^2
\label{confrel}
\end{equation}
where $T$ is the temperature of the system. In our example this would correspond to the Hawking temperature of the Kerr AdS black hole. Now using (\ref{M2}) we can easily calculate $dS_{W}^2$ and from that we can get $dS_{R}^2$ by using (\ref{confrel}) and (\ref{temp}). An explicit form of $dS_{R}^2$ is given by
\begin{equation}
dS_{R}^2=g_{SS}dS^2+2g_{SJ}dSdJ+g_{JJ}dJ^2
\label{metricth}
\end{equation}
where
\begin{eqnarray}
g_{SS} &=&- \frac{1}{2}\left(-\frac{3}{S}-\frac{1}{\pi+S}-\frac{S(2\pi+3S)}{4J^2\pi^3+S^2(\pi+S)}+\frac{4S(\pi^2+6\pi S+6S^2)}{-4J^2\pi^4+S^2(\pi+S)(\pi+3S)}\right)\nonumber\\
g_{SJ} &=& g_{JS}=-4J\pi^3\left(-\frac{1}{4J^2\pi^3+S^2(\pi+S)}+\frac{2\pi}{4J^2\pi^4-S^2(\pi+S)(\pi+3S)}\right)\label{compon}\\
g_{JJ} &=& -\frac{4\pi^3S^3(\pi+S)^2}{(4J^2\pi^3+S^3+S^2\pi)(-4J^2\pi^4+S^2(\pi+S)(\pi+3S))}\nonumber
\end{eqnarray}
By definition Ruppeiner curvature is constructed exactly like the Riemannian curvature and for two dimensions the curvature scalar is given by \cite{Ruppeiner:1995zz},
\begin{equation}
R=-\frac{1}{\sqrt g}\left[\frac{\partial}{\partial S}\left(\frac{g_{SJ}}{\sqrt{g}g_{SS}}\frac{\partial g_{SS}}{\partial J}-\frac{1}{\sqrt g}\frac{\partial g_{JJ}}{\partial S}\right)+\frac{\partial}{\partial J}\left(\frac{2}{\sqrt g}\frac{\partial g_{SJ}}{\partial J}-\frac{1}{\sqrt g}\frac{\partial g_{SS}}{\partial J}-\frac{g_{SJ}}{{\sqrt g}g_{SS}}\frac{\partial g_{SS}}{\partial S}\right)\right]
\label{curv}
\end{equation}
Considering the metric (\ref{compon}) we now calculate $R$ as a function of $S$ and $J$. Finally using the relation (\ref{26}), one can find $R$ as a function of $S$ and $\Omega$. Since these expressions are quite lengthy we do not give those results here, instead we plot $R$ with $S$ for different values of $\Omega$, as shown in figure 6.3. 
\begin{figure}[h]
\centering
\includegraphics[angle=0,width=14cm,keepaspectratio]{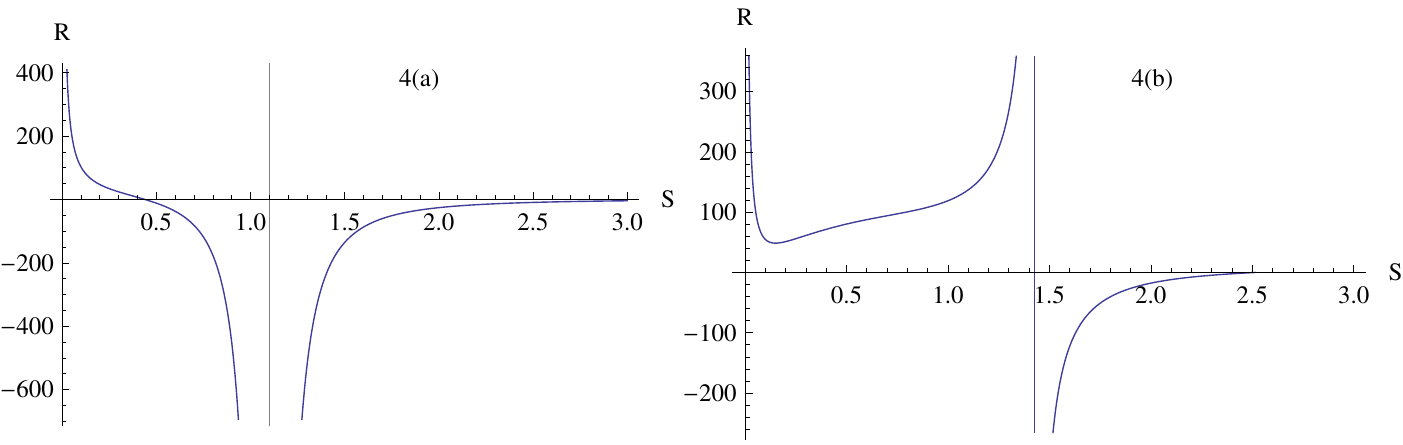}
\caption[]{{Thermodynamic scalar curvature ($R$) vs entropy ($S$) plot: $\Omega=0.3$ in 6.3(a) and $\Omega=1.5$ in 6.3(b).}}
\label{figure 5}
\end{figure}

Let us now explain the different plots one by one:\\
(i) In figure 6.3(a) we take $\Omega=0.3$ and this curve shows a divergence at $S_{\textrm{crit}}=1.0976$ which is exactly the case where $C_{\Omega}$ was discontinuous. Since we have already shown that the nature of this phase transition is second order it confirms the common belief that the divergence of $R$ means occurrence of second order phase transition.\\
(ii) In figure 6.3(b) we see that $R$ is also divergent at $S=1.4$ for $\Omega=1.5$. Note that this choice for $S$ and $\Omega$ is compatible with the consistency condition (\ref{cond}). But we do not find any discontinuity in $C_{\Omega}$ at this point and therefore there is no phase transition occurring here. This clearly shows that for this case it is incorrect to associate the divergence of $R$ with a phase transition.\\
Though $R$ is divergent in both figures 6.3(a) and 6.3(b), it is clear that the nature of divergences are completely different. While in figure 6.3(a) the divergent sector of the curve is symmetric with respect to the $R$ axis shifted at the singular point ($S_{\textrm{crit}}=1.0976$), in the other figure it is antisymmetric at the singular point. Thus from our analysis we conclude that a divergence of $R$ that is similar to figure 6.3(a) signals a second order phase transition in the Kerr AdS black hole.

\section{\label{discpt}Discussions}
To summarise, in this chapter we adopted the standard formalism, based on Clapeyron's and Ehrenfest's scheme used in conventional thermodynamic systems, for black holes. According to Clapeyron's formalism a discontinuity in entropy is necessary for a first order phase transition. However the semiclassical entropy of the Kerr-AdS black hole was continuous and thus the possibility of a first order phase transition was absent. The discontinuity in the specific heat ($C_{\Omega}$) of the Kerr-AdS black hole suggested the possibility of a higher order phase transition and motivated us to check the validity of Ehrenfest's equations. We derived such elations for rotating black holes. For a true second order phase transition these equations must be satisfied. We discovered that for all allowed values of the angular velocity ($0<\Omega<1$), where the critical point lied in the physical domain, the Ehrenfest's equations were satisfied.

The differences in the application of Ehrenfest's scheme to black holes vis-a-vis conventional thermodynamical systems were highlighted. The infinite divergences appearing on the phase transition curves were discussed. We developed a method for checking the validity of Ehrenfest's relations infinitesimally close to the critical point. As a remarkable fact it was found that the infinite divergences of various physical quantities ($C_{\Omega},~\alpha,~k_T$) cancel each other eventually leading to a confirmation of both Ehrenfest's relations.

Another aspect of our work was an attempt to connect the state space geometry with the phase transition of the Kerr AdS black hole. We calculated the Ruppeiner curvature scalar ($R$) and investigated its behaviour at the critical point of a second order phase transition ($\Omega < 1$). It was found that $R$ always diverged at this critical point. This was compatible with a pattern, suggested by individual studies carried out for various thermodynamic systems (for a review see \cite{Ruppeiner:1995zz}), that the divergence of $R$ is a characteristic of a second order phase transition. However, a divergence in $R$ was also noted for $\Omega\geq 1$ where no phase transition occurs. This indicated a deviation from the afore-stated pattern where a divergence in $R$ signalled a phase transition. But it must be stressed that the nature of the divergence of $R$ in the latter ($\Omega\geq 1$) case was found to be completely different from the previous ($\Omega < 1$) case. Thus our study revealed that only a particular type of singularity (symmetric type) in $R$ can correctly locate the phase transition point for the Kerr-AdS black hole.

\chapter{\label{chap:conclusions} Conclusions}
This thesis is dedicated to study some aspects of black hole thermodynamics by using the concepts of quantum mechanics and conventional thermodynamics. In our journey we derived some known results using new techniques and also found some new results with fresh insights. In the following we summarize our findings for each of the last five chapters.

In the second chapter, we provided a brief introduction to Komar conserved charges.  Then we generalised the original work of Smarr \cite{smarr} to obtain the mass formula for arbitrary dimensional Reissner-Nordstrom black holes. However this approach could not be generalised to the rotating case which motivated us to perform an alternative analysis. In this analysis we evaluated the Komar conserved charges corresponding to appropriate Killing vectors at any arbitrary point on or outside the event horizon. This may be compared with conventional methods where these conserved charges are calculated either at the horizon or at infinity. Using these results, we found the conserved quantity ($K_{\chi^{\mu}}$) at the event horizon for the null Killing vector ($\chi^{\mu}$) which was a combination of two vectors which are, respectively,  asymptotically time-like and space-like. Interestingly, $K_{\chi^{\mu}}$ satisfies the remarkable relation $K_{\chi^{\mu}}=2ST$ for any dimensions. The striking nature of $K_{\chi^{\mu}}=2ST$ was that in spite of the explicit dimensional dependence of all individual quantities ($K_{\chi^{\mu}}$ (\r{lhs}), $S$ (\r{entr}), $T$ (\r{temp})) the relation itself was completely dimension independent. Furthermore, no explicit black hole parameters (like mass, angular momentum or charge) appeared in this relation. As an application of this identity we also derived the generalized Smarr formula for charged Myers-Perry black holes in any dimension. This new identity is local and thus can be interpreted as a local version of the inherently non-local Smarr formula. Several consistency checks were performed which reassured the validity of this new formula. Using Euler's theorem on homogeneous functions the compatibility of the generalised Smarr formula with the first law of black hole thermodynamics was demonstrated. There are three appendices presented at the very end of the second chapter. In first two appendices (\ref{appendix2A1}, \ref{appendix2B}) we provide relevant information about the Myers-Perry and Kerr-Newman spacetimes. In the final appendix (\ref{appendix2C}) dimensional reduction ($N+1$ dimensions to two dimensions) of the Myers-Perry black holes was briefly discussed.

 In the third chapter, we discussed Hawking effect by using the method of quantum tunneling. Following a WKB type approach we derived Hawking temperature by considering tunneling of scalar particles and fermions. This particular method, for studying Hawking effect, was provided by Padmanabhan {\it et al} (\cite{Paddy}). We extend this methodology by applying it to a axisymmetric, non-static (KN) spacetime and also by going beyond the standard semi-classical approximation. In order to do the latter we used the method of Banerjee and Majhi (\cite{Majhibeyond}) and generalized it for the above spacetime. We include the higher order in $\hbar$ terms in the WKB ansatz which resulted into higher order corrections to the Hawking temperature with some unknown coefficients that were subsequently discussed in details in chapter \ref{chap-entropy} 
 
We then followed a density matrix technique and derived the blackbody radiation spectrum for both black hole (event) horizon and cosmological horizon in arbitrary dimensions. This method was first provided in (\cite{Majhiflux}) and was concerned only to (3+1) dimensional black hole (event) horizons. There are three appendices at the end of the third chapter. In appendices (\ref{appendix3A}, \ref{Appendix3B}) we provided Kruskal extensions of black hole event horizons and cosmological horizons respectively. In the last appendix (\ref{Appendix3C}) an outline for identifying various modes was provided.

In the fourth chapter, we introduced a new and simple approach to derive the ``first law of black hole thermodynamics'' from the thermodynamical perspective where one does not require the ``first law of black hole mechanics''. The key point of this derivation was the observation that ``black hole entropy'' is a {\it {state function}}. In the process we obtained some relations involving black hole entities, playing a role analogous to {\it {Maxwell's relations}}, which must hold for any stationary black hole. Based on these relations, we presented a systematic calculation of the semi-classical Bekenstein-Hawking entropy taking into account all the ``work terms on a black hole''. This approach is applicable to any stationary black hole solution. The standard semi-classical area law was reproduced. An interesting observation that emerged from the calculation was that the work terms did not contribute to the final result of the semi-classical entropy.

We then extended our method for calculating entropy in the presence of higher order corrections to the Hawking temperature found in chapter \ref{chap:hawking-effect}. However this result involved a number of arbitrary constants. Demanding that the corrected entropy be a state function it was possible to find an appropriate form of the corrected Hawking temperature. By using this result we explicitly calculated the entropy with higher order corrections. In the process we again found that work terms on black hole did not contribute to the final result of the corrected entropy. This analysis was done for the Kerr-Newman spacetime and it was trivial to find the results for other stationary spacetimes like (i) Kerr, (ii) Reissner-Nordstrom and (iii) Schwarzschild by taking appropriate limits. It is important to note that the functional form for the corrected entropy is same for all the stationary black holes. The logarithmic and inverse area terms as leading and next to leading corrections were quite generic up to a dimensionless pre-factor.

It was shown that the coefficient of the logarithmic correction was related with the trace anomaly of the stress tensor and explicit calculation of this coefficient was also done. This was a number ($\frac{1}{90}$) for both Schwarzschild and Kerr black hole. The fact that both Kerr and Schwarzschild black holes have identical corrections was explained on physical grounds (the difference between the metrics being purely geometrical and not dynamical) thereby serving as a nontrivial consistency check on our scheme. It may be noted that the factor ($\frac{1}{90}$) was also obtained (for the Schwarzschild case) in other approaches \cite{Hawkzeta, Fursaev} based on the direct evaluation of path integrals in a scalar background. For the charged spacetime (Reissner-Nordstrom and Kerr-Newman) the coefficients were not pure numbers, however in the $Q=0$ limit they reproduced the expressions for the corresponding charge-less versions.  

The fifth chapter was devoted to study the thermodynamics of non-commutativity inspired Schwarzschild black hole. Using a graphical analysis we derived the modified area law which is valid for the entire nonextremal regime. Our result was valid up to all order effects of the non-commutative parameter ($\theta$). The higher order corrections due to $\theta$ involved exponentials, as well as error functions. We then found a more general expression for the entropy of NC Schwarzschild black hole by considering both NC and quantum effect (only upto leading orders) terms. We again found a logarithmic correction in the leading order, however, the coefficient of this correction was modified due to the NC effect. In the $\theta\rightarrow 0$ limit all known results for the Schwarzschild black hole in commutative space were recovered.

Finally, in the sixth chapter we presented a new formulation to exhibit and classify phase transitions in black holes. It was based on an application of Clapeyron's and Ehrenfest's ideas to black hole systems.  According to these well known methods for a first order phase transition first order derivatives (entropy and volume) of Gibbs energy is discontinuous and they satisfy the Clausius-Clapeyron equation. A famous example of this is liquid to vapour phase transition. Similarly, for a second order phase transition, the second order derivatives of $G$ ({\it i.e.} specific heat, volume expansion coefficient and compressibility) are discontinuous and two Ehrenfest's relations are satisfied \cite{stan, zeman}.
 
We first exploited a gibbsian approach and derived Ehrenfest's relations for rotating black holes. We then explicitly considered the Kerr black holes defined in AdS spacetimes. It was shown that for Kerr-AdS black holes there exists a phase transition from the lower to higher mass branch, which is not first order. We established the validity of Ehrenfest relations. Thus this phase transition is genuine second order. An alternative method, based on thermodynamic geometry was also executed, which reconfirmed the occurrence of the above mentioned phase transition. Our analysis thus strongly suggested that black holes are indeed governed by the concepts of thermodynamics.  
   
There are some issues which are worthy for working out in future. For example, generalizing our method for deriving an appropriate mass formula (similar to Smarr mass formula) and first law for black holes in more general theories of gravity (e.g. Lanzcos-Lovelock (LL) gravity) could be possible. For that we first need to find out the analog of Komar conserved charges in LL gravity. From that we would also be able to check whether or not the identity $K_{\chi^{\mu}}=2ST$ is valid for LL theory. Another important project is the study of black hole phase transition in more detail. Particularly, finding an order parameter description and connecting it with the AdS/CFT framework are important future prospects.

\backmatter




\end{document}